\documentclass[twocolumn]{aastex63}
\usepackage[T1]{fontenc}
\usepackage{times,color,natbib,graphicx,multirow,amsmath,amsfonts}

\def\idm#1{{\mbox{\scriptsize #1}}}
\def\vec#1{{\boldsymbol{#1}}}
\newcommand{\mJ}{\mbox{m}_{\idm{Jup}}}
\newcommand{\mb}{\mbox{m}_{\idm{b}}}
\newcommand{\mc}{\mbox{m}_{\idm{c}}}
\newcommand{\md}{\mbox{m}_{\idm{d}}}
\newcommand{\me}{\mbox{m}_{\idm{e}}}
\newcommand{\Pb}{P_{\idm{b}}}
\newcommand{\Pc}{P_{\idm{c}}}
\newcommand{\Pd}{P_{\idm{d}}}
\newcommand{\Pe}{P_{\idm{e}}}
\newcommand{\au}{\mbox{au}}
\newcommand{\Mmean}{\mathcal{M}}
\newcommand{\phase}{t_{\idm{phase}}}
\newcommand{\rotation}{\omega_{\idm{rot}}}
\newcommand{\posterior}{\mathcal{P}}
\newcommand{\rinner}{r_{\idm{inner}}}
\newcommand{\D}{\mathcal{D}}

\newcommand{\Mb}{\mathcal{M}_{\idm{b}}}
\newcommand{\Mc}{\mathcal{M}_{\idm{c}}}
\newcommand{\Md}{\mathcal{M}_{\idm{d}}}
\newcommand{\Me}{\mathcal{M}_{\idm{e}}}
\newcommand{\M}{\mathcal{M}}
\newcommand{\eb}{e_{\idm{b}}}
\newcommand{\ec}{e_{\idm{c}}}
\newcommand{\ed}{e_{\idm{d}}}
\newcommand{\ee}{e_{\idm{e}}}
\newcommand{\omegab}{\varpi_{\idm{b}}}
\newcommand{\omegac}{\varpi_{\idm{c}}}
\newcommand{\omegad}{\varpi_{\idm{d}}}
\newcommand{\omegae}{\varpi_{\idm{e}}}
\newcommand{\yr}{\mbox{yr}}
\newcommand{\Ym}{\langle Y \rangle}
\newcommand{\msun}{\mbox{M}_{\odot}}

\newcommand{\mj}{\,\mbox{m}_{\idm{Jup}}}

\def\hr8799{{HR~8799}}

\def\methods{Appendix}
\def\gpi{GPI}
\def\gravity{GRAVITY}
\def\gaia{GAIA}
\def\mcoa{{MCOA}}
\def\megno{{MEGNO}}
\def\alma{{ALMA}}
\def\po{PO}
\def\keck{KECK}
\def\hst{HST}
\def\rms{\mbox{RMS}}
\def\sphere{VLT/SPHERE}
\def\lbt{LBT}
\def\subaru{SUBARU}
\def\mcmc{MCMC}
\def\kepler{Kepler}

\definecolor{myred}{rgb}{0.9,0.1,0.1}
\definecolor{mybrown}{rgb}{0.24,0.18,0.06}

\newcommand\corr[1]{{\color{mybrown}\bfseries #1}}
\renewcommand\corr[1]{#1}

\begin{document}

%
\shorttitle{An exact Laplace resonance in the HR~8799 planetary system}
\title{An exact, generalised Laplace resonance in the HR~8799 planetary system}
\shortauthors{Go\'zdziewski \& Migaszewski}
\author[0000-0002-8705-1577]{Krzysztof Go\'zdziewski}
\author[0000-0003-2546-6328]{Cezary Migaszewski}
\affil{
Faculty of Physics, Astronomy and Informatics, Nicolaus Copernicus Univ., Grudzi\c{a}dzka 5, Toru\'n, Poland
}
%
\begin{abstract}
A system of four super-Jupiter planets around \hr8799{} is the first multi-planet configuration discovered via the direct imaging technique. Despite over decade of research, the system's architecture remains not fully resolved. The main difficulty comes from still narrow observing window of $\sim 20$~years that covers small arcs of orbits with periods from roughly $50$ to $500$~years.  Soon after the discovery it became clear that unconstrained best-fitting astrometric configurations self-disrupt rapidly, due to strong mutual gravitational interactions between $\simeq 10$-Jupiter-mass companions. Recently, we showed that the \hr8799{} system may be long term stable when locked in a generalized Laplace 8:4:2:1 mean motion resonance (MMR) chain, and we constrained its orbits through the planetary migration. Here we qualitatively improve this approach by considering the MMR in terms of an exactly periodic configuration. This assumption enables us to construct for the first time  the self-consistent $N$-body model of the long-term stable orbital architecture, \corr{using only available astrometric positions of the planets relative to the star. We independently determine planetary masses, which are consistent with thermodynamic evolution, and the parallax overlapping to $1\sigma$ with the most recent GAIA DR2 value}. We also determine the global structure of the inner and outer debris discs in the [8, 600]~\au{} range, consistent with the updated orbital solution. 
\end{abstract}


%
\section{Introduction}
%
Several approaches are being used to detect extrasolar planets. Indirect methods, such as the radial velocity \citep{Mayor1995}, transits \citep{Henry2000}, timing \citep{Wolszczan1992}, classic astrometry \citep{Muterspaugh2010} rely on studying radiation of the central star, while the planets themselves are not observed. The imaging technique detects the planets directly, given their own infra-red (IR) radiation. This method, limited by the contrast, stability and resolution of the images  is sensitive for massive and young planets in wide orbits. Therefore, even for a nearby star \hr8799{} located $\sim 40$~pc from the Sun \citep{Gaia2018} it is only possible to detect the planets with long periods of $10^2$--$10^3$ years, as discovered by \cite{Marois2008, Marois2010}. This makes the orbits determination a very difficult task. The measured astrometric positions of the planets relative to the star are typically uncertain to a few $0.001''$ (mas), but this is still not sufficient to uniquely constrain the orbits. Qualitatively different architectures are consistent with the present observations  \citep[e.g.][]{Wertz2017}. \corr{Astrometric (purely geometric) orbital models are strongly unstable \citep[e.g.,][]{Konopacky2016,Wang2018}, however, there are also reported dynamically tuned configurations} which, though hardly chaotic in the Lyapunov exponent sense, can survive for hundreds of Myrs \citep{Gotberg2016}, comparable with the age of \hr8799{} of $\simeq 30$ -- $60$~Myr \citep{Marois2010,Wilner2018}.

On the other hand, a resonant or near resonant system resulting from the convergent migration was shown to explain the observations as well \citep[][furthermore, GM14 and GM18, respectively]{Gozdziewski2014,Gozdziewski2018}. We justified a rigorously stable 8:4:2:1 MMR as the most likely architecture on the dynamical grounds, consistently with the recent studies in \citep{Konopacky2016,Wang2018}. They independently found, imposing \mcmc{} dynamical priors, that coplanar orbits near the 8:4:2:1 MMR result in orders of magnitude more stable orbits than any other scenario, and provide adequate fits to the measurements. In the present work, we extend the MMR hypothesis by linking the putative resonance chain with the planetary $N$-body periodic solutions \citep[][see also citations therein]{Hadjidemetriou1976,Hadjidemetriou1981}. In the later paper, they computed families of periodic orbits (POs) for the 4-body planetary system, and applied the results to the Galilean moons of Jupiter. This follows de Sitter's mathematical theory of the Laplace resonance in a Newtonian framework. De Sitter found a family of stable \po{}s as the Poincaré orbits of the second kind \citep{Broer2016}. These authors also propose that librations (quasi-periodic solutions) near these \po{}s may provide realistic explanation of the observations. We apply a similar reasoning to the \hr8799{} system.

Because the 8:4:2:1 MMR chain in the \hr8799{} system generalizes the Laplace resonance \citep{Papaloizou2015}, there is a fairly obvious link of this prior resonant model found in GM14 and GM18,  with a \po{} interpreted as the MMR center. Here, similarly to \citep{Hadjidemetriou1981}, we consider a planetary system as a \po{} when the osculating orbital elements of the orbits are periodic in time w.r.t. non-uniformly rotating reference frame tied to the osculating apsidal line of a selected planet. The present study is inspired by our finding that the 8:4:2:1 MMR configurations fitting the observations of \hr8799{} are in fact very close to an exact \po{} of the 5-body system.

Periodic configurations are known to result from smooth convergent migration in systems of two and more planets \citep[e.g.][]{Beauge2006,Migaszewski2015}.  In our numerical simulations of \corr{convergent} migration (see GM14 and GM18), three-- and four--planet MMR chains of the Laplace resonance appear naturally, in wide ranges of the migration time scales and planet masses.  The migration quickly drives planets to long-term stable systems, typically in a few Myr time scale. This indicates that the 8:4:2:1 MMR capture \corr{may be} an efficient process, weakly dependent on physical conditions in the protoplanetary disk. Simultaneously, stable systems are confined to tiny, isolated islands in the orbital parameter space, as narrow as $\simeq 0.5$~\au{} in semi-major axes and $\simeq 0.05$ in eccentricity (GM14 and GM18), and that reflects deterministic character of the migration. Also the increased eccentricities of the innermost planets observed in the best-fitting, stable solutions and in our simulations is consistent with an early evolutionary period of convergent inward migration of all four planets, trapping them pairwise in 2:1~MMRs and pumping of the orbital eccentricities while in resonance lock \citep{Wang2018,Yu2001}. 

Our new method improving the migration constrained optimisation (\mcoa{}) in GM14 and GM18 relies on two attributes of periodic configurations. When modelling the data with stable \po{}, the long-term dynamical stability is guaranteed per-se. Also, similarly to \mcoa{}, instead of exploring large, $n$-dimensional parameter space, where the number of free parameters $n=23$ for a co-planar system of four planets, including their masses and the parallax, we may limit the optimisation to a $p$-dimensional manifold embedded in this space; here, as explained below, $p=11$, or -- if the masses and the parallax are fixed (given a'priori) -- $n=18$ vs. $p=6$. Therefore the MMR (periodic) constraint makes it possible to substantially reduce the number of free parameters characterising the orbital configuration, and to avoid degeneracy caused by a small ratio of the data points to the degrees of freedom.

But the advantage of the \po{}--constrained method over the standard orbit fitting lies not only in the reduction of the parameter space. In modeling a generic planetary configuration, the orbital elements as well as the planetary masses must be all free parameters, independent one of another. Therefore, the masses cannot be determined from the astrometric observations, unless they sufficiently map the orbits or are sufficiently precise to make it possible to detect mutual gravitational perturbations. In turn,
the orbital elements of a periodic (exactly resonant) solution strongly depend on the planet masses  and the total angular momentum of the system, assuming the linear scale of the system and the central star's mass to be given \citep{Hadjidemetriou1976,Hadjidemetriou1981}.  Therefore these critical parameters may be derived with relatively short orbital arcs, and independently of the planets' cooling theory.  As we also show below, because the \po{} impose tight timing on the orbital evolution, it is already possible {\em to indirectly measure the system parallax}. The planet masses and parallax determined from the relative astrometry which are self-consistent with the astrophysically fixed stellar mass, establish a test-bed and a benchmark for our hypothesis. 

We describe the results of the \po{} model of the \hr8799{} system in Sect.~\ref{section:laplace}, the global structure of debris discs in Sect.~\ref{section:discs}, and main conclusions in Sect.~\ref{section:summary}. The details and supplementary material are given in \methods{}.

\begin{table*}
\label{table:ic}
\caption{
The best-fitting, strictly periodic model of the \hr8799{} planetary system. Median values and the $1\sigma$ uncertainties resulting from the DE-MC sampling are given in top row of each parameter, while values in the bottom row with more significant digits, in order to closely reproduce the \po{} solution, correspond to the best-fitting parameters in terms of the $\chi^2$ statistics.  \corr{Gaussian planet mass priors from the hot-start evolutionary models are $(5.8 \pm 0.5)\,\mj$ for \hr8799{}b, and $(7.2 \pm 0.7)\,\mj$ for all other planets \citep{Wang2018}, and the parallax prior is $(24.22\pm 0.09)$~mas \citep{Gaia2018}.} The first part of the Table is for the primary fit parameters $\vec{x}$, and the last four rows are for the inferred osculating, astrocentric Keplerian elements at the epoch of $t_0=1998.829$, the first measurement in \citep{Lafreniere2009,Soummer2011}. The mass of the parent star $(1.52\pm 0.15)\,\msun$ \citep{Konopacky2016} is fixed at its nominal value, in order to avoid the mass--orbital scale correlation. For the model with $p\equiv \mbox{dim}\,\vec{x} =11$ free parameters, $\chi^2=142.93$ and an $\rms{}\simeq 6.7$~mas. 
}
\label{tab:tab1}
\centerline{
\begin{tabular}{l c c c c}
\hline
parameter/planet & \hr8799{}e & \hr8799{}d & \hr8799{}c & \hr8799{}b\\
\hline
\multirow{2}{*}{Planet mass, $m (\mJ)$} & $7.4 \pm 0.6$ & $9.1 \pm 0.2$ & $7.8 \pm 0.5$ & $5.7 \pm 0.4$ \\
 & $7.34688506$ & $8.97059370$ & $7.78986828$ & $5.85290522$\\
\hline
\multirow{2}{*}{Longitude of ascending node, $\Omega$ (deg)} & \multicolumn{4}{c}{$61.85 \pm 0.45$} \\
 & \multicolumn{4}{c}{$62.02658660$}\\
 \hline
\multirow{2}{*}{Inclination, $I$ (deg)} & \multicolumn{4}{c}{$26.4 \pm 0.3$} \\
 & \multicolumn{4}{c}{$26.55235715$}\\
\hline
\multirow{2}{*}{Parallax, $\Pi$ (mas)} & \multicolumn{4}{c}{$24.3 \pm 0.1$} \\
 & \multicolumn{4}{c}{$24.36337601$}\\
\hline
\multirow{2}{*}{Period ratio, $P_d/P_e$} & \multicolumn{4}{c}{$1.985 \pm 0.002$} \\
 & \multicolumn{4}{c}{$1.983096$}\\
\hline
\multirow{2}{*}{Scale factor, $\rho$} & \multicolumn{4}{c}{$1.054 \pm 0.002$} \\
 & \multicolumn{4}{c}{$1.05177401$}\\
\hline
\multirow{2}{*}{Relative Phase, $\phase$ (yr)} & \multicolumn{4}{c}{$331.42 \pm 0.13$} \\
& \multicolumn{4}{c}{$331.39813970$}\\
\hline
\multirow{2}{*}{Rotation angle, $\rotation$ (deg)} & \multicolumn{4}{c}{$157 \pm 1$} \\
 & \multicolumn{4}{c}{$156.38496284$}\\
\hline
\hline
\multirow{2}{*}{Semi-major axis, $a (\au)$} & $16.25 \pm 0.04$ & $26.67 \pm 0.08$ & $41.39 \pm 0.11$ & $71.6 \pm 0.2$ \\
 & $16.21068245$ & $26.59727940$ & $41.27484337$ & $71.42244964$\\
\hline
\multirow{2}{*}{Eccentricity, $e$} & $0.1445 \pm 0.0013$ & $0.1134 \pm 0.0011$ & $0.0519 \pm 0.0022$ & $0.016 \pm 0.001$ \\
 & $0.14421803$ & $0.11377309$ & $0.05273512$ & $0.01587597$\\
\hline
\multirow{2}{*}{Argument of pericentre, $\omega$ (deg)} & $111.2 \pm 0.6$ & $29 \pm 1$ & $92.7 \pm 0.7$ & $42.0 \pm 2.2$ \\
 & $111.05649395$ & $28.35795849$ & $92.75709369$ & $41.35621703$\\
\hline
\multirow{2}{*}{Mean anomaly, $\Mmean$ (deg)} & $-23.5 \pm 0.7$ & $60.8 \pm 0.8$ & $145.3 \pm 0.9$ & $-48.2 \pm 2.1$\\
 & $-23.88996589$ & $60.75229572$ & $144.93313134$ & $-47.73005711$\\
\hline
\end{tabular}
}
\end{table*}

\begin{figure*}
\centerline{
\hbox{
\includegraphics[height=0.4\textheight]{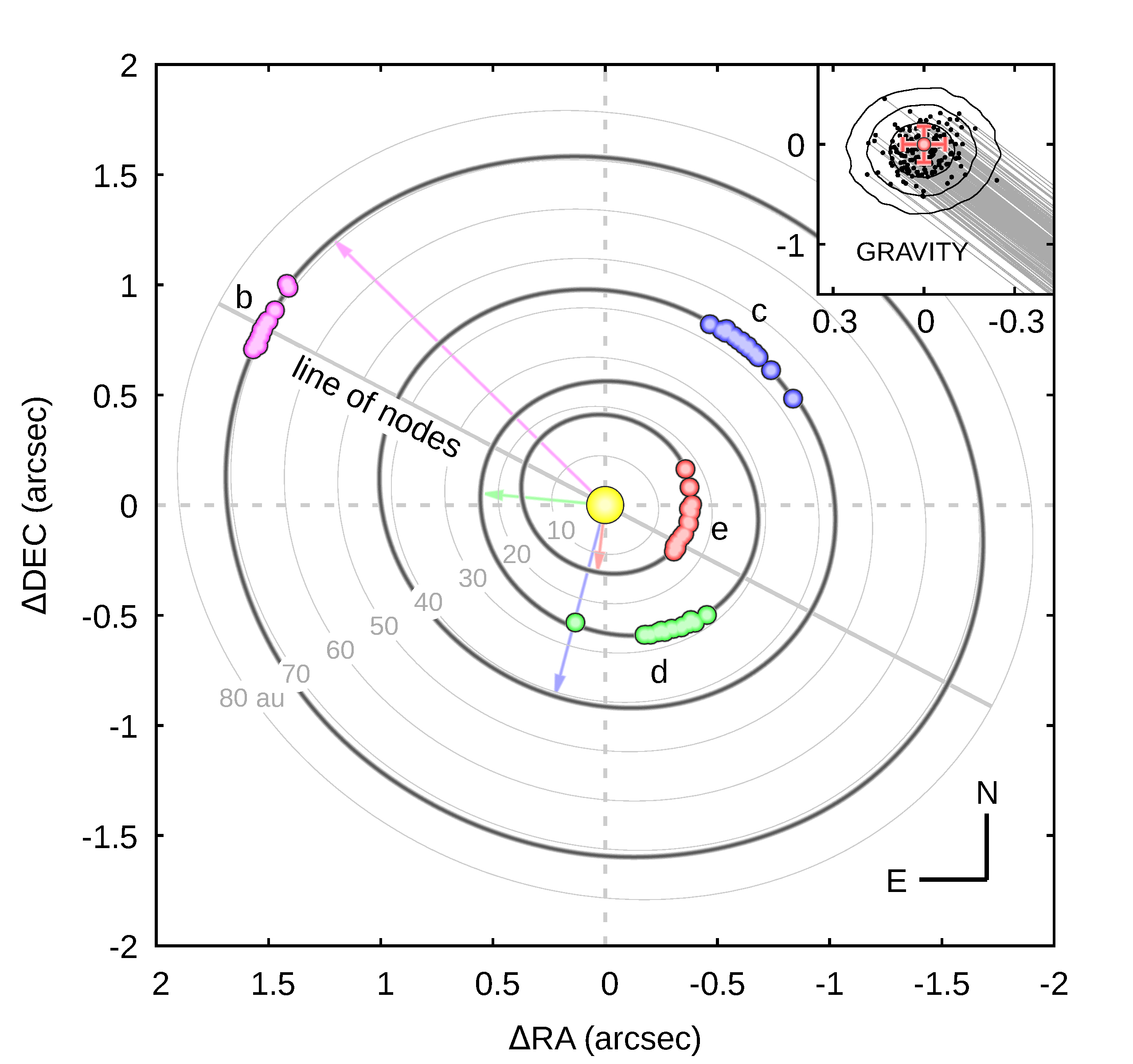}
\includegraphics[height=0.4\textheight]{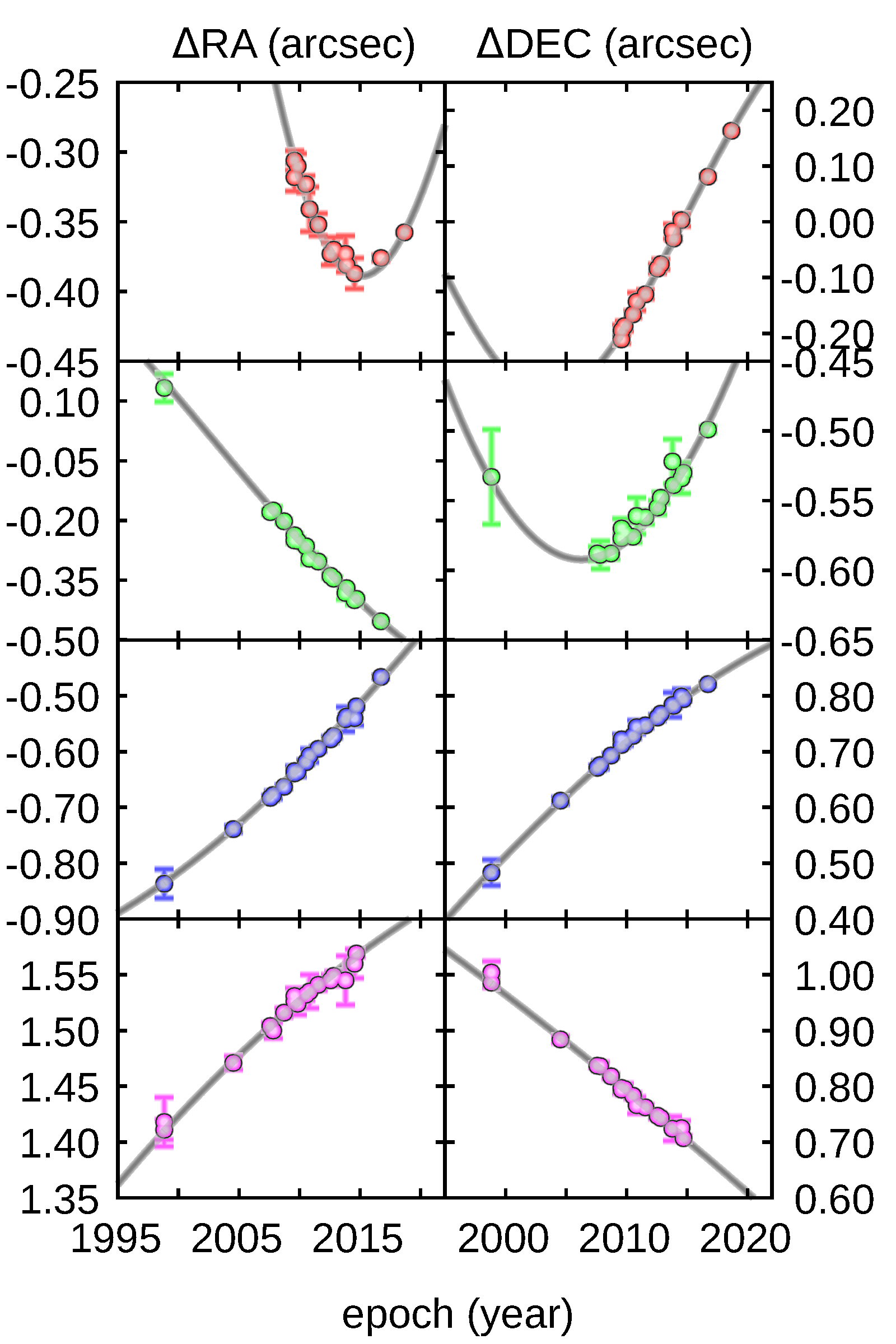}
}
}
\caption{
Astrometric observations (red, green, blue and magenta points for planets \hr8799{}e,d,c,b, respectively) in the set used for the analysis. The grey thick curves illustrate the best-fitting orbits (Tab.~\ref{tab:tab1}). The light grey curves in the left-hand panel mark referencing circular orbits of radii $10, 20, 30, \dots, 80\,\au$ in the orbital plane of the system. The red, blue, green and magenta arrows point to the periastron of each orbit. The close-up of the \gravity{} datum is illustrated in a small panel in the top-right corner of the left plot. Grey curves represent $200$ randomly chosen orbits from the DE-MC sampling, the black dots mark positions at the orbits in the epoch of the \gravity{} observation ($2018.656$). The graph is centred at the datum, the axes are expressed in mili-arcseconds (mas). Black curves are for $1\sigma$, $2\sigma$ and $3\sigma$ confidence intervals for the model position at the right ascension--declination, (RA, DEC)--plane at the epoch of $2018.656$, derived from the DE-MC sampling. The right-hand panel illustrates the observations and the model orbits as the time-functions of RA and DEC.
}
\label{fig:fig1}
\end{figure*}

\begin{figure}
\centerline{
\includegraphics[width=0.45\textwidth]{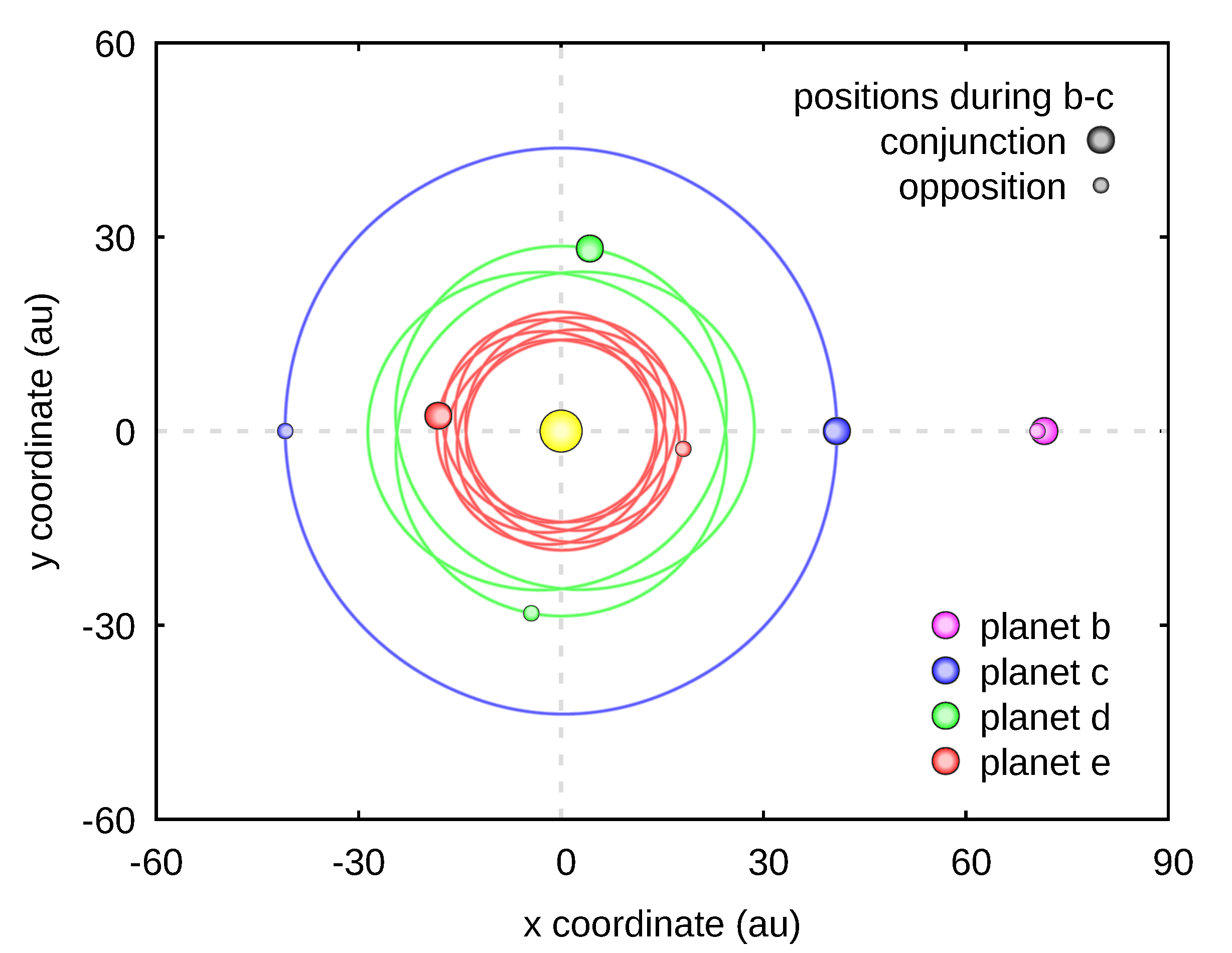}
}
\caption{
Astrocentric positions of the \hr8799{} planets over $1\,$Gyr $N$-body integration presented in the orbital plane co-rotating with \hr8799{}b. The red, green and blue curves illustrate the orbits of \hr8799{}e,~d~c, respectively. Big red, green, blue and~magenta symbols mark the positions of planets e, d, c and~b during the conjunction of planets b and~c, while smaller symbols denote the positions during their opposition. The yellow symbol at the origin marks the parent star.
}
\label{fig:fig2}
\end{figure}

\begin{figure*}
\centerline{
\vbox{
\hbox{
\includegraphics[height=0.28\textwidth]{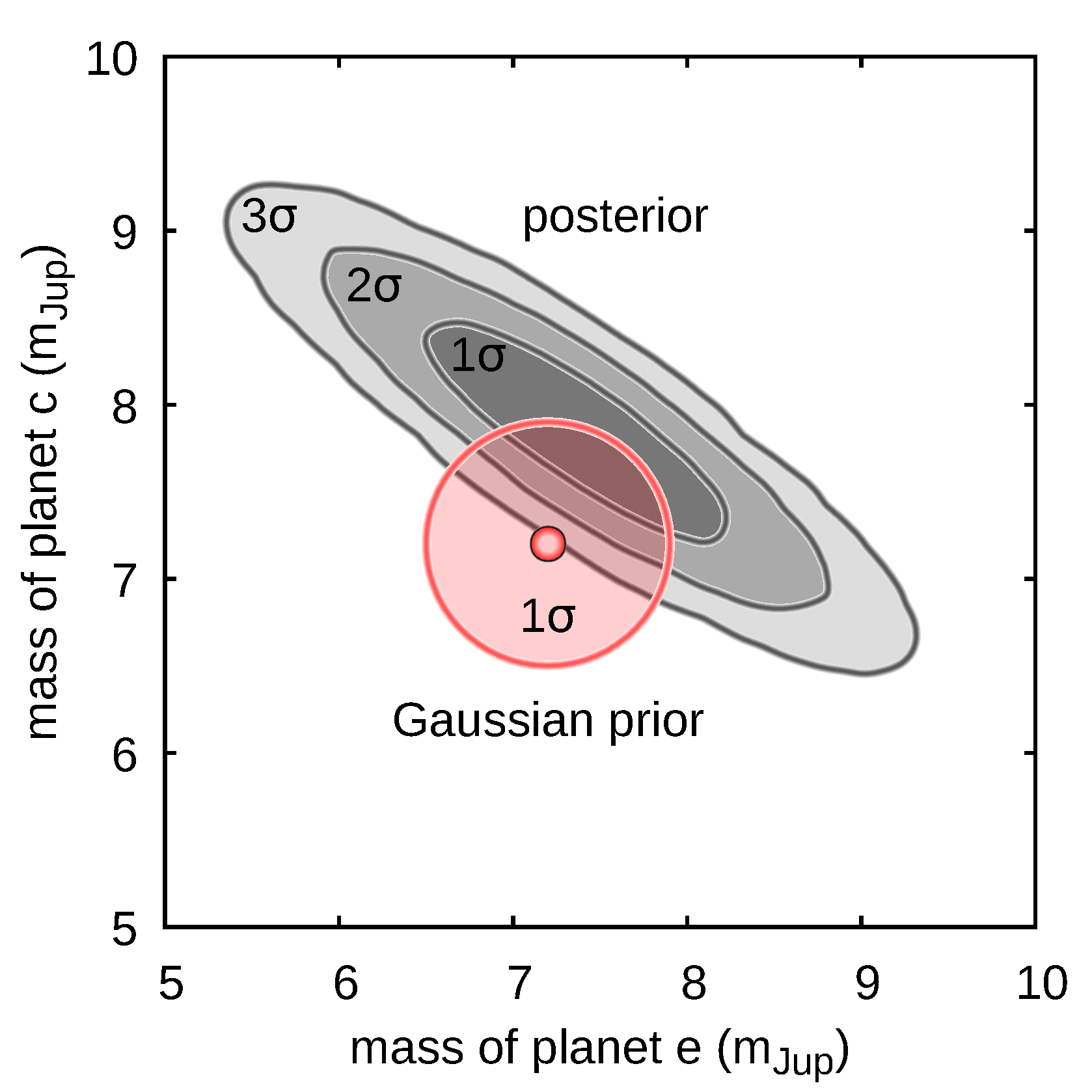}
\includegraphics[height=0.28\textwidth]{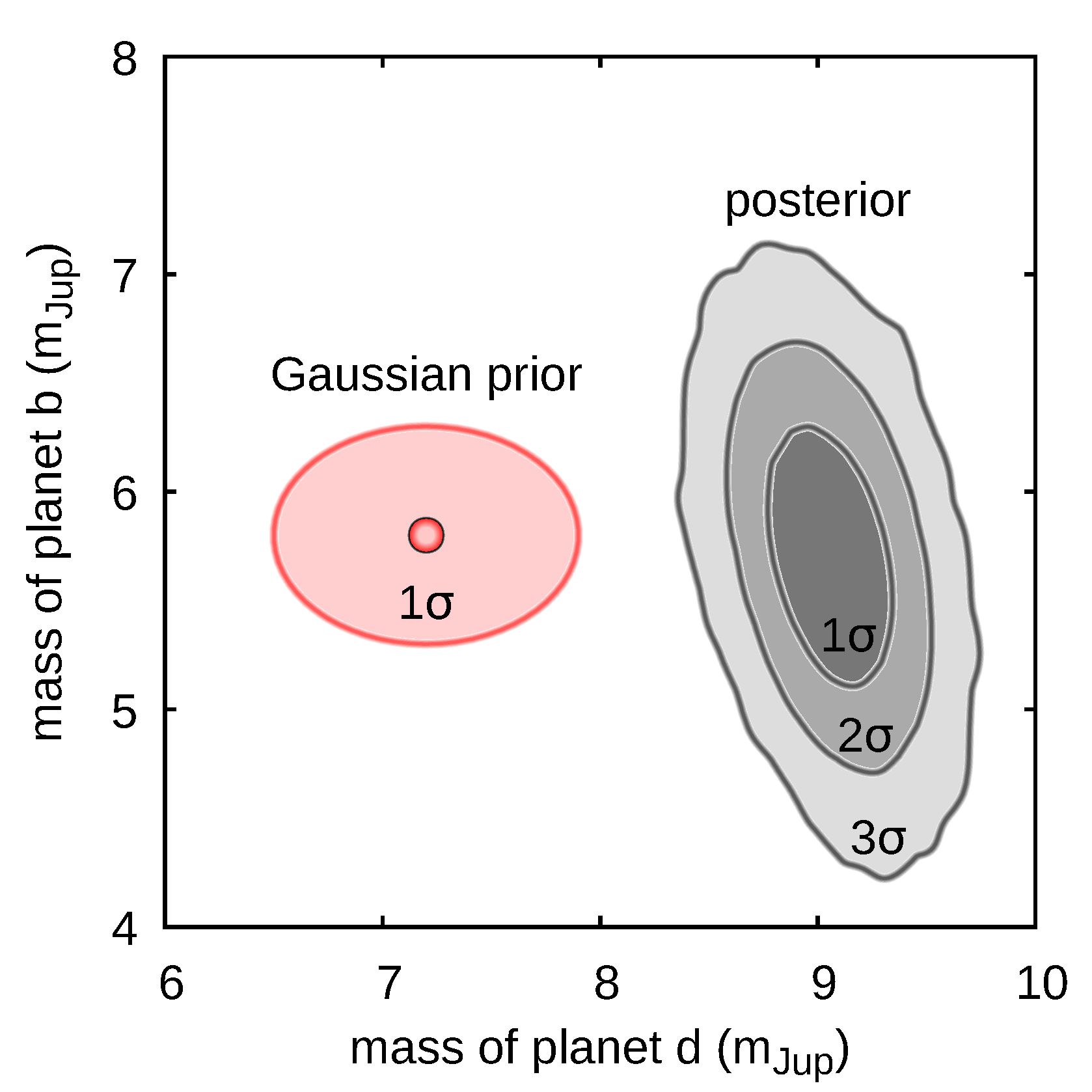}
}
\hbox{
\includegraphics[height=0.28\textwidth]{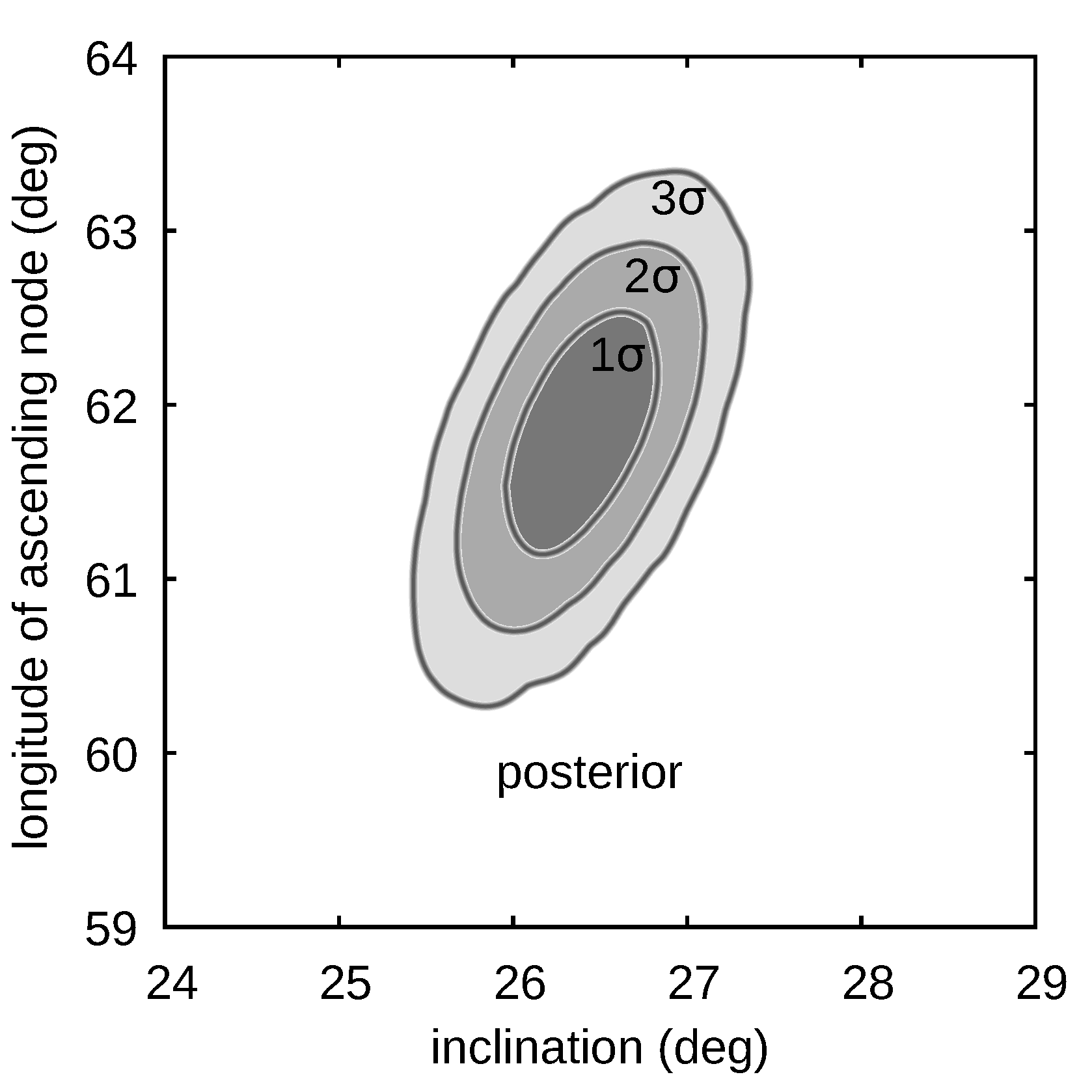}
\includegraphics[height=0.28\textwidth]{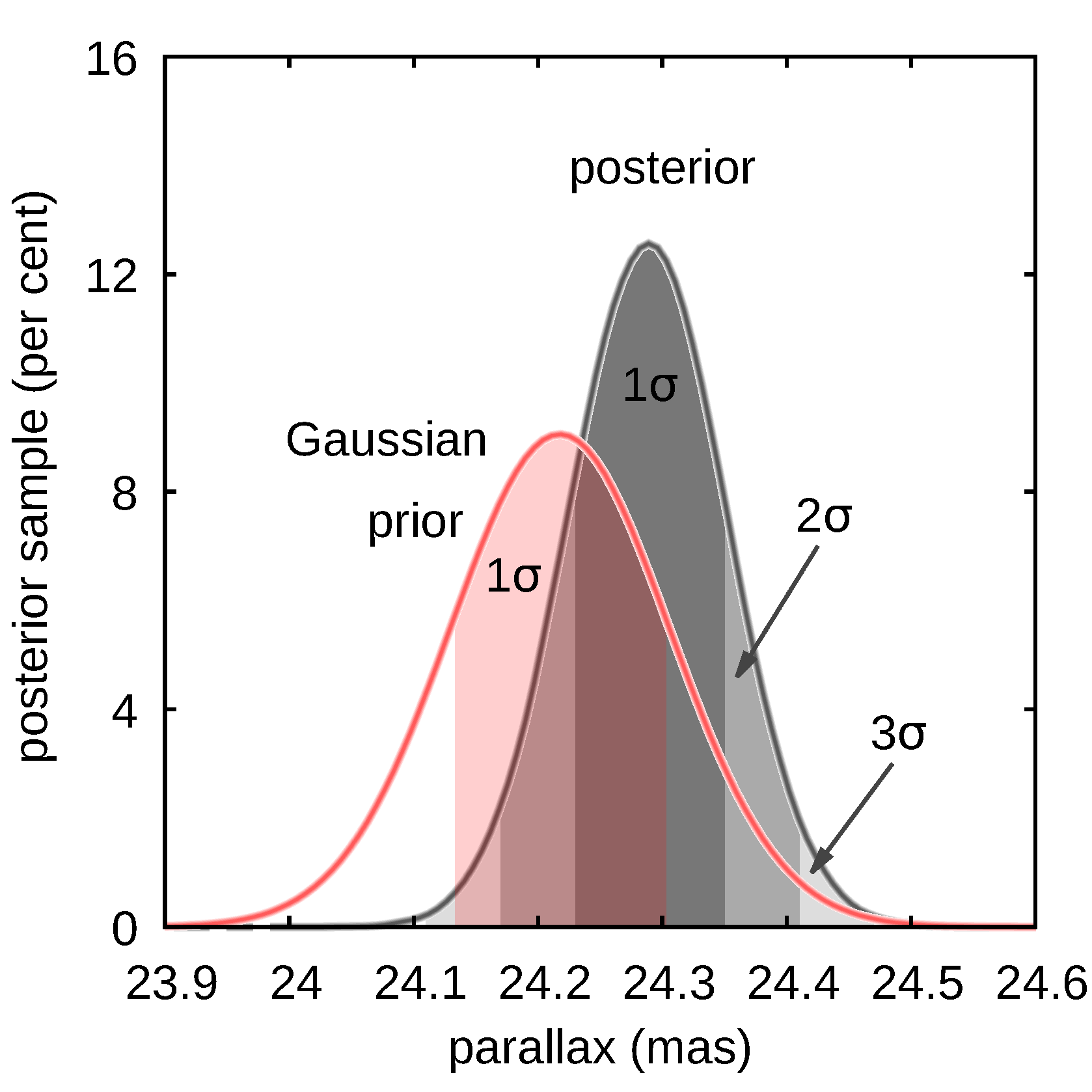}
}
}
}
\caption{
Posterior probability distributions of the planets' masses, inclination, the longitude of the ascending node, and the parallax. The shades of grey indicate the $1\sigma, 2\sigma$ and~$3\sigma$ confidence ranges of the parameters. Red symbols with circle/ellipse contours in the top panels show the astrophysical mass constraints \citep{Wang2018}, which are the priors in the DE-MC sampling. The red curve in the bottom-right panel is the prior put on the parallax, according to the GAIA DR2 catalogue \citep{Gaia2018}.
\corr{See also Table~\ref{tab:tab1} and its caption.}
}
\label{fig:fig3}
\end{figure*}

%
\section{Fitting the exact Laplace resonance}
%
\label{section:laplace}

In order to test the \po{} hypothesis, we used the most early \hst{} observations in \citep{Lafreniere2009,Soummer2011}, a homogeneous, uniformly reduced data in \citep{Konopacky2016}, as well as the most recent refined Gemini Planet Imager (\gpi{}) observations in \citep{DeRosa2020}, and the most accurate detection of \hr8799{}e in \citep{Gravity2019} with the \gravity{} instrument. This primary set does not contain all observations available in the literature, and we limited the data in order to reduce possible observational biases due to different instruments and pipelines, but to extend the observational window as much as possible. This approach follows our earlier work GM18 and also \cite{Wang2018}. The data set ${\cal D}$ consists of $N_\idm{obs}=65$ astrometric  planet positions (the right ascension RA$_i$ $\equiv \alpha_i$,  and declination DEC$_i$ $\equiv \delta_i$, $i=1,\ldots,N_\idm{obs}$) relative to the star with the mean uncertainty $\simeq 8$~mas. However, there is a particular datum from the \gravity{}  with $(\alpha,\delta)$ errors as small as 0.07~mas and 0.2~mas, respectively. This precision detection with the optical interferometry seems to be critical for constraining the best-fitting solutions. As the epoch $t_0$ of the osculating initial condition we chose the date of the first \hst{} observation $t_0=1998.829$ in \citep{Lafreniere2009,Soummer2011}. We also tested the \po{} model against all $N_{\idm{obs}}=127$ measurements available in the literature, as listed in \citep{Wertz2017}, and updated with newer or re-reduced \keck{}, \gpi{} and \gravity{} points (Appendix, Sect.~\ref{appendix:pogrid}). In both cases, the fitting results and conclusions closely overlap. 

Fitting a \po{} to the astrometric data is similar to our approach in GM14 and GM18, in which a migration-constrained co-planar solution is appropriately 
transformed to be consistent with the observations. Here, the optimisation process is essentially deterministic (fully reproducible), better constrained, regarding the masses and parallax as free parameters of the dynamical model, and CPU-efficient --- computations may be performed on a single workstation. Instead of simulating the migration, for given masses $\me, \md, \mc, \mb$ and $C$ or, equivalently, the period ratio  $\kappa = \Pd/\Pe$ of the inner pair of planets, we find a strictly periodic, co-planar resonant solution (see \methods{} for details), however with some arbitrary relative phases of the planets. Then, using the $N$-body dynamics scale invariance, the inferred ``raw'' semi-major axes are linearly re-scaled with the factor $\rho$, the orbital plane rotated to the sky plane by three Euler angles $(I, \Omega, \rotation)$, and planets propagated along the \po{} with the $N$-body integrator to epoch $\phase$ equivalent to the first observation epoch $t_0$. For given observation \corr{epochs} $t_i$, $i=1,\ldots,N_{\idm{obs}}$, Cartesian coordinates of the planets are re-scaled by the parallax $\Pi$ in order to obtain the angular positions
$(\alpha_i,\delta_i)$. Then the ephemeris may be compared with the observations. To quantify that, we construct the merit function, e.g., the common $\chi^2$ or other goodness-of-fit measure, such as the maximum likelihood function ${\cal L}$ or the Bayesian posterior $\posterior$. In the most general settings, the merit function depends on $11$ free parameters,
$\vec{x} \equiv (\me, \md, \mc, \mb, \kappa, \rho, I, \Omega, \omega, \phase, \Pi$), i.e., masses of the planets, period ratio of two inner planets, linear scale factor, Euler angles, the epoch and the parallax. At this point the data modelling becomes almost the standard optimization --- almost, and not really trivial, since we now must seek the best-fitting $\vec{x}$ constrained to a manifold of a stable \po{} family, representing the particular MMR chain.  One needs to find an extremum of the merit function, as well as to estimate uncertainties of the best-fitting parameters \citep[e.g.,][]{Gregory2010}. 

Details and variants of our experiments, regarding optimisation on the \po{} manifold, are described in \methods{}. Here, we quote  the final best-fitting parameters in Table~\ref{tab:tab1}. The first part of this Table shows the primary fit parameters $\vec{x}$, and the bottom part is for the derived, osculating astrocentric Keplerian elements at the epoch $t_0=$1998.829. Uncertainties are estimated with the help of the Differential Evolution Markov Chain (DE-MC) method \citep{TerBraak2006}. The astrometric data together with the best-fitting model are illustrated in Fig.~\ref{fig:fig1}. The left-hand panel shows the (RA, DEC)-diagram, with a close-up of \gravity{} datum \citep{Gravity2019}. In this zoom, $200$ randomly chosen synthetic orbits are shown with grey curves, while black points denote the positions of the synthetic solutions in the epoch of the observation. Grey oval contours mark $1\sigma$, $2\sigma$ and $3\sigma$ confidence intervals stemming from the DE-MC sampling. The right-hand panel illustrates the RA and DEC of the model and observations as functions of the epoch. The \po{} described in Table~\ref{tab:tab1} yields the reduced $\chi^2_\nu \simeq 1.24$ for $p=11$ free parameters, $N_{\idm{obs}}=65$, $\nu=119$ and the $\rms{} \simeq 6.7$ mas compares to the mean uncertainty of the measurements $\simeq 8$~mas. It adequately explains the data in a statistic sense. In particular, the time-- and sky-plane-- synchronisation of the model with the \gravity{} datum (left panel in Fig.~\ref{fig:fig1}) is apparently perfect.

The orbital evolution of this best-fitting system, integrated for $1$~Gyr, is presented in Fig.~\ref{fig:fig2}. This figure shows orbits of \hr8799{}e, \hr8799{}d and~\hr8799{}c in a reference frame co-rotating with \hr8799{}b. All trajectories are closed, consistent with the periodic evolution of the system. The positions of the planets are shown only in epochs of conjunctions between \hr8799{}b and \hr8799{}c (big filled circles) as well as their oppositions (small circles). Both the conjunctions and the oppositions repeat in the same pattern. The system is then an exact 8:4:2:1 MMR chain, consisting of triple two-body 2:1~MMRs of subsequent pairs of planets, with librations of the critical angle of the zero-th order 4-body MMR $\phi_{8:4:2:1} = \lambda_{\idm{e}}-2\lambda_{\idm{d}}-\lambda_{\idm{c}}+2\lambda_{\idm{b}}$ (where $\lambda{\idm{b,c,d,e}}$ are the mean longitudes of the planets)  with a small amplitude $\simeq 4$~degrees
(Fig.~\ref{fig:figA2}). It is worth noting that while the mean orbital osculating period ratios are $\simeq 2.03$, $\simeq 2.08$ and $\simeq 2.17$ for the innermost to outermost pairs of planets, respectively, the canonical (proper) mean motion frequencies \citep{Morbidelli2001} ratios are equal to $1/2$, indicating exact 2-body 2:1 MMRs. Therefore the MMR chain is understood as the generalized Laplace resonance. 
\corr{In order to illustrate the long term stability of the model}, we computed dynamical maps in terms of the Mean Exponential Growth factor of Nearby Orbits \citep[\megno{} aka $\Ym$,][]{Cincotta2003} for each planet. The integration interval of 10~Myrs translates to $\simeq 20,000$ outermost orbits, sufficient to detect short-term, MMR-induced instability. Remarkably, the maps (Fig.~\ref{fig:figA3}) are similar to our earlier Fig.~9 in GM14, illustrating the \mcoa{} model build upon much narrower data window, and still consistent with the updated periodic model of the system.

Uncertainties of the parameters are illustrated in Fig.~\ref{fig:fig3} (also Figs.~\ref{fig:figA8} and \ref{fig:figA9} in \methods{}). Two top panels are for the mass--mass diagrams. Red points with shaded ellipses indicate Gaussian priors imposed on the masses consistent with the hot-start cooling theory \citep[][]{Wang2018}, while grey filled contours denote $1\sigma$, $2\sigma$ and $3\sigma$ confidence intervals of the posterior probability distributions. Apart of the \hr8799{}d mass, the posterior closely fits with the astrophysical constraints. The bottom-left panel shows the posterior distribution of the orbital inclination and the longitude of ascending node. These parameters exhibit substantial correlations, yet much reduced thanks to the priors. The bottom-right panel is for the parallax, nominally agreeing to $\simeq 0.3\%$ with the GAIA DR2 value.  The Gaussian prior as the \gaia{} parallax \citep[][the red curve]{Gaia2018} closely overlaps with the DE-MC posterior.  

%
\section{Resonant structure of debris discs}
%
\label{section:discs}

\begin{figure}
\centerline{ 
\includegraphics[width=0.48\textwidth]{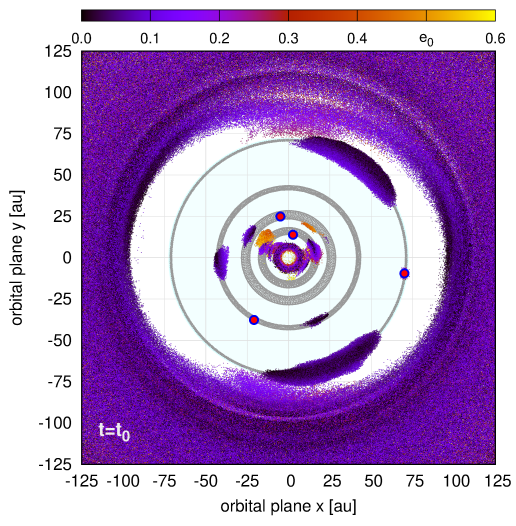}
}
\caption{\corr{
The inner part of the outer debris disk revealed by $\simeq 3.3\times 10^6 \Ym$-stable orbits found in the whole debris discs simulation, illustrated as a snapshot of astrocentric coordinates $(x,y)$ and osculating orbital eccentricities $e_{\rm 0}$ of these orbits at the initial epoch, color-coded and labeled in the top bar. Initial positions of planets are marked with filled circles. Gray rings illustrate their orbits integrated in a separate run for 10~Myr.}
}
\label{fig:fig4}
\end{figure}
The orbits of the planets likely share the common plane with the outer debris disk \citep{Matthews2014,Booth2016,Read2018, Wilner2018}.  Determination of the debris disk structure with the infra-red and millimetre observations is still not fully conclusive, both in terms of the orientation as well as the inner edge $\rinner$ of the disk \citep{Booth2016}. They argue that the structure of the disk might be a footprint of a fifth, yet unseen planet beyond \hr8799{}b. \cite{Read2018} proposed such an additional planet \hr8799{}f with the mass and semi-major axis of $0.1\mj$ and 138~\au{} that could predict the outer belt's edge and explain the \alma{} observations. Later, \cite{Wilner2018}, with observations at the Submillimeter Array at 1340 $\mu$, detected  the inner edge of the debris disk at $104 (+8,-12)\,\au$, and the disk extending to $\simeq 500\,\au$. They also constrained the mass of outer planet \hr8799{}b to $\simeq 6 \mj$. Remarkably, it is close to our best-fitting value. Furthermore, \cite{Geiler2019} found that a single, wide planetesimal disk does not reproduce the observed emissions and proposed a two-population model, comprising a Kuiper-Belt-like structure of low-eccentricity planetesimals and a scattered disk comprising of high-eccentricity population of comets. 

With the new, strictly resonant configuration of the four planets, including their updated masses and the parallax, we conducted preliminary $N$-body simulations resulting in $3.3\times 10^6$ small-mass asteroids, that reveal the global dynamical structure of the debris discs (see Appendix, Sect.~\ref{sec:calibration} for details). 
The inner border of the \corr{outer disk (Figs.~\ref{fig:fig4} and \ref{fig:figA15})} is significantly non-symmetric, with non-uniform density of asteroids, which may bias the disk orientation angles derived from simple models assuming the axial symmetry. The inner edge from our simulations agrees with the observational model of \cite{Wilner2018}. Moreover, we found a ring of high-eccentricity asteroids at $\simeq 140-160\,\au$ \corr{(Fig.~\ref{fig:figA15})}, close to the inner edge reported in \citep{Booth2016,Read2018}, which results in locally increased velocity dispersion. The velocity dispersion could impose higher dust production rate and  stronger emission, making the disk radial intensity profile no longer consistent with 
a simple power law. 

%
\section{Discussion and future research}
%
\label{section:summary}

Under the \po{} hypothesis, which is justified on the dynamical and the system formation grounds, the present astrometric data of the \hr8799{} planets make it possible to determine not only the parallax but also their masses, independently from the cooling theory. In order to illustrate this prediction, we simulated new synthetic observations around the best-fitting model in Table~\ref{tab:tab1} with fixed $\me = 7\,\mJ$, and Gaussian noise equal to the \gravity{} datum uncertainty.  We performed the $\chi^2$ minimisation without the planets' masses priors, adding new synthetic measurements after the last epoch of each planet. The resulting time-series of the best-fitting  $\me$ and its $1\sigma$ range indicate (Fig.~\ref{fig:figA1}) that with merely one more epoch $\simeq 2020.5$, all masses become meaningfully constrained \corr{without prior information}. If the \hr8799{} system is indeed represented by a \po{}, or a nearby stable resonant configuration, then it may be possible to determine the planets' masses basing solely on the relative astrometry. This could be a test-bed for the cooling theory of \hr8799{}--like, massive planets, and possibly other multiple planetary systems discovered via the direct imaging. The deterministic \po{} model may serve as a reference configuration useful for the astrometric and physical characterisation of such resonant or close to resonant systems.

\corr{The \po{} hypothesis may be naturally confronted with compact multiple \kepler{} and super-Earth systems which are predominately close to, but not actually inside of MMRs \citep[e.g.][]{Fabrycky2014}. The planetary migration might easily generate resonant states, but does not preferentially retain small planets in such states. From this perspective, the \po{} of \hr8799{} might not be necessarily preferred over near-MMR (possibly weakly chaotic) configuration, with the Lagrangian (geometric) stability timescale exceeding the age of the system. But the tight observational constrains invoked here seem to contradict that. Moreover, \cite{Ramos2017} argue that 2:1~MMR systems relatively distant from the star, such as HD 82943~and \hr8799{} are characterized by very small resonant offsets, while higher offsets are typical of short-period \kepler{} systems. Achieving an exact MMR versus near-MMR state likely depends on the differing efficacy of resonant retention of four enormous giant planets vs. much smaller \kepler{} planets and different formation of such systems. Wide-orbit systems require long formation timescales, furthermore inconsistent with type II migration characteristic for massive planets. Alternatively, pebble accretion initially accompanying type I migration \citep{Johansen2017} or new paradigms of type II migration \citep{Ida2018} may explain the putative MMR chain. Therefore the confirmed \po{} of the \hr8799{} planets could be the border condition and a footprint of the system migration, shedding more light on its uncertain origin.
}

\corr{As the bottom line we note that the self-consistent model of the \hr8799{} system and our predictions may be verified shortly, during the next few years}.

%
\section*{Acknowledgments}
%
We are very grateful to the anonymous reviewer whose comments improved the manuscript.
We thank the staff of the Pozna\'n Supercomputer and Network Centre (PCSS, Poland) for the generous long-term support and computing resources (grant No.~313).

\bibliographystyle{aasjournal}
\bibliography{ms}
\label{lastpage}
%

%
\appendix
%
\setcounter{section}{0}
\setcounter{figure}{0}
\renewcommand\thesection{A\Roman{section}}
\renewcommand\thefigure{A\arabic{figure}} 

\begin{figure}
\centerline{
\includegraphics[width=0.5\textwidth]{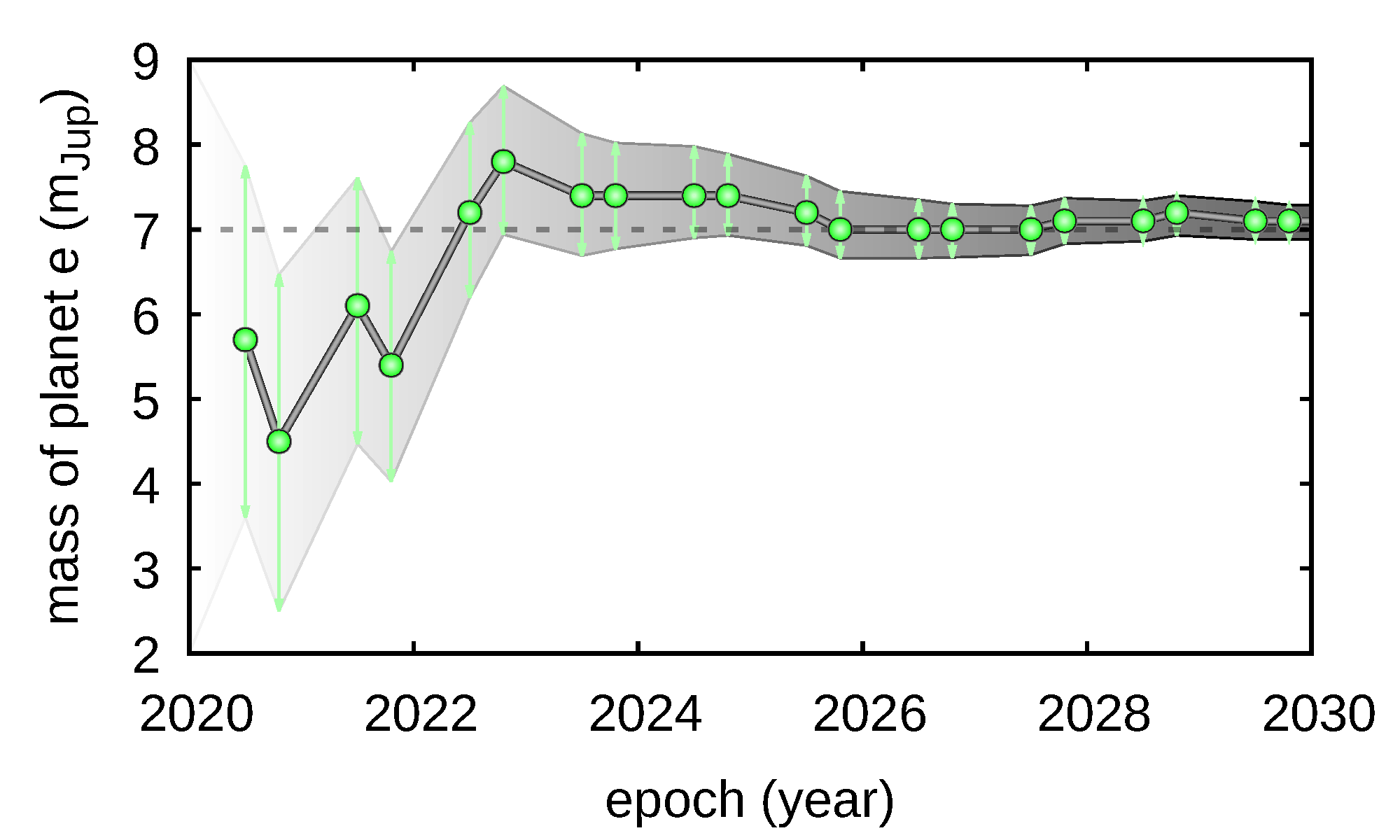}
}
\caption
{
The best-fitting value of the innermost planet with $1\sigma$ uncertainty for the data set with additional synthetic measurements given in subsequent epochs up to the year of $2030$. See the main text (Sect. 4) for details.
}
\label{fig:figA1}
\end{figure}

\begin{figure*}
\centerline{
\vbox{
\hbox{
\hbox{\includegraphics[height=0.2\textheight]{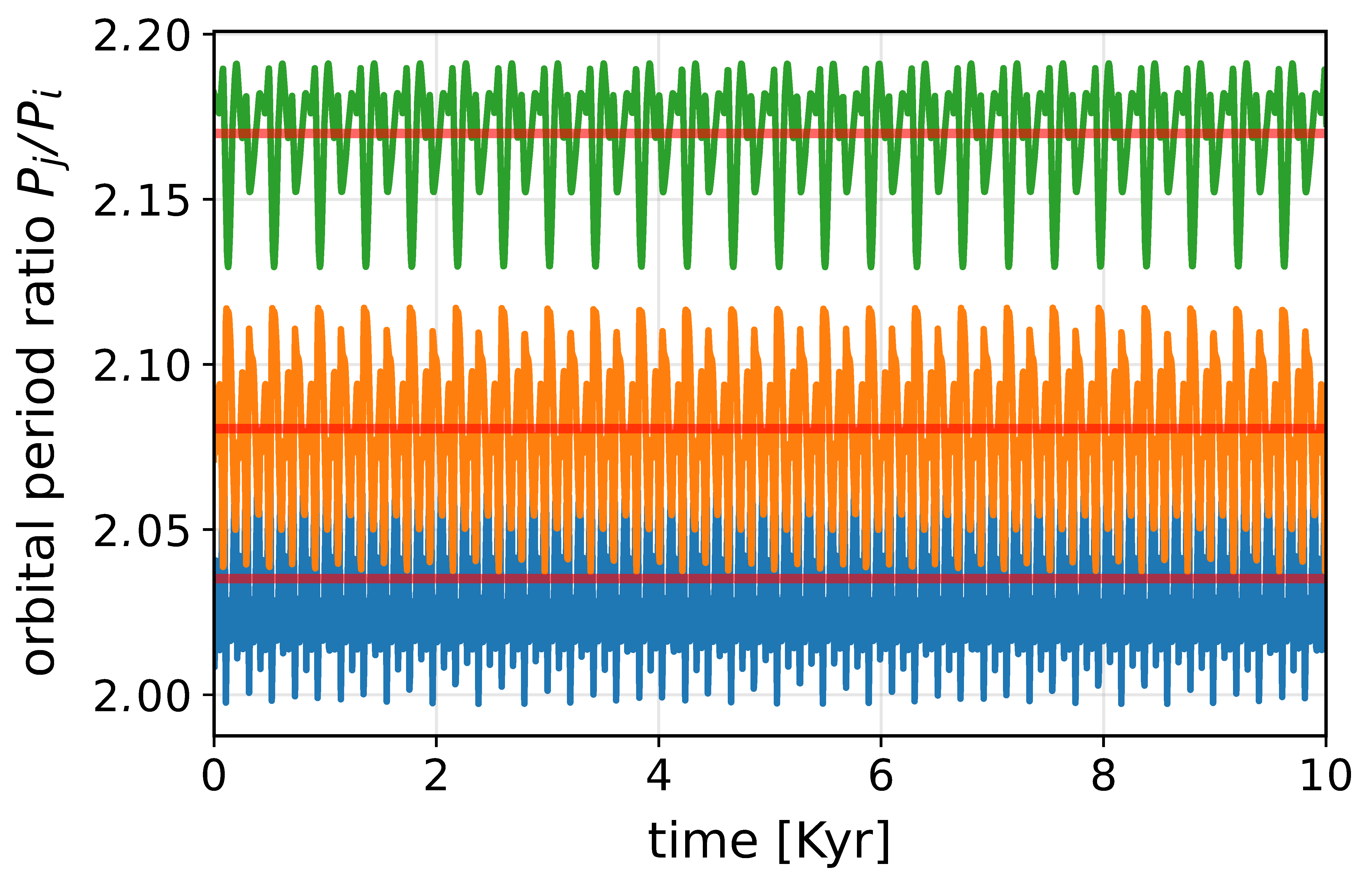}}
\quad
\hbox{\includegraphics[height=0.2\textheight]{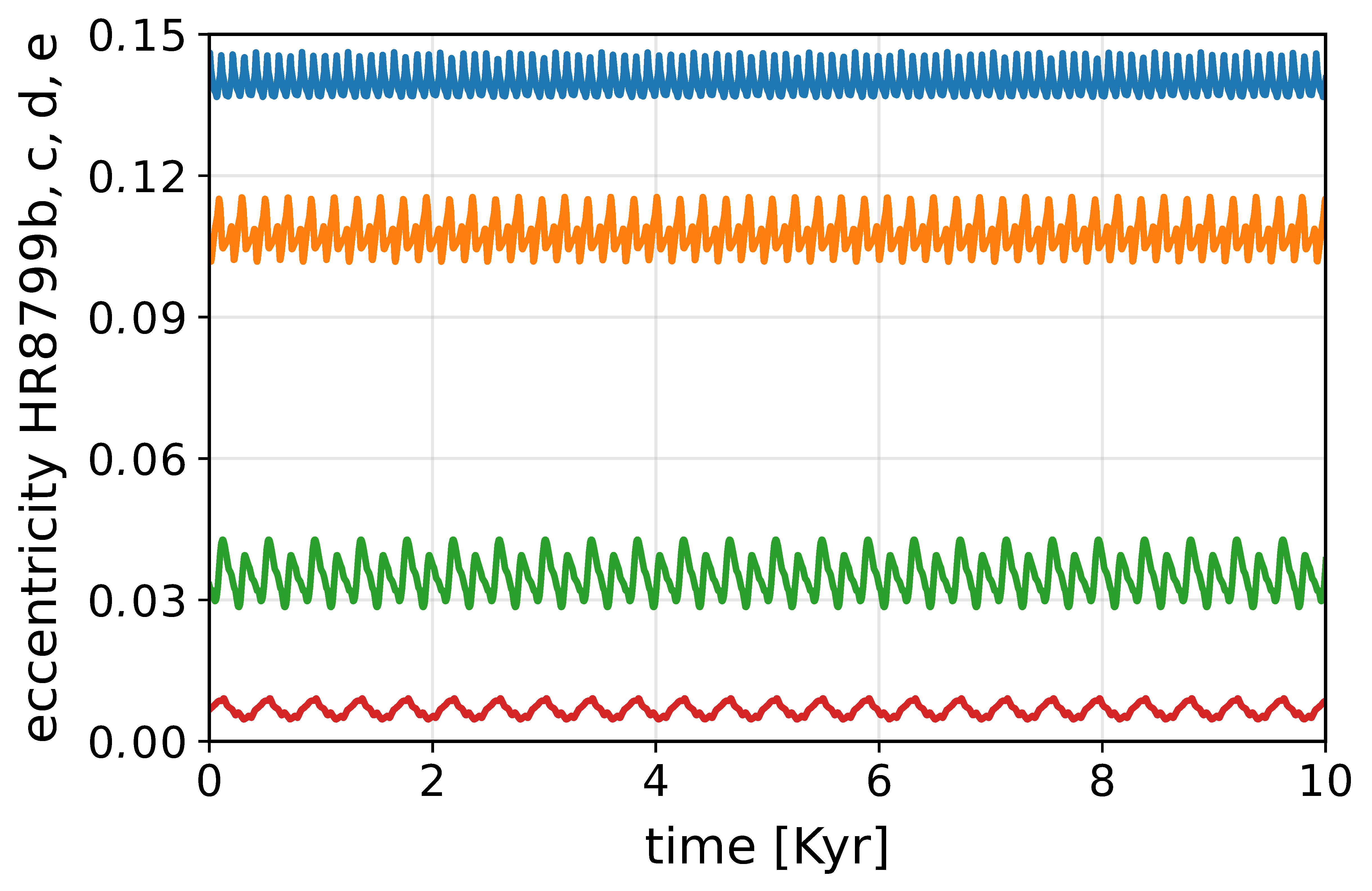}}
}
\hbox{
\hbox{\includegraphics[height=0.2\textheight]{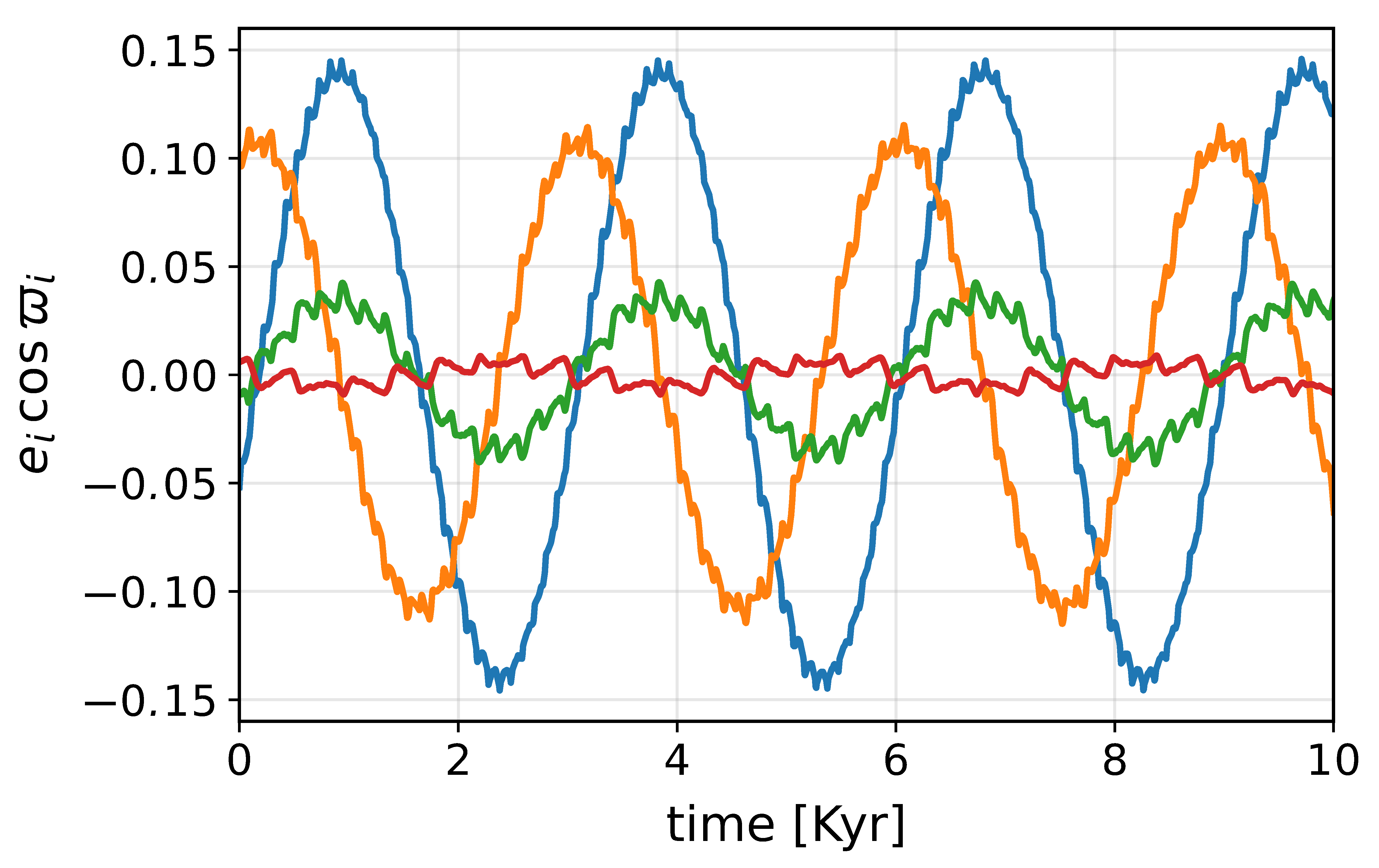}}
\quad
\hbox{\includegraphics[height=0.2\textheight]{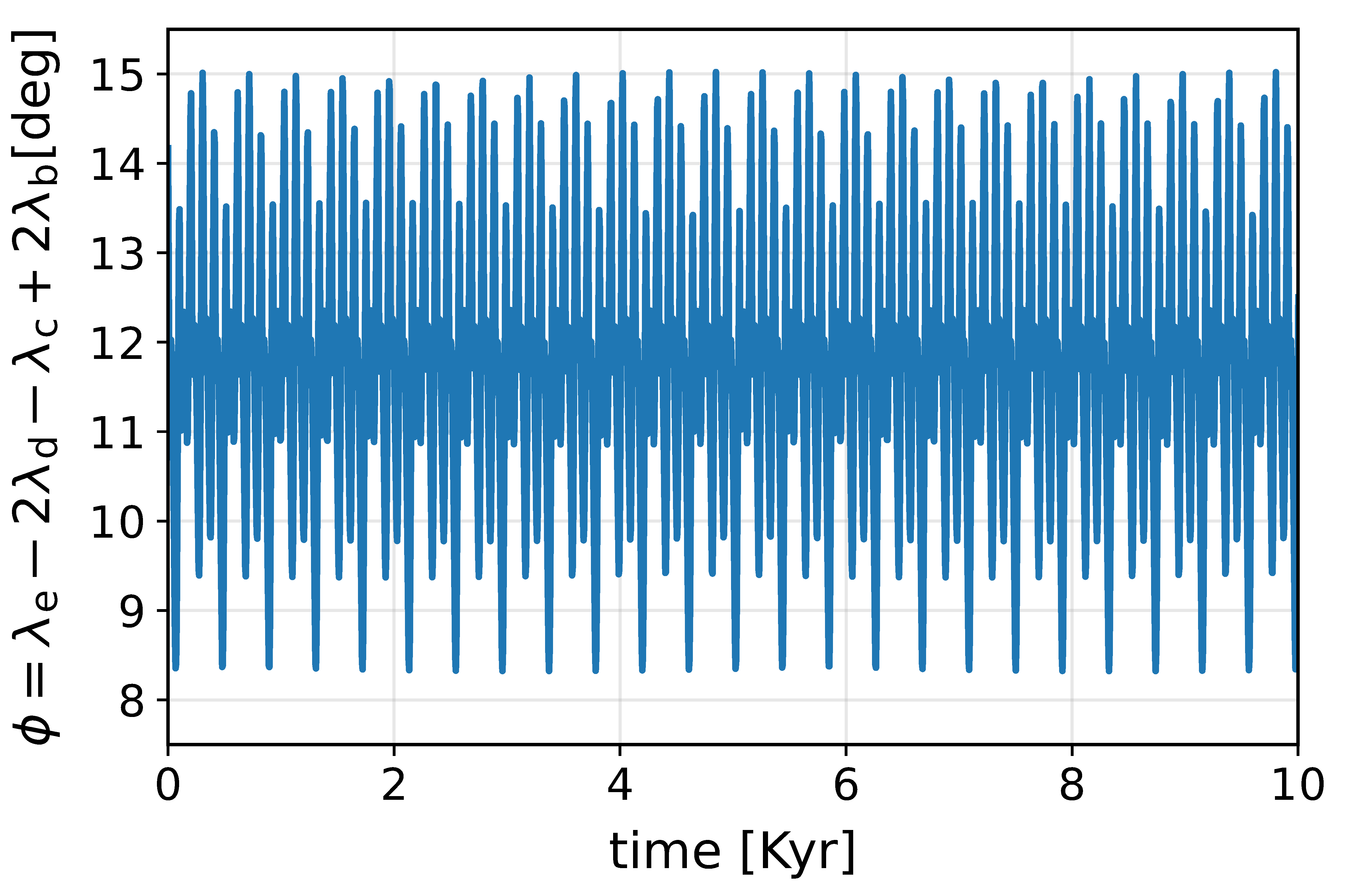}}
}
}
}
\caption{
Temporal evolution, for the first 10~Kyr, of the canonical, osculating orbital elements, expressed in the Jacobian reference frame, for  the \po{} configuration in Tab.~\ref{tab:tab1}.
Top-left panel: 
osculating period ratios for subsequent pair of planets, and their mean values (horizontal red lines) from top to bottom, $P_\idm{b}/P_\idm{c}\simeq 2.170$, $P_\idm{c}/P_\idm{d}\simeq 2.081$, and $P_\idm{d}/P_\idm{e}\simeq 2.035$. We note that the proper orbital periods, in the sense of the mean motions as fundamental frequencies \citep{Morbidelli2001}, expressed in Julian years of 365.25~days, are $52.36995$, $104.73989$, $209.47987$, and $418.96324$ for planets \hr8799{}e,d,c,b, respectively, forming an exact 2:1, 2:1, 2:1 MMR chain.
Top-right panel: eccentricities of the planets \hr8799{}e,d,c,b from top to bottom, respectively.
Bottom-left panel: one of the   elements $x_i \equiv e_i\cos\varpi_i$, ($i=$\hr8799{}b,c,d,e), used to compute the secular frequency of the apsides rotation. The second component of the quasi-periodic signal (not shown) is $y_i \equiv e_i\sin\varpi_i$. 
Bottom-right panel: the critical argument of the zero-th order, four-body generalized Laplace resonance for the same initial condition that librates around $\simeq 12$~degrees. 
} 
\label{fig:figA2} 
\end{figure*}

\begin{figure*}
\centerline{
\vbox{
\hbox{
\includegraphics[width=0.45\textwidth]{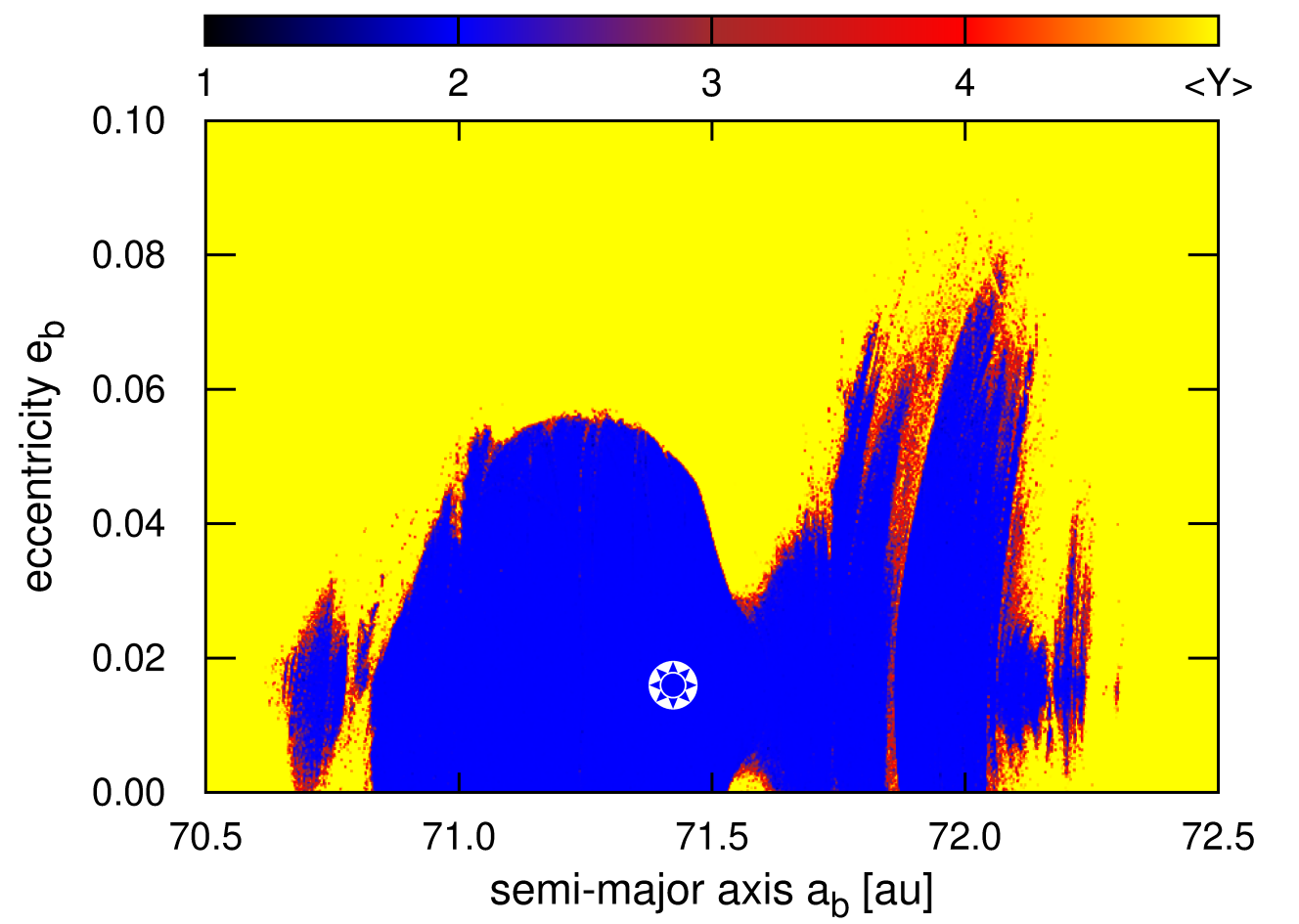}
\includegraphics[width=0.45\textwidth]{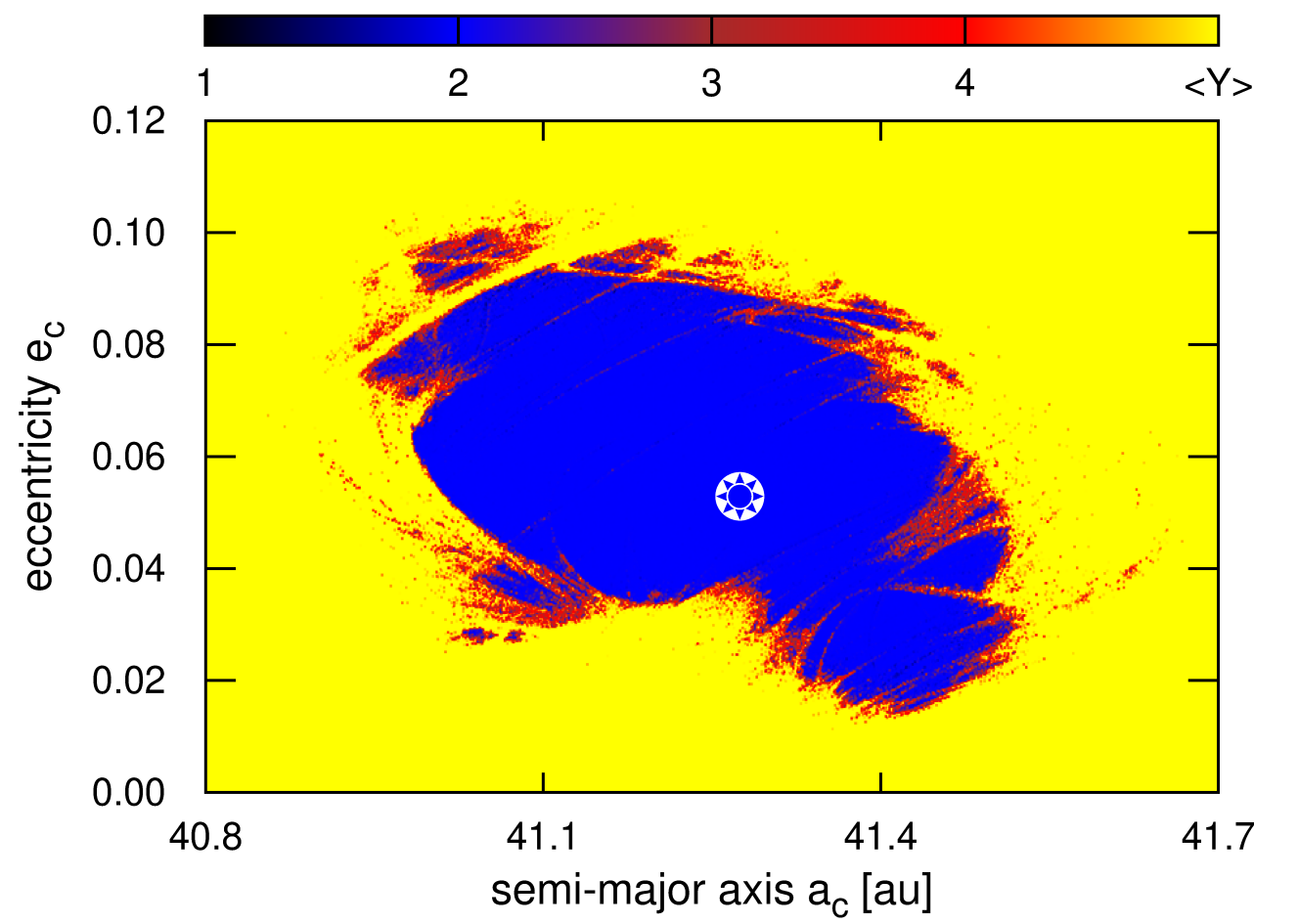}
}
\vspace*{0.3cm}
\hbox{
\includegraphics[width=0.45\textwidth]{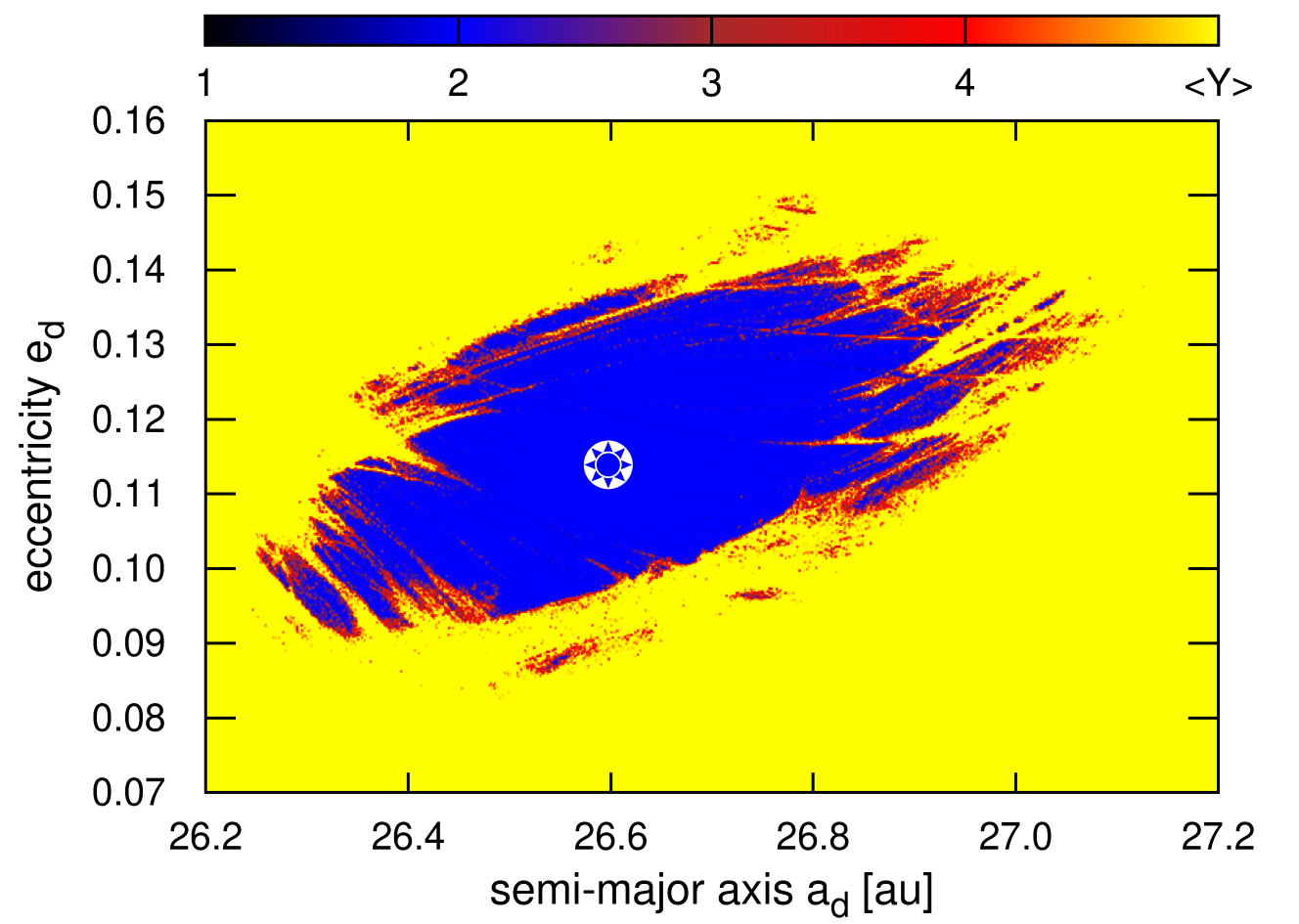}
\includegraphics[width=0.45\textwidth]{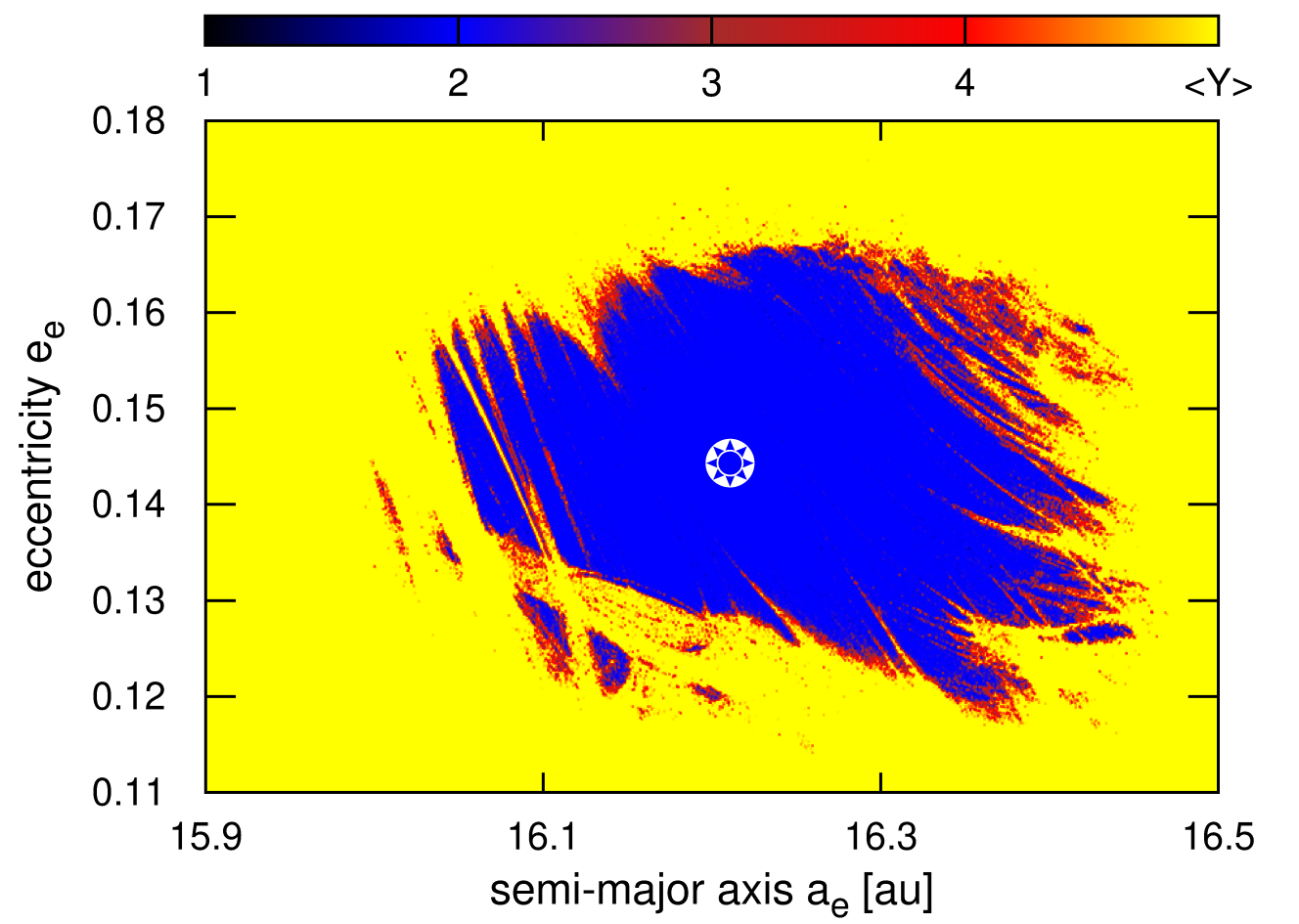}
}
}
}
\caption{
Dynamical maps in the semi major-axis--eccentricity plane,  in terms of the \megno{} fast indicator. Each point in the maps was integrated for 10~Myrs, equivalent to more than 20,000 orbits of planet \hr8799{}b. Stable systems are confined to $\Ym \simeq 2$ (blue) and yellow colour marks strongly unstable configurations typically self-disrupting in less than 1~Myr. The star symbol marks the nominal osculating elements in  the initial condition in Table~1, for each planet in subsequent panels. The resolution of each map is $800\times480$~pixels.
}
\label{fig:figA3}
\end{figure*}
%
\section{Numerical setup and algorithms}
%
\subsection{Searching for periodic configurations}
A co-planar orbital configuration of a planetary system is determined by a vector containing positions and velocities of the planets $\lbrace x_i, y_i, u_i, v_i \rbrace$ or, equivalently, by a vector consisting of astrocentric Keplerian elements of the orbits $\lbrace a_i, e_i, \varpi_i, \M_i \rbrace$, \corr{i.e., the semi-major axis, eccentricity, pericenter longitude, and  the mean anomaly, respectively, where
$i=$e,~d,~c,~b or, equivalently, $i=1,2,3,4$}. Both the state vectors are given at a particular epoch and the state of the system at another epoch may be obtained through propagating the initial condition with the numerical integration of the $N$-body equations of motion. A configuration of the planetary $N$-body problem is called periodic if after time interval $T$ (called the period of \po{}) it returns to its initial state in the reference frame rotating non-uniformly with the temporal (osculating) pericenter of a selected planet \citep{Hadjidemetriou1976}. For the Keplerian representation of orbits we have the boundary conditions: $a_i(T) = a_i(0)$, $e_i(T) = e_i(0)$, $\Mmean_i(T) = \Mmean_i(0)$ for all orbits. Since the angular momentum $C$ of the system must be conserved, the pericenter longitudes $\varpi_i(T) \neq \varpi_i(0)$. However, $\Delta\varpi_{i,j} \equiv \varpi_i - \varpi_j (i \neq j)$ must be the same after $T$, $\Delta\varpi_{i,j}(T) = \Delta\varpi_{i,j}(0)$. This means that the relative configuration of the planets remains fixed after the period, while the whole system rotates by a certain angle around its angular momentum vector aligned with the $z$-axis \citep[e.g.][]{Lithwick2012}.

Here, we consider the evolution of the Keplerian elements
periodic in a reference frame co-rotating with the apsidal line of the innermost orbit
--- the reference orbit can be chosen freely. Equivalently, when considering the Cartesian coordinates, one needs to search for configurations whose positions and velocities expressed in a reference frame co-rotating with one of the planets ($\bar{x}$, $\bar{y}$, $\bar{u}$, $\bar{v}$) fulfil the periodicity conditions $\bar{x}_i(T) = \bar{x}_i(0)$, $\bar{y}_i(T) = \bar{v}_{x,i}(0)$, $\bar{v}_{y,i}(T) = \bar{v}_{y,i}(0)$ \citep{Hadjidemetriou1976}.  We used this Cartesian representation, common in the literature, only for an illustration (see Fig.~\ref{fig:fig2}); note that in this case we chose the outermost planet as the reference one.

For given planet masses, there exist families of periodic configurations parameterised by total angular momentum or a value of the osculating period ratios of one of the planet pairs in a chosen phase of the evolution \citep{Hadjidemetriou1976,Hadjidemetriou1981}. To select a particular family, we fix the period ratio of the innermost pair at the epoch in which the innermost planet resides in its pericentre. We denote this period-ratio as parameter $\kappa = (\Pd/\Pe)|_{\Me=0}$. Formally, for a chain of the 8:4:2:1 MMR, there are $8$ different epochs in which the innermost planet is in the pericentre. We select one of them and keep this choice when continuing a given family w.r.t. other parameters of the solution, denoted as a generic parameters vector
$\vec{x}$.

After testing various parameterisations of the \po{} in terms of numerical efficiency and reliability, we decided to use the following set of components of the state vector $\vec{X}$, each of which being a function of the astrocentric, osculating Keplerian elements:
$X_1 = \log_{10} \ee$, $X_2 = \log_{10} \ed$, $X_3 = \log_{10} \ec$, $X_4 = \log_{10} \eb$, 
$X_5 = \Pc/\Pd - (\Pc/\Pd)_{\idm{nom}}$, $X_6 = \Pb/\Pc - (\Pb/\Pc)_{\idm{nom}}$, 
$X_7 = \omegae - \omegad$, $X_8 = \omegad - \omegac$, $X_9 = \omegac - \omegab$,  
$X_{10} = \Md$, $X_{11} = \Mc$, $X_{12} = \Mb$, where 
\begin{equation*}
(\Pc/\Pd)_{\idm{nom}} = \Big[1 + C_0 \lbrace 1 -
\left(\Pd/\Pe\right)\rbrace\Big]^{-1}, \quad
(\Pb/\Pc)_{\idm{nom}} = \Big[1 + C_1 \lbrace 1 -
\left(\Pc/\Pd\right)_{\idm{nom}}\rbrace\Big]^{-1},\quad
C_0 = \frac{j\,q}{i\,(p + j)}, \quad
C_1 = \frac{k\,p}{j\,(r + k)}.
\end{equation*}
The nominal values of the period ratios $(P_{\idm{c}}/P_{\idm{d}})_\idm{nom}$
and $(P_{\idm{b}}/P_{\idm{b}})_\idm{nom}$  corresponds to a chain of exact MMRs \citep{Delisle2017}, $\Pd/\Pe \approx (q + i)/q$, $\Pc/\Pd \approx (p + j)/p$, $\Pb/\Pc \approx (r + k)/r$. Therefore, for the case of the 2:1, 2:1, 2:1 MMRs chain, both the factors $C_0 = C_1 = 1/2$. Although the relations given above were designed for weakly interacting systems, whose evolution is well described with the averaging approach \citep{Delisle2017}, we found that such a representation enables appropriately controlling the period ratios.

For a given (fixed) period ratio $\kappa$ and masses $\me, \md, \mc$, and $\mb$, parameterising a given family of \po{}, its member is being searched for with the Newton method for nonlinear equations \citep{Press2002}, in 12-dimensional $\vec{X}$ space. Since $\Delta X_i \equiv X_i(T) - X_i(0)$, where $T$ is fixed so the innermost planet completes exactly $8$ full revolutions (counted from its pericentre to pericentre), the set of non-linear equations to be solved reads as $\Delta X_i (X_1, X_2, ..., X_{12}) = 0$, where $i=1, \dots, 12$. At first, the starting point is drawn randomly, yet around $X_5,X_6 \simeq 0$ (or close to an approximate solution, which we already know, such as the 8:4:2:1~MMR fits found in GM14 and GM18) until the algorithm finds a solution with $\Delta X_i \approx 0$. Next, this solution can be continued for changed $\kappa$ and the planet masses. In general settings, the continuation of \po{} is a complex problem, since many stable and unstable families may exist, in different parameter ranges \citep[][and references therein]{Hadjidemetriou1981}. 

\subsection{Data fitting on the parametric grid of \po{}}
\label{appendix:pogrid}
At the first attempt, we performed the optimisation similarly to the \mcoa{} variant in GM18.
Here, instead of CPU-demanding migration simulations, which result in resonant but not necessarily periodic systems, we continued \po{}s in a $5$-dimensional $\vec{Z}$ grid ($Z_1 = \me, Z_2 = \md, Z_3 = \mc, Z_4 = \mb, Z_5 = \kappa$), and we found $\sim 10^7$ configurations covering the interesting region of the 8:4:2:1 MMR chain. In this sense, we obtained the exactly resonant (periodic) configurations which might fit the observations as well, each being the 8:4:2:1 MMR center for different masses and the inner orbits period ratio. From this point, the analysis is essentially similar to GM18. In order to fit a given \po{} to the measurements, one needs to find a minimum of $\chi^2$ in a few-dimensional space of model parameters. We performed the optimisation experiments using the same parameters, as in GM18: the scale parameter $\rho$, the phase of a periodic configuration corresponding to the reference epoch  $\phase$, and the 3-1-3 Euler angles $(I, \Omega, \rotation)$, \corr{fixing orientation of the orbital plane w.r.t. the sky (observer) frame}. In such settings, $\chi^2$ (equivalently, the maximum likelihood ${\cal L}$) depends on $\vec{Y}$, whose components are $Y_1 = \rho$, $Y_2 = \phase$, $Y_3 = \Pi$, $Y_4 = I$, $Y_5 = \Omega$, $Y_6 = \rotation$. We also updated the parameter vector $\vec{Y}$ by the system parallax $\Pi$, and the so called {\em error floor} $\sigma_{\alpha,\delta}$, re-scaling the nominal uncertainties, in order to account for possibly underestimated errors and biases of the observations. 

The merit function $\cal L(\vec{Y})$ is defined for this variant of parametrisation as follows:
\begin{eqnarray}
\label{eq:logL}
 \ln {\cal L}(\vec{Y}) & = &
-\frac{1}{2} \chi^2(\vec{Y}) 
-\frac{1}{2}\sum_{i=1}^{N_{\idm{obs}}}\left[ \ln \theta^2_{i,\alpha} + \ln \theta^2_{i,\delta} \right]- N_{\idm{obs}} \ln (2\pi),  \\ 
\label{eq:chi2}
\chi^2(\vec{Y}) &=& \sum_{i=1}^{N_{\idm{obs}}}
\left[ \frac{[\alpha_i-\alpha(t_i,\vec{Y})]^2}{\theta^2_{i,\alpha}} +
       \frac{[\delta_i-\delta(t_i,\vec{Y})]^2}{\theta^2_{i,\delta}} \right],
\end{eqnarray}
where $(\alpha_i,\delta_i)$ are the measurements at time $t_i$,  $\alpha(t_i,\vec{Y}),\delta(t_i,\vec{Y})$ are the ephemeris values,  $\theta^2_{i,\alpha}$ and $\theta^2_{i,\delta}$ are the nominal measurements uncertainties in RA and DEC scaled in quadrature with the error floor, $\theta^2_{i,\alpha}=(\sigma^2_{i,\alpha} + \sigma^2_{\alpha,\delta} )$ or  $\theta^2_{i,\delta}=(\sigma^2_{i,\delta} + \sigma^2_{\alpha,\delta} )$, for each datum, respectively, and $N_{\idm{obs}}$ is the number  of observations. Also, $N=2 N_\idm{obs}$, since RA and DEC are measured in a single detection. The $\ln{}{\cal L}$ function in Eq.~\ref{eq:logL} is defined in the way that assuming the uncertainties as Gaussian and uncorrelated, the resulting best-fitting models should yield the reduced $\chi_\nu^2 \simeq 1$. The merit function was optimized with the help of genetic and evolutionary algorithms \citep{Izzo2012}.

In order to illustrate the results of the \po{}-grid approach, we invoke a particular experiment in which we fitted the $\ln {\cal L}$ function in Eq.~\ref{eq:logL} to all measurements available at the moment in the literature: the early \hst{} data in \citep{Lafreniere2009} and \citep{Soummer2011}, a homogeneous data set in \citep{Konopacky2016}, \gpi{} data in \citep{Wang2018}, and the \gravity{} measurement in \citep{DeRosa2020} updated with mostly
\sphere{}, \subaru{} and \lbt{} data, collected in \citep{Wertz2017}, from
\cite{Metchev2009,Hintz2010,Currie2011,
Currie2012,Currie2014,Bergfors2011,Pueyo2015,Galicher2011,Esposito2013,
Maire2015,Zurlo2016}. This set consists of $N_{\idm{obs}}=127$ (RA, DEC) observations, some of them clearly deviating from any astrometric model. Because preliminary \po{} and \mcoa{} fits indicated  the reduced $\chi_\nu^2 \simeq 2.5$ for the best-fitting models, we introduced the error floor $\sigma_{\alpha,\delta}$ in order to account for possible data biases and un-modeled errors. 

\begin{figure*}
\centerline{
\vbox{
\hbox{
\includegraphics[width=0.45\textwidth]{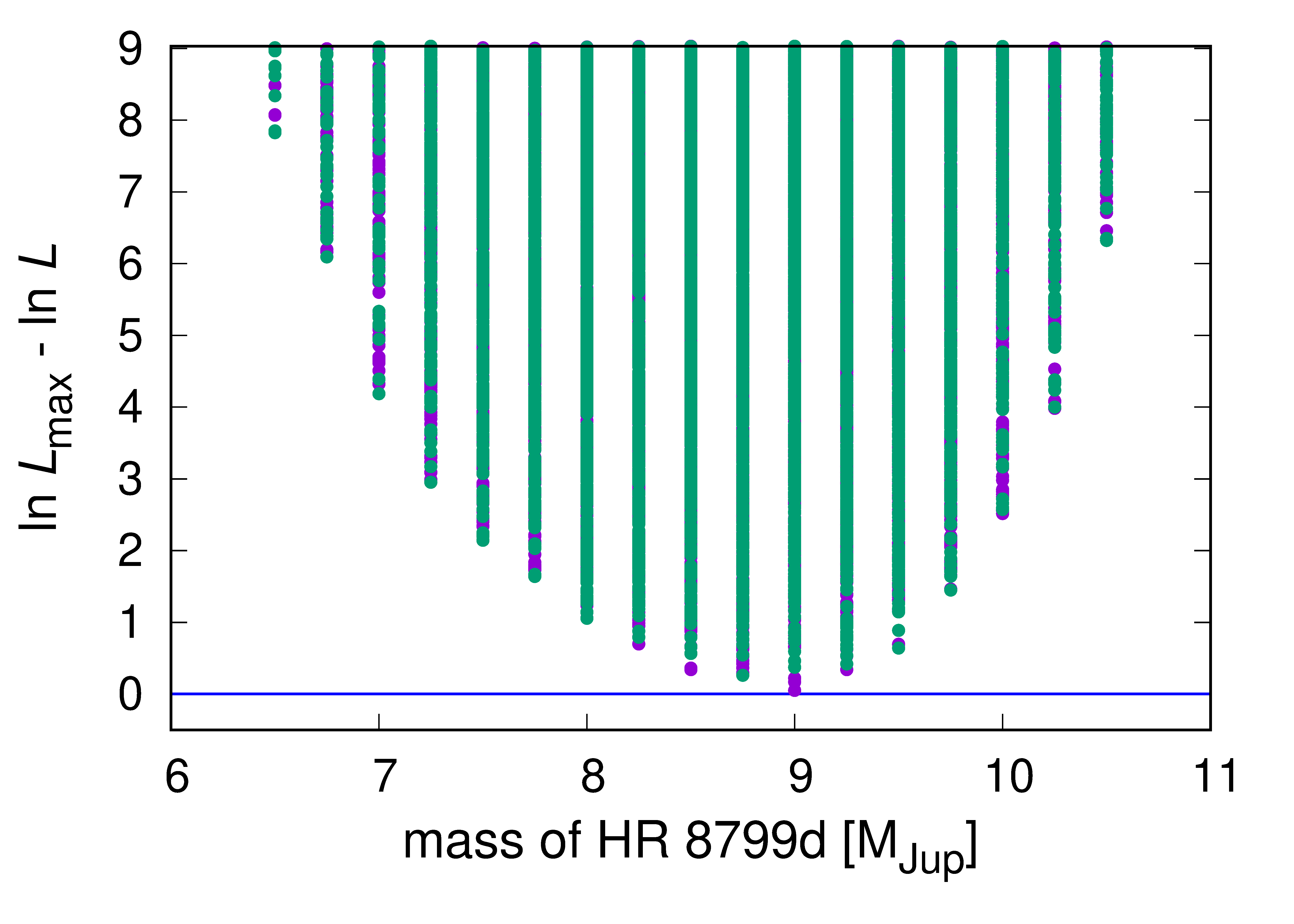}
\includegraphics[width=0.45\textwidth]{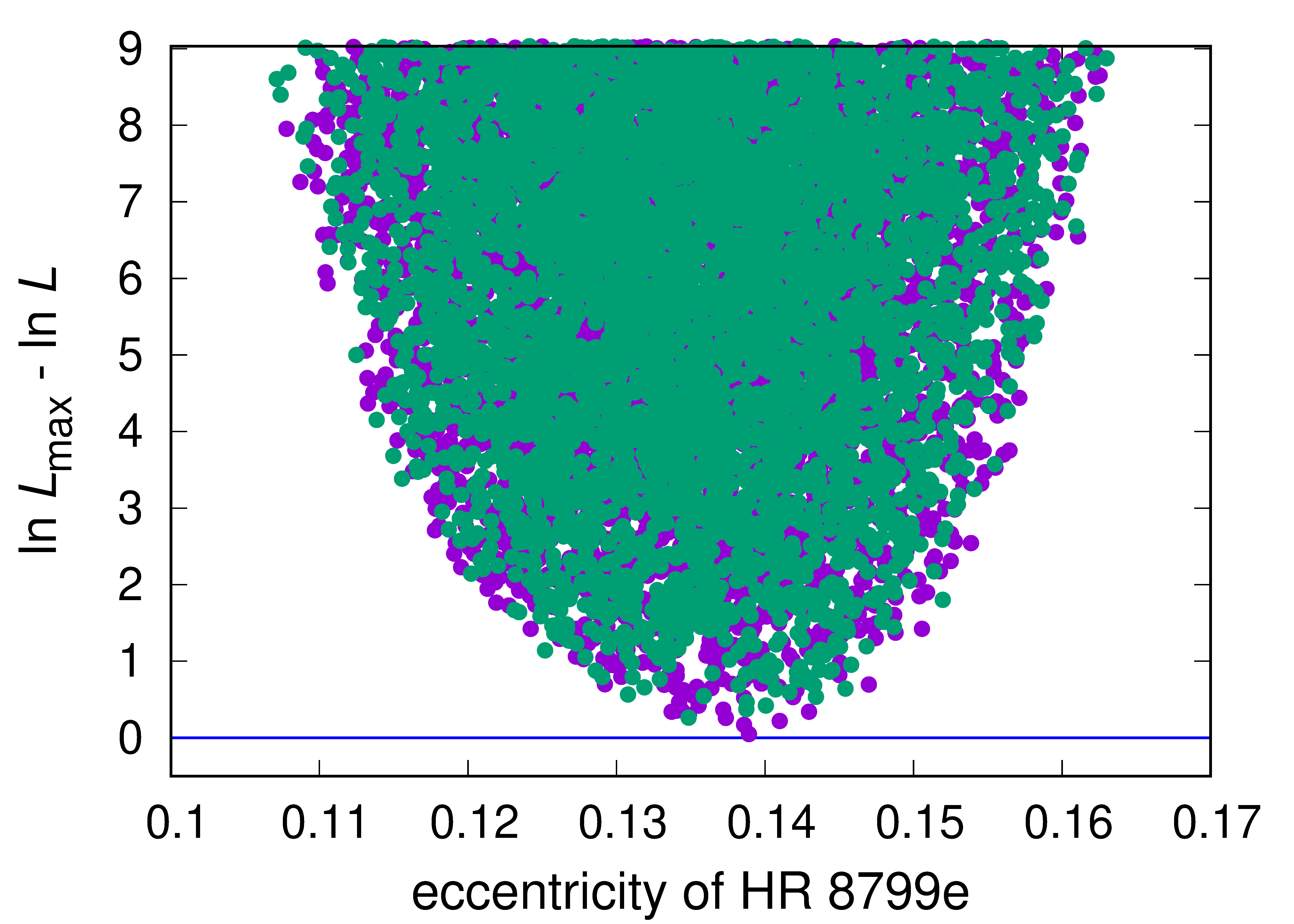}
}
\hbox{
\includegraphics[width=0.45\textwidth]{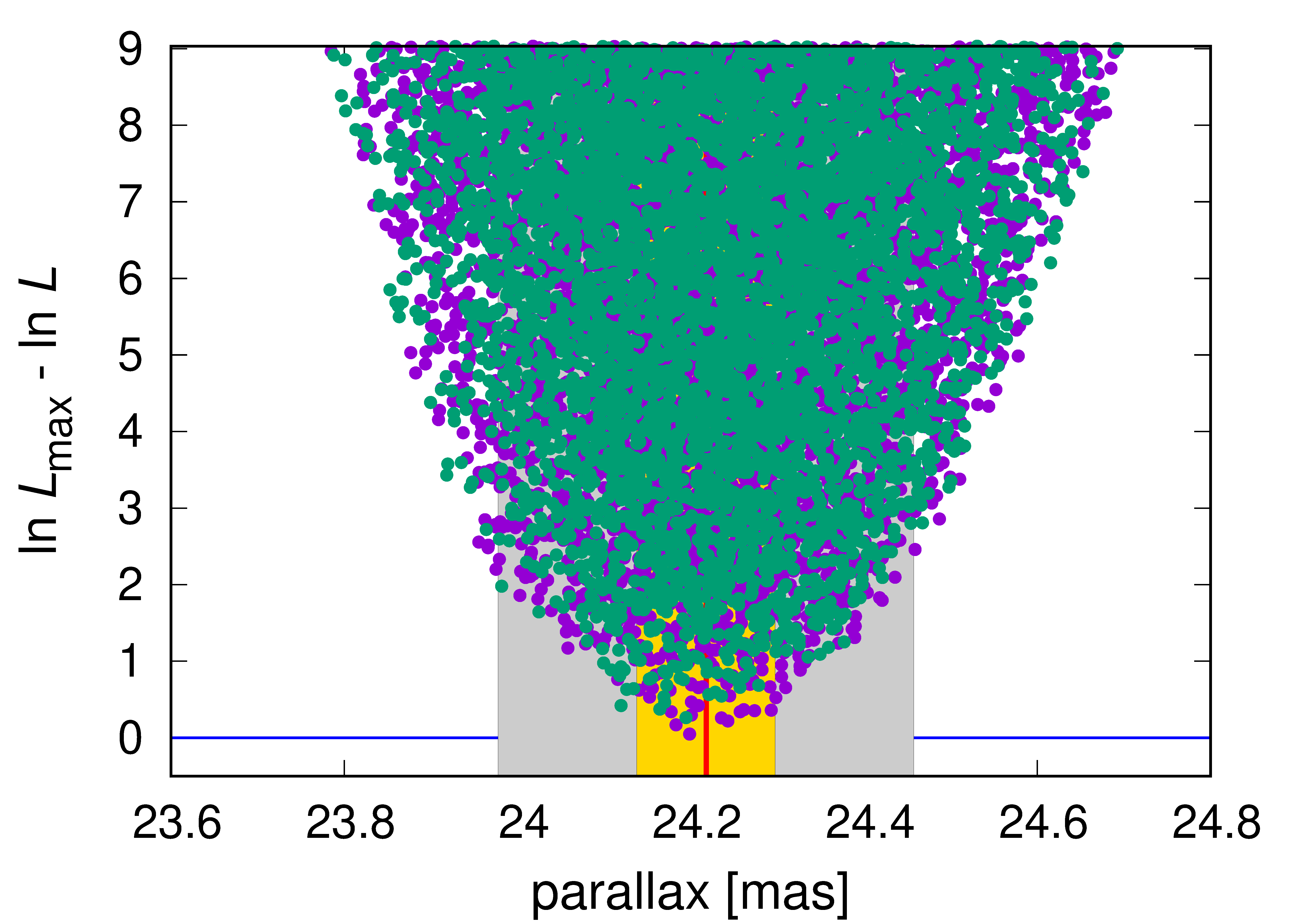}
\includegraphics[width=0.45\textwidth]{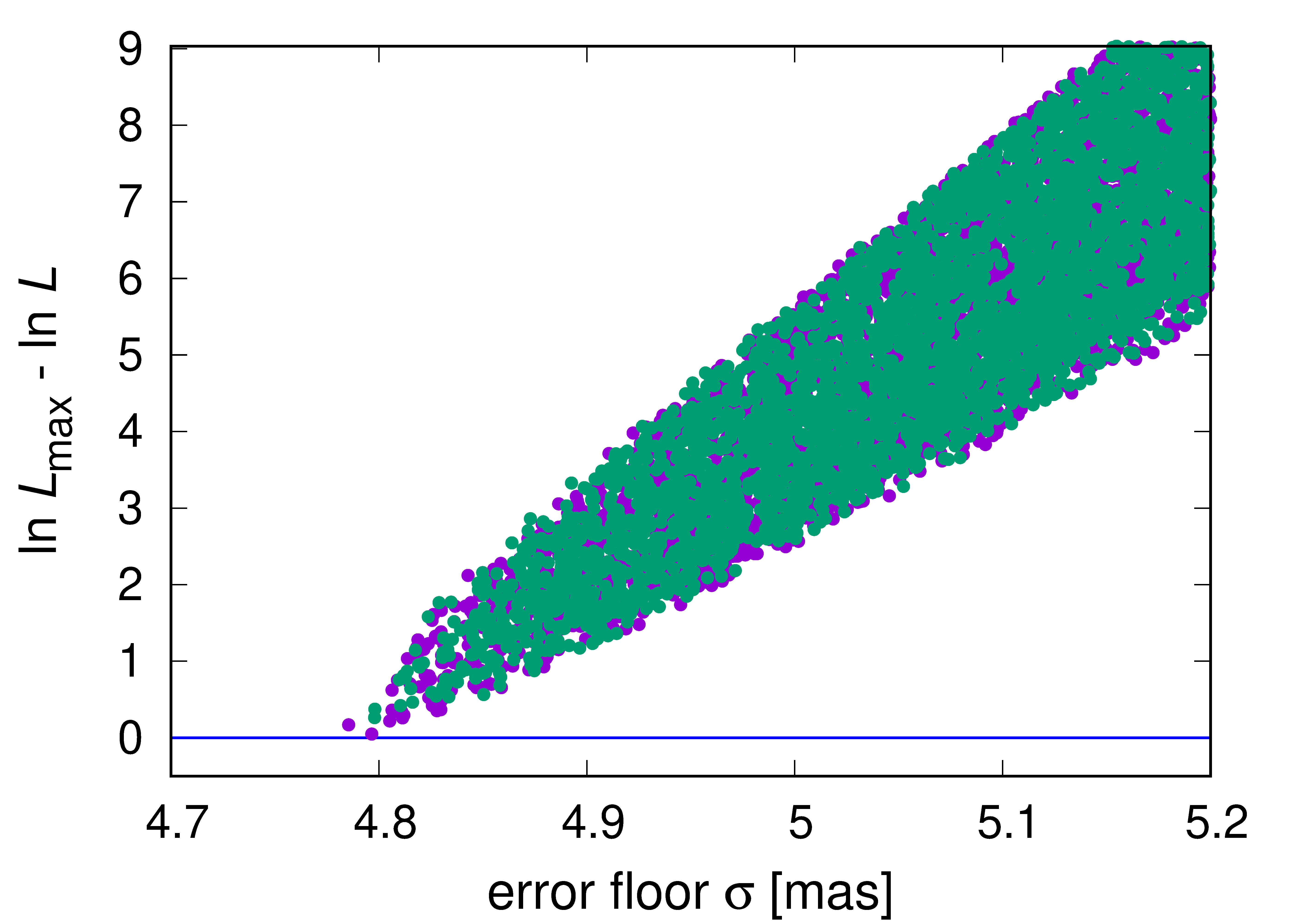}
}
}
}
\caption{
Plots illustrating the \po{}--grid search for the best fitting parameters to
astrometric measurements of the \hr8799{} system. In this experiment, we fitted seven free parameters, including the orbit scale $\rho$, three Euler angles, the initial epoch,
system parallax $\Pi$, and the error floor $\sigma_{\alpha,\beta}$,  to all measurements
gathered in the literature, $N_{\idm{obs}}=127$ data points. We optimised the likelihood function $\ln {\cal L}$ with the error floor term, and without any priors. The extremum of $\ln {\cal L} \simeq -834.4$, the error floor $\simeq 4.8$~mas. Fitted systems in two Monte-Carlo sampling runs on the \po{} grid, are marked with filled circles and different colors.
Top-left panel: mass of \hr8799{}d with a clear extremum.
Top-right panel: inferred eccentricity of \hr8799{}e from the \po{} models.
Bottom-left panel: system parallax, with the GAIA DR2 nominal value (red line) marked with its $1\sigma$ (gold rectangle) and $3\sigma$ (grey rectangle) confidence intervals.
Bottom-right plot: the error floor $\sigma_{\alpha,\delta}$, as the global correction
factor of the measurements uncertainty, yielding the reduced $\chi_\nu^2\simeq 1$.
}
\label{fig:figA4}
\end{figure*}

The results are illustrated in Fig.~\ref{fig:figA4}, which shows selected fitted parameters, which are gathered on the pre-computed \po{} grid and plotted vs. $\Delta{}\ln{\cal L} \equiv \ln{\cal L}_{\idm{max}}- \ln{\cal L}$,  relative to the best-fitting value $\ln{\cal L}_\idm{max}$ found in the search. Most of the primary $\vec{Y}$ parameters, such as the mass of \hr8799{}d (top-left panel) and \hr8799{}b (not shown), the error floor $\sigma_{\alpha,\delta}$ (the bottom-right panel), and the system parallax (the bottom-left panel) exhibit clear extrema.  Also the \po{}--constrained eccentricities (such as for \hr8799{}e in top-right panel) are quasi-parabolically bounded. We found particularly surprising that the masses of \hr8799{}d and \hr8799{}b could be potentially constrained, although the $\ln {\cal L}$ extremum is apparently shallow. That also regards the parallax $\Pi$, which, {\em as the free parameter of the astrometric model}, overlaps with the GAIA DR2 trigonometric parallax $\Pi \simeq (24.22 \pm 0.09)$~mas within its formal $1\sigma$ uncertainty -- our best-fitting value $\Pi \simeq 24.25$~mas is accurate to the fourth significant digit \corr{($\simeq
0.1\%$, in other fits up to $\simeq 0.3\%$)}. We consider this as a meaningful benchmark of the self-consistency of the astrometric model, the parallax and the physical characteristics of the system: the derived masses of the planets and the adopted stellar mass of 1.52$\msun$ \citep{Konopacky2016}. We note that the stellar mass must be fixed or tigthly constrained, since otherwise it would introduce a strong mass--period (linear scale) correlation through the Keplerian law (actually, the $N$-body dynamics scaling).

\begin{figure*}
\centerline{
\vbox{
\hbox{
\includegraphics[width=0.38\textwidth]{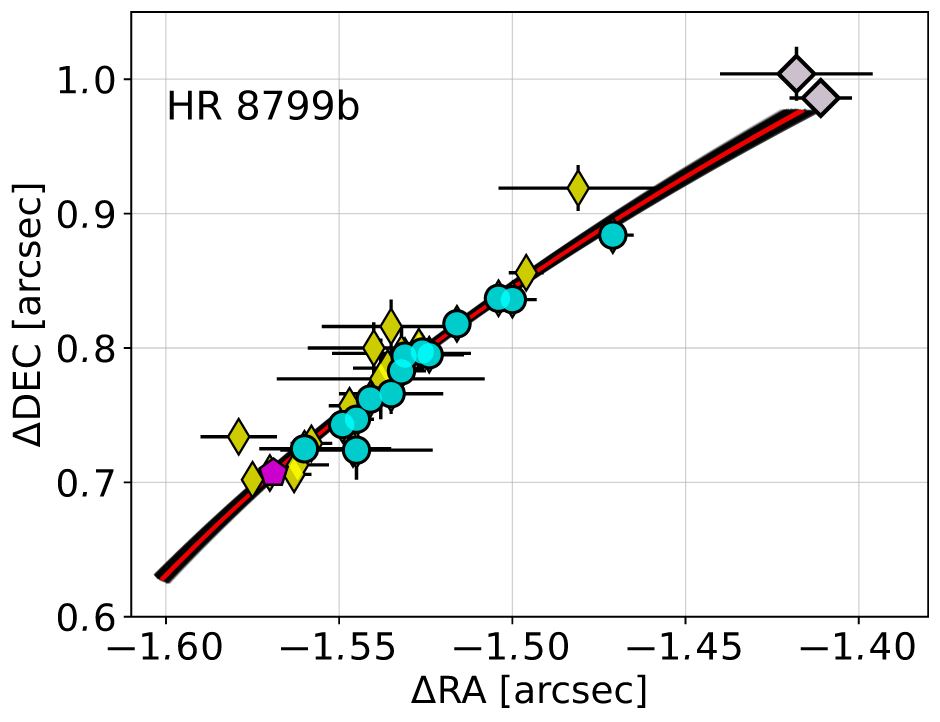}
\hspace*{0.42cm}
\includegraphics[width=0.38\textwidth]{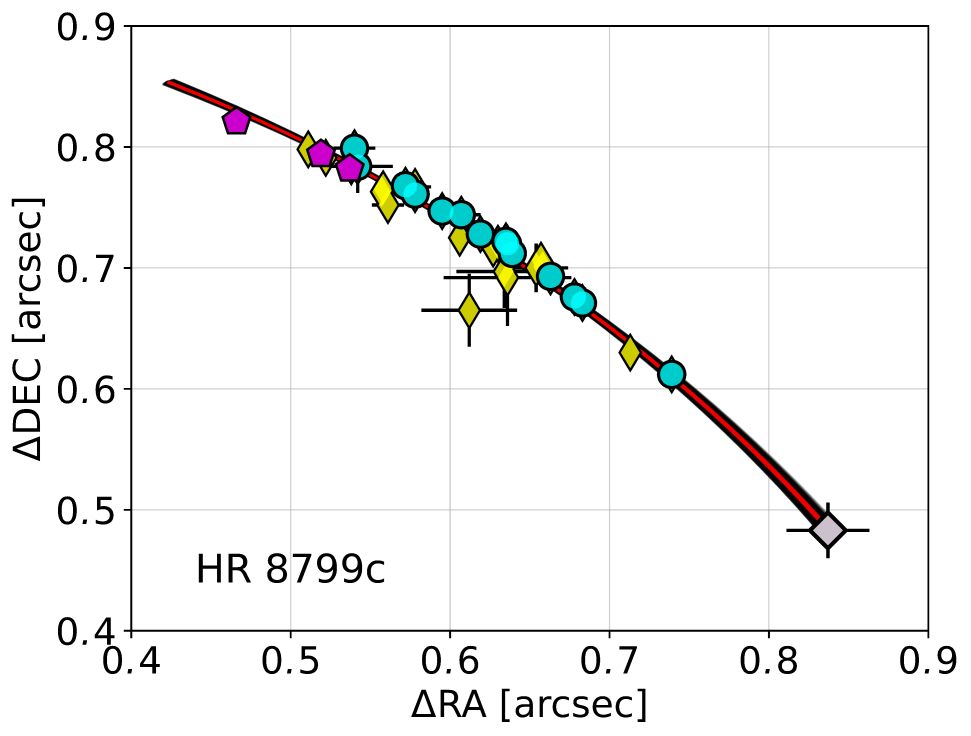}
}
\hbox{
\includegraphics[width=0.38\textwidth]{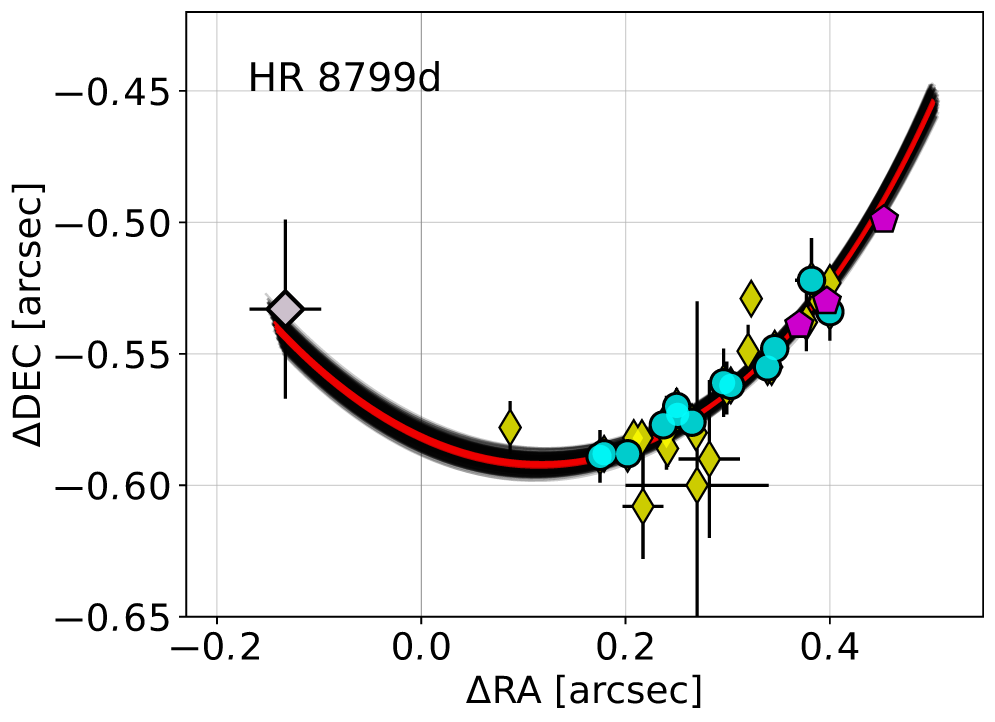}
\hspace*{0.15cm}
\includegraphics[width=0.38\textwidth]{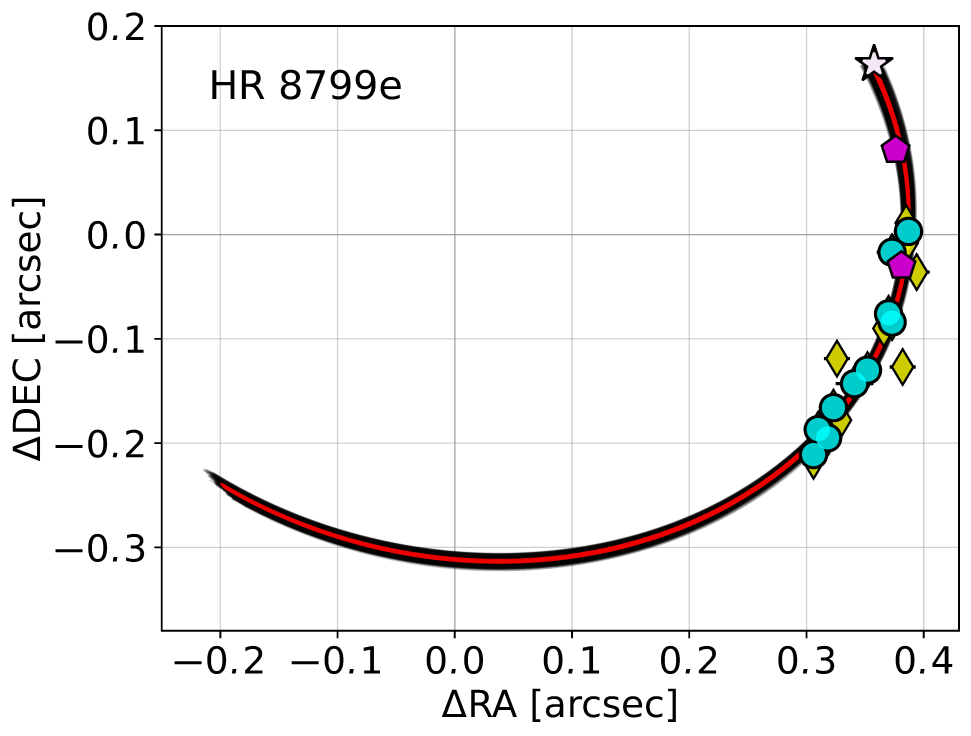}
}
}
}
\caption{
A \po--model to all \hr8799{} observations in the literature (see the text for details). Big rotated rectangles are for the earliest \hst{} data in \citep{Lafreniere2009,Soummer2011}, circles are for the uniformly reduced data set in \citep{Konopacky2016}, the star symbol is for the \gravity{} measurement in \citep{DeRosa2020}, pentagons are for \gpi{} data in \citep{Wang2018}, and diamonds are for the \sphere{}, \subaru{} and \lbt{} data  collected in \citep{Wertz2017}.
Black curves illustrate synthetic orbits in the sky-plane, derived from the \po{} grid search, and plotted between epochs $t_0=1998.829$ and $t=2018.654$ of the \gravity{} datum, for $\ln {\cal L}<-825.0$, and the red curves are for solutions providing $\ln {\cal L}<-834.0$, see Fig.~\ref{fig:figA4}. Subsequent panels are for close-ups of the data and orbital arcs for each planet, respectively. \corr{Note that $\Delta{}$RA is labeled negative w.r.t. the RA direction.}
}
\label{fig:figA5}
\end{figure*}

The orbital geometry of the best-fitting models is illustrated in Fig.~\ref{fig:figA5}.
Panels are for the close-ups of the sky-plane for subsequent  planets, with all data points marked with symbols for different sub-sets, gray curves for $250$ random solutions drawn within $\simeq 1$~mas range around the best-fitting models which are illustrated with the red curves,  and yielding $\ln {\cal L}$ marginally worse than the extremum value. The orbits are plotted for the time interval between $t_0=1998.829$ and the last \gravity{} epoch. The $\Delta\ln{\cal L}$--range in Fig.~\ref{fig:figA4} translates to a substantial spread of the models, which is best visible for \hr8799{}d. Although we did not compute formal confidence intervals, the spread indicates sensitivity of $\Delta\ln{\cal L}$ to a variation of the parameters.

\begin{figure*}
\centerline{
\hbox{
\includegraphics[width=0.527\textwidth]{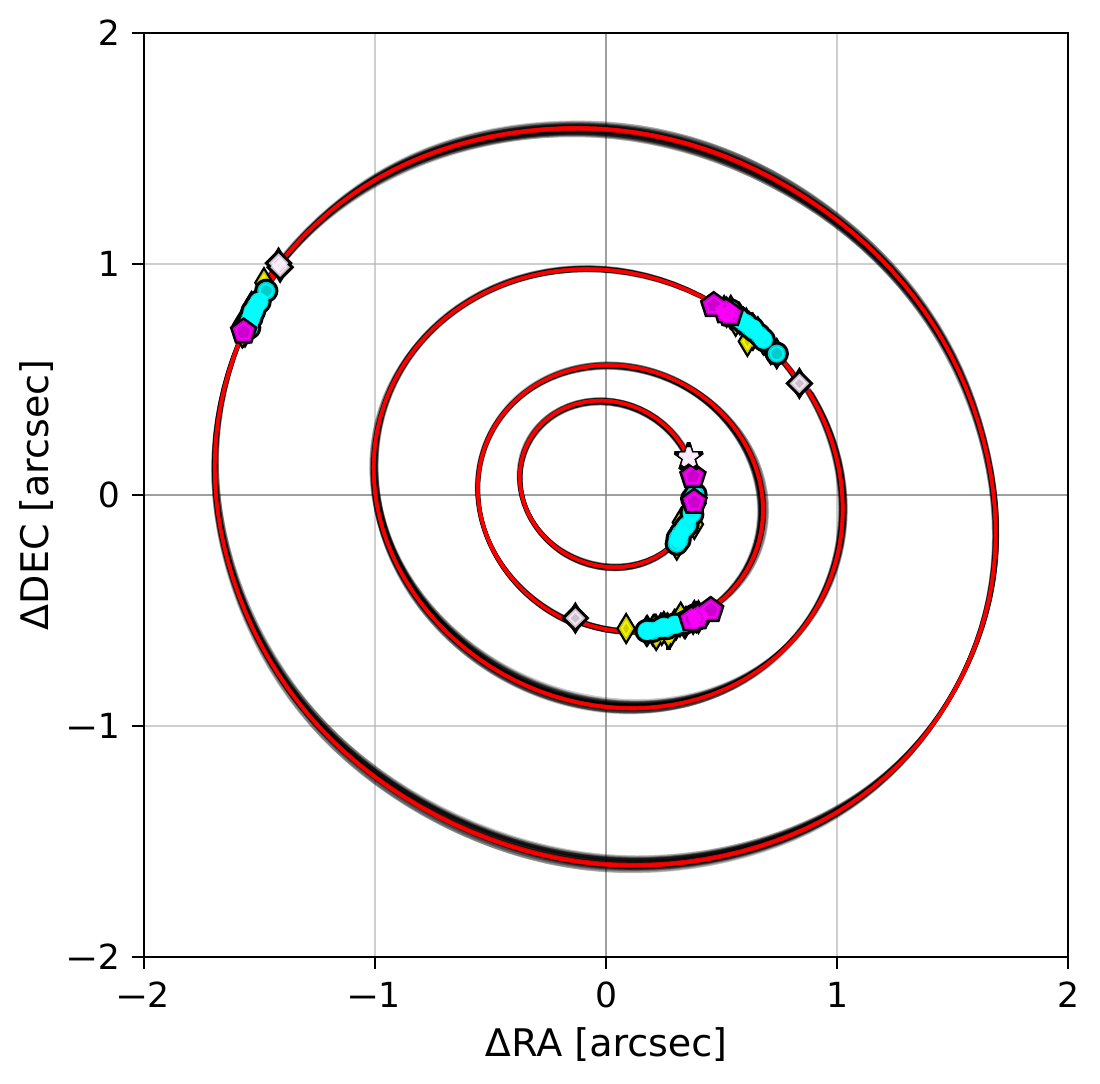}
}
}
\caption{
Global view of the \hr8799{} system geometry in the plane of the sky 
for all data in literature, and the astrometric model derived in the \po{}--grid search,
see caption to Fig.~\ref{fig:figA5}.
}
\label{fig:figA6}
\end{figure*}

The orbits plotted globally (Fig.~\ref{fig:figA6}), for the osculating orbital period of each planet separately, appear well bounded, and this is particularly apparent for the innermost, fast moving planets. The model might predict {their geometric} positions close to the best-fitting \po{} motion for a long time, in spite of the parameter uncertainties.

The bad message received from the \po{}--grid fitting is a strong anti-correlation between the masses of \hr8799{}c,e ($\mc$, $\me$); moreover,  the best-fitting configurations exhibit very small innermost mass $\me \simeq 0.1 \mJ$. The grid method also depends on the resolution, which should be individually tuned for each parameter in 5-dim space, as illustrated in the \hr8799{}d mass scan (top-left panel in Fig.~\ref{fig:figA4}). We found that the error floor does not help in eliminating nor even reducing the mass correlation.

These preliminary \po{}--grid experiments provide interesting and useful hints for the final approach, described below (Sect. \ref{appendix:optpo}), yet we could not consider them as fully conclusive. Unfortunately, the parametric grid approach is tedious and introduces large CPU-overhead. The \po{} continuation and sampling must be multi-dimensional and that implies not only the need of computing huge sets of  the solutions but also optimising the initial conditions one by one --- although we rather performed the Monte Carlo search on the grid. Determining the best-fitting model to the present observations is also difficult due to the ($\mc$, $\me$) anti-correlation. Getting rid of that degeneracy needs additional, prior information, such as the planet masses estimated on the grounds of the thermodynamical evolution and cooling tracks. 

In order to address these issues, we improved the grid MMR-constrained optimisation in ways making it CPU-efficient and independent on the grid resolution, described below. In the final experiments, we examined the reduced data set (${\cal D}$ from hereafter) comprising of \hst{} data in \citep{Lafreniere2009,Soummer2011}, a homogeneously reduced data set in \citep{Konopacky2016}, and \gpi{} data in \citep{Wang2018} derived in a similar instrument, as well as the \gravity{} datum in \citep{DeRosa2020}. We used the \hst{} and \gravity{} data to extend the time--window of the observations. For this set, we did not consider the error--floor, since the \po{} models with the basic $p=11$ free parameters yield $\chi^2_\nu \simeq 1.25$ and an $\rms{} \simeq 7$~mas, consistent with the mean uncertainty of the measurements in the reduced set, $\simeq 8$~mas. 

\subsection{Optimisation on a manifold of periodic configurations}
\label{appendix:optpo}

From the mathematical point of view, our goal is to find the best-fitting \po{} in the space of vectors $\vec{Z} = (\me, \md, \mc, \mb, \kappa)$. We note that the single $\kappa$ period ratio of the innermost pair of planets is sufficient to identify the required \po{}, since we seek the \po{} in a small range around the nominal, remaining period ratios, $X_5$ and $X_6$ with fixed MMR factors $C_0, C_1$. As explained before, in the \mcoa{}-like optimisation the space of vectors $\vec{Z}$ is explored in a grid of pre-computed \po{}. Unfortunately, even in the 5-dimensional space of $\vec{Z}$, the number of solutions to be data--fitted becomes huge. Moreover, the mass anti-correlation with $\me$ tending toward very small and non-realistic values implies a difficulty in estimating parameters ranges of the grid as well as its resolution. Even if we reach the neighborhood of the merit function's extremum, it is difficult to find and tune the proper grid resolution for all parameters of the model, and the 5-dim \po{} grid must be updated in subsequent iterations. Still, the method is useful in investigating the parameter space in wide ranges, and provides a good starting point (solutions) for a more refined and accurate method.

Clearly, a mathematically correct algorithm must explore the model parameter sub-space (a manifold) fixed by the requirement of \po{}. In order to implement this manifold fitting, for a given (prescribed) $\vec{Z}$, a co-planar \po{} is  searched for, resulting in the Cartesian state vector $\vec{X}$. Next, we select the best-fitting parameters as a vector $\vec{Y}$, in the orbital and geometric elements space. In this way, for the given $\vec{Z}$, the objective function, such as $\chi^2 = \chi^2(Z_1, Z_2, Z_3, Z_4, Z_5)$ is optimised under assumptions that i) the model orbits are periodic, ii) the \po{}s are optimised through the linear scaling, time-phasing,  its distance (parallax), and spatial orientation ($\vec{Y}$). We closed the whole algorithm in a single procedure, being numerical implementation of the merit function for the optimisation of $\vec{Z}$. Both steps that consist of continuing the \po{} and its final fitting to the data, in the $\vec{Y}$ and $\vec{Z}$ spaces equivalent to the $11$-dim vector of sampled parameters $\vec{x} = (Z_1, Z_2, Z_3, Z_4, Z_5, Y_1, Y_2, Y_3, Y_4, Y_5, Y_6)$, explicitly 
$\vec{x} = (\me, \md, \mc, \mb, \kappa, \rho, I, \Omega, \rotation, \phase, \Pi)$, are being done with a help of the Levenberg--Marquardt (LM) algorithm \citep{Press2002}. The iterative scheme enables us to find the best-fitting \po{} in a small number of steps, typically a few tens iterations.

\subsection{DE-MC sampling and uncertainties of the best-fitting parameters}
In order to assess realistic uncertainties, and to investigate possible parameter correlations, we performed the Differential Evolution Markov Chain (DE-MC) sampling \citep{TerBraak2006}. 
Recalling some well known elements of the Bayesian statistics and the Markov Chain Monte Carlo sampling \citep[e.g.][]{Gregory2010}, we consider the posterior probability distribution $\posterior(\vec{x}|\D) \sim {\cal L}(\D|\vec{x}) P(\vec{x})$, where $\D$ denotes the data set,  ${\cal L}(\D|\vec{x})$ represents a probability that parameters $\vec{x}$ explain the data set $\D$, and $P(\vec{x})$ is the prior information imposed on $\vec{x}$. We define the $\ln {\cal L}$ function the same, as in Eq.~\ref{eq:logL},
\begin{equation}
\ln {\cal L}(\D|\vec{x})  = -\frac{1}{2} \chi^2(\vec{x})  
-\sum_{i=1}^{N_{\idm{obs}}}\left[ \ln \sigma_{i,\alpha} + \ln \sigma_{i,\delta} \right]- N_{\idm{obs}} \ln (2\pi) \equiv -\frac{1}{2} \chi^2(\vec{x}) + \mbox{const},
\end{equation}
but skipping the error floor, since we performed the \mcmc{} sampling on the reduced data set ${\cal D}$ described in Sect.~\ref{section:laplace}, and for these measurements the best-fitting models yield $\chi_\nu^2 \sim 1$ -- there is no need to account for the uncertainties correction.

The DE-MC sampling, which is a variant of the canonical Metropolis-Hastings algorithm, occurs according to the probability of moving from a starting point $x_1^{(i)}$ to a new point $x_2^{(i)}$ in the parameter space, $p(x_2^{(i)}|x_1^{(i)})$, which is a product of $q(x_2^{(i)}|x_1^{(i)})$ and $\alpha(x_1^{(i)}, x_2^{(i)})$, where $q(x_2^{(i)}|x_1^{(i)})$ is a probability of choosing a candidate point $x_2^{(i)}$ when starting from $x_1^{(i)}$. The superscript denotes the $i$-th chain from a population of $n=100$ chains which are evolved in parallel. The candidate point of the $i$-th chain is chosen according to:
\begin{equation}
\label{eq:mh}
x_2^{(i)} = x_1^{(i)} + \gamma \left( x_1^{(j)} - x_2^{(k)} \right) + \mbox{Uniform}\left( -\Delta^{(i)}, +\Delta^{(i)} \right),
\end{equation}
where the chains $j$ and $k$ ($j \neq k$ and $j, k \neq i$) are chosen randomly, while $\Delta^{(i)}$ is chosen individually for each parameter. In order to obtain $\simeq 50\%$ acceptance rate we chose $\gamma = 0.3$ and  $\Delta^{(i)}$ were 
$(10^{-4}\,\mJ$, $1.7 \times 10^{-4}\,\mJ$, $1.5\times 10^{-4}\,\mJ$, $7\times 10^{-4}\,\mJ$, $5\times 10^{-7}$, $5\times 10^{-5}\,\yr$, $1.2\times 10^{-6}$, $10^{-4}$~mas, $1.2\times10^{-5}\,\deg$, $5\times 10^{-6}\,\deg$, $5\times 10^{-6}\,\deg)$, for subsequent components of $\vec{x}$ 
(see the previous subsection). The second term in Eq.~\ref{eq:mh} is the Metropolis-Hastings ratio:
\begin{equation}
\alpha(x_1^{(i)}, x_2^{(i)}) = \min\bigg[1, \frac{\posterior(x_2^{(i)})}{\posterior(x_1^{(i)})}\bigg],
\end{equation}
which denotes the acceptance probability of $x_2^{(i)}$ when starting from $x_1^{(i)}$. Importantly, the DE-MC algorithm propagates a number of Markov chains in parallel, starting from different initial positions in the parameters space, and introduces mixing of the solutions in the chains through the Differential Evolution \citep{Price2005}. That makes this algorithm both simple and computationally efficient. We also note that the DE-MC approach is crucial for our optimisation problem, given the need of computationally complex \po{} continuation w.r.t. model parameters, since the \po{} cannot be updated sufficiently freely, as required by the Markov chain propagation.

Priors $P(\vec{x})$ for the masses were set as Gaussian with mean values and standard deviations \corr{according to \cite{Wang2018}, from the hot-start evolutionary models,  to $(5.8 \pm 0.5)\mj$ for  \hr8799{}b, and $(7.2 \pm 0.7)\mj$ for all other planets. Similarly, the parallax Gaussian prior is $\Pi=( 24.22\pm0.09)${}\,mas \citep{Gaia2018}.} For the $6$ remaining parameters, the prior distributions were uniform, \corr{in sufficiently wide ranges}.

\begin{figure*}
\centerline{
\vbox{
\hbox{\includegraphics[width=0.6\textwidth]{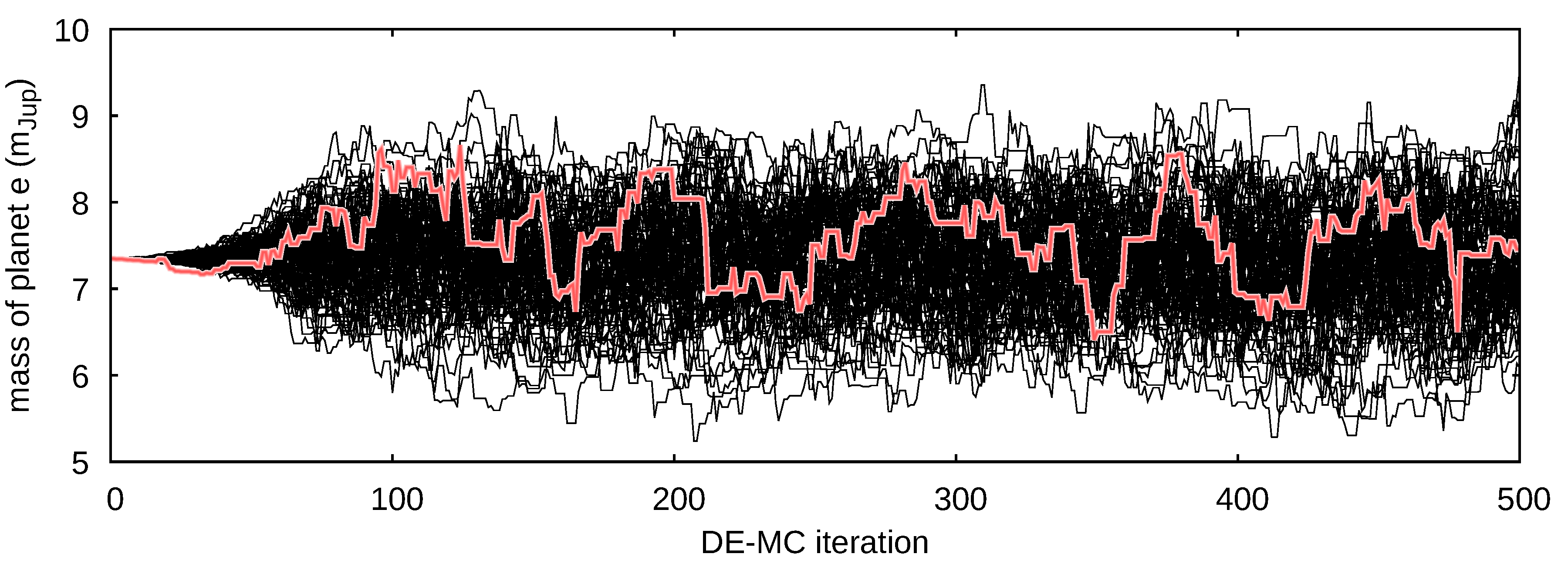}}
\hbox{\includegraphics[width=0.6\textwidth]{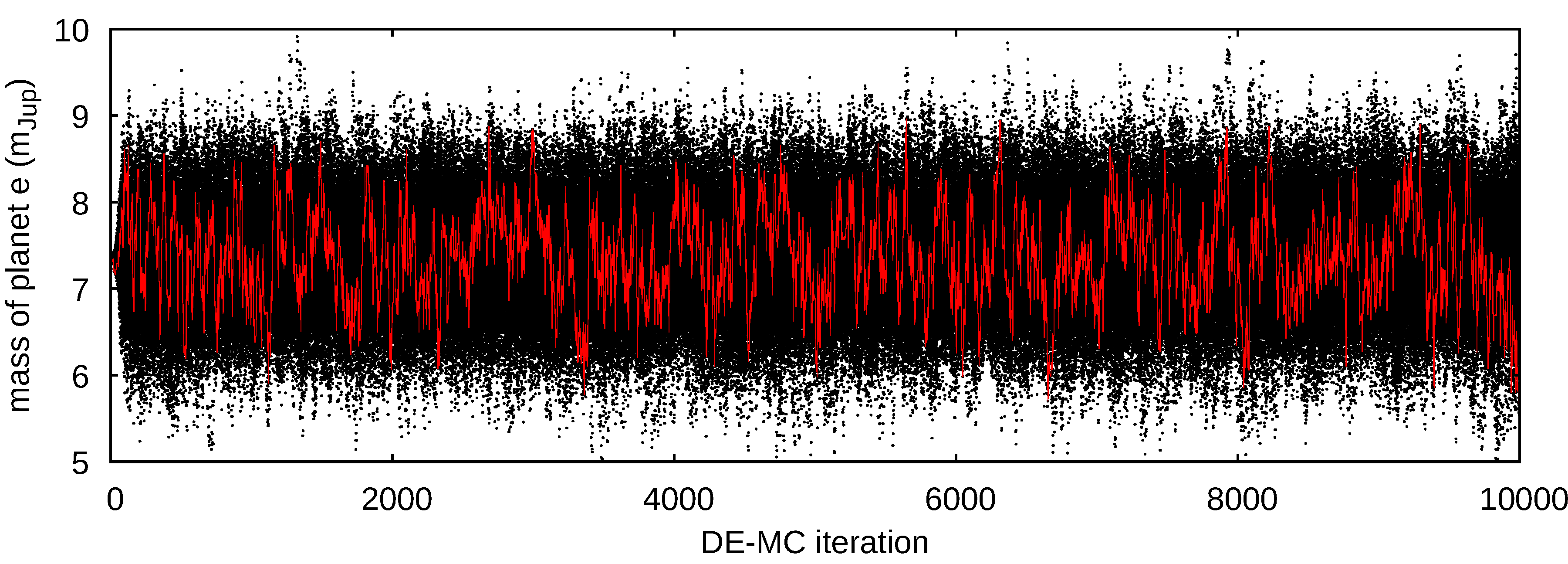}}
}
}
\caption{
The DE-MC sampling for the mass of \hr8799{}e. Black curves indicate the evolution of all $100$ individual chains, while the red curve illustrates the behaviour of one chosen chain. First $200$ steps are treated as burn-in steps. The total number of iterations is $10000$, in the top panel first $500$ steps were shown for clarity.
}
\label{fig:figA7}
\end{figure*}

We initiated the DE-MC sampling by choosing $100$ solutions from the vicinity of the best-fitting model in Table~\ref{tab:tab1}. The evolution of the whole population of Markov chains is illustrated in Fig.~\ref{fig:figA7} with black curves, while one selected, example chain is depicted with the red colour. At the beginning (first $\sim 100$ iteration steps) all the chains evolve closely to the initial condition. Since the differences $x_1^{(j)} - x_2^{(k)}$ increase, the sampling begins to occur over wider part of the parameter space. After $\sim 200$ steps the chain is already burnt-out. Those first $200$ steps were not included in the final statistics of solutions obtained after $10000$ iterations. In this DE-MC experiment we did not estimate the auto-correlation time for the Markov chains, since clearly the relatively small number of iterations already leads to a smooth approximation of the posterior. Also, as illustrated in Fig.~\ref{fig:figA7}, each of the chains quickly reach the random-walk state and explore the whole parameter space. Remarkably, this behaviour is much different from the \mcmc{} sampling with the full Keplerian or even the $N$-body models \citep[e.g.][GM18, and references therein]{Konopacky2016,Wertz2017,Wang2018}, that notoriously exhibit parameter correlations and long auto-correlation times $\sim 10^5$, due to multi-modal posteriors and ill-constrained optimisation problem implied by a small ratio of the measurements to the number of free parameters and narrow time window of the data.

According to the final results of the DE-MC sampling, as well as to the scans of $\chi^2$ function, the best-fitting configuration is meaningfully constrained w.r.t. all parameters. In particular, the bottom-right panel in Fig.~\ref{fig:fig3} (also Fig.~\ref{fig:figA8}) illustrates the Gaussian prior as \gaia{} parallax \citep{Gaia2018} (the red curve) over-plotted with the DE-MC posterior. The distributions closely overlap.  We also recall the grid-based experiments indicating that the best-fitting parallax may be determined independently of the \gaia{} measurements. The $1$-dim posterior probability distributions of all the free parameters determined with the DE-MC sampling are shown in Fig.~\ref{fig:figA8}, while $2$-dimensional contour plots of the posteriors for the Keplerian elements are illustrated in Fig.~\ref{fig:figA9}. The parameters uncertainties derived from the sampling are listed in Tab~\ref{tab:tab1}.

\section{Masses w.r.t. data biases}
\label{appendix:massbiases}

\corr{
With the improved \po{} algorithm described in Sect.~\ref{appendix:optpo}, we systematically explored the $\chi^2$ minimum in 2-dimensional planet mass planes, without (Fig.~\ref{fig:figA10}) and with independently determined astrophysical priors (Fig.~\ref{fig:figA11}), regarding  the measurements set ${\cal D}$ with $N_{\idm{obs}}=65$ observations, and also reducing it by particular \gpi{} points (data set ${\cal D}_1$ from hereafter), as explained below. For a given, fixed point in selected 2-dim masses plane, the remaining two masses and the $\kappa$ period ratio are optimised in terms of the best-fitting $\chi^2$. 
In the $\chi^2$ optimisation, the mass priors may be included as additional terms in the $\chi^2$ function (Eq.~\ref{eq:chi2}): $(m_i - m_{i, \idm{cooling}})^2/(\sigma_{i, \idm{cooling}})^2$, where $i = $e, d, c,~b and $m_{i, \idm{cooling}}$ denotes the planet $i$ mass constraint (prior),  while $\sigma_{i, \idm{cooling}}$ is its $1\sigma$ uncertainty. \corr{We set the mass priors after \citep{Wang2018}, the same as in the DE-MC experiments.}
The $\chi^2$-scans in the  mass planes for this enhanced model are shown in Fig.~\ref{fig:figA11}. We note that in the $\chi^2$ experiments the parallax was treated as a free parameter with no prior.
}

In the first experiment for data set ${\cal D}$ and without considering mass priors, we found the best-fitting $\me \lesssim 0.1 \mJ$. Also, the best-fitting and astrophysical masses are significantly different, as marked in the top row of Fig.~\ref{fig:figA10}, particularly for \hr8799{}d. In this figure, the mass estimates from the cooling theory are shown for a reference.  Masses of \hr8799{}e and \hr8799{}c are strongly anti-correlated and not bounded at all, since the best-fitting $\me$ converges towards very small and non-realistic values. When fixing the inner planet's mass at $7\,\mJ$, the anti-correlation disappears, and masses $m_\idm{d}$ and $m_\idm{b}$ become much better constrained, but their values are still significantly shifted with respect to the astrophysical priors (Fig.~\ref{fig:figA11}, top row).

In order to explain the discrepancy between the prior and posterior estimates, especially significant for the mass of \hr8799{}d, revealed also by the MCMC sampling, we searched for possible data biases. The left-hand panel of Fig.~\ref{fig:figA12} illustrates the RA and DEC residuals of the best-fitting model in Table~\ref{tab:tab1} derived for the ${\cal D}$ set -- rows from top to bottom are for subsequent planets. While the most precise \gravity{} datum is modelled apparently perfectly, there are precision \gpi{} observations \citep{Wang2018} significantly  deviating from the astrometric model, compared to the uncertainties. In the next experiment, we temporarily removed these points from the data set and we found a new best-fitting model for this modified set ${\cal D}_1$, with residuals shown in the right panel of  Fig.~\ref{fig:figA12}. The \gpi{} points, over-plotted with bigger grey symbols, reveal systematic shifts w.r.t. this best-fitting model. 

Figure~\ref{fig:figA13} illustrates the residuals in the (RA, DEC)-plane. All the \gpi{} points exhibit a systematic positive RA shift with respect to the model, and apart from one point, all of them have negative DEC deviations (the top-left panel). Moreover, most of the  data points deviate from the model by more than $3\sigma$ (the bottom-left panel in \ref{fig:figA13}). Observations ${\cal D}_1$, without the \gpi{} data, are distributed uniformly in the (RA, DEC)--residuals plane, as expected for a statistically valid solution (the right column). This may suggest a bias in the \gpi{} data w.r.t. the other measurements, yet of an unknown origin.

\begin{figure*}
\centerline{
\hbox{
\includegraphics[width=0.6\textwidth]{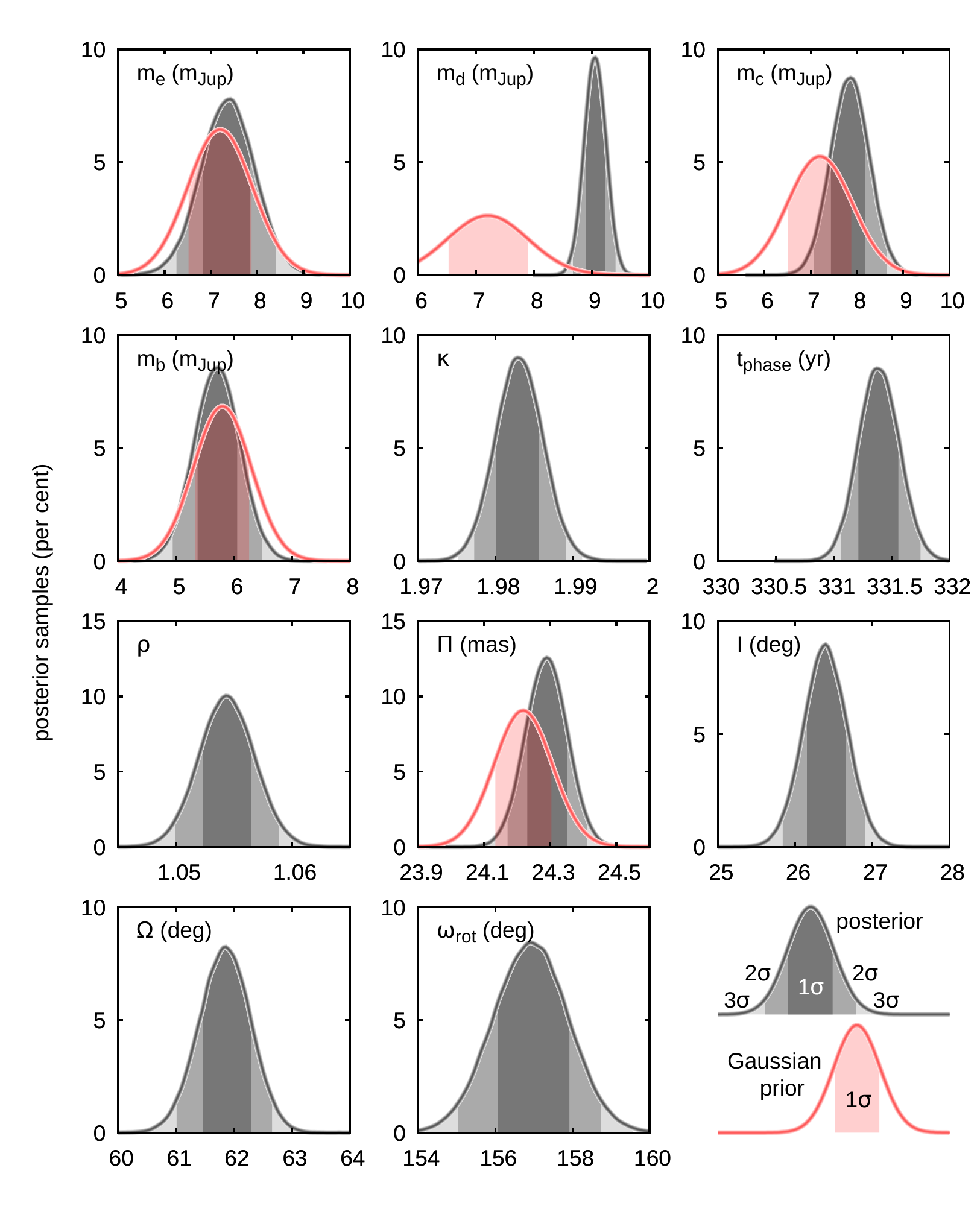}
}
}
\caption{
Posterior probability distributions for the free parameters of the \po{} model of the \hr8799{} system  (grey curves). The shades of grey are for $1\sigma, 2\sigma$ and $3\sigma$ confidence intervals. Red curves mark the Gaussian priors set for $5$ out of $11$ model parameters. The red areas under the curves indicate $1\sigma$ ranges of each parameter. The remaining $6$ parameters of the model have the uniform (non-informative) priors, not shown in the plot. \corr{See Table~\ref{tab:tab1} and the text in Appendix}.
}
\label{fig:figA8}
\end{figure*}

\begin{figure*}
\centerline{
\hbox{\includegraphics[width=0.45\textwidth]{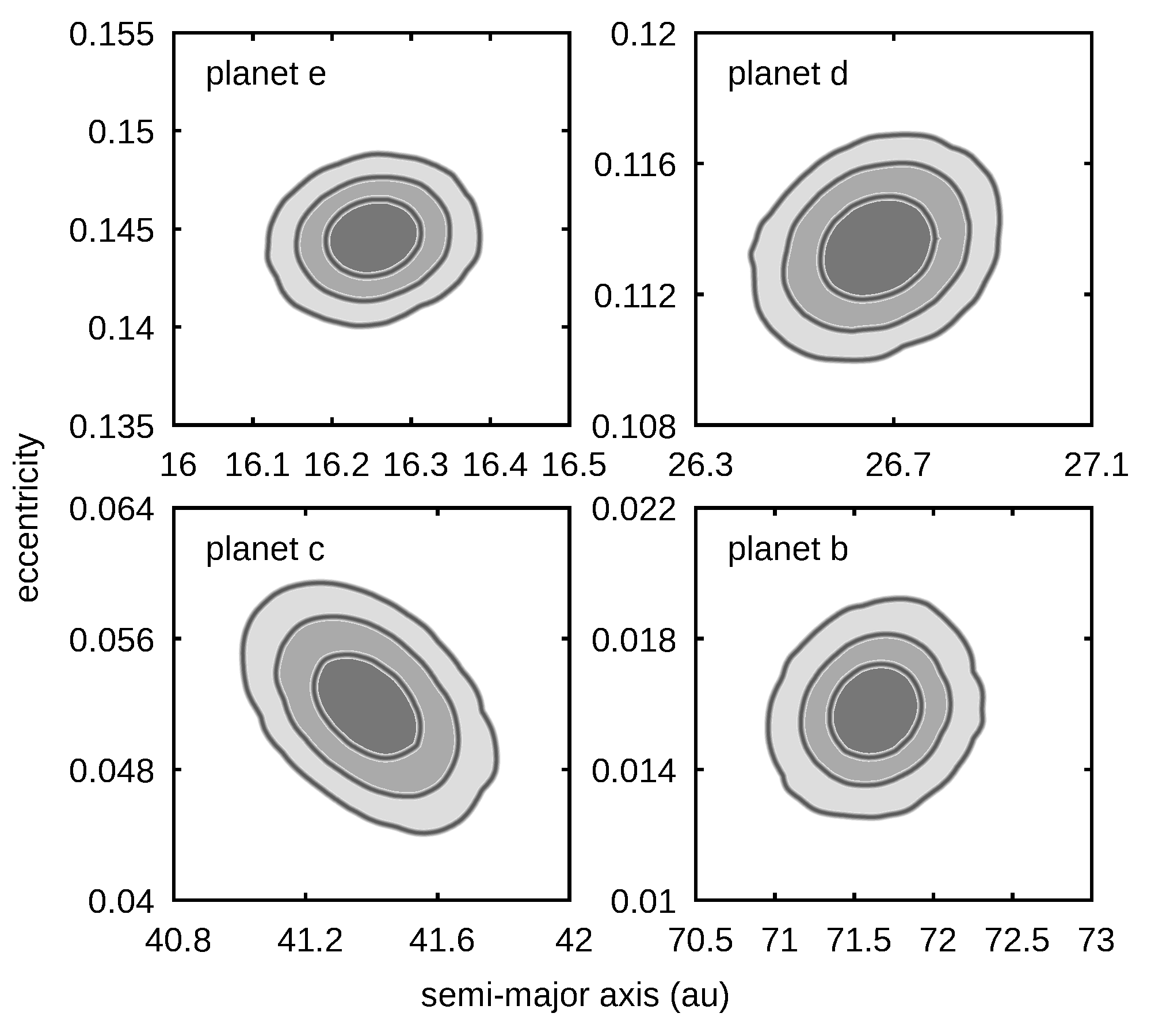}}
\qquad
\hbox{\includegraphics[width=0.45\textwidth]{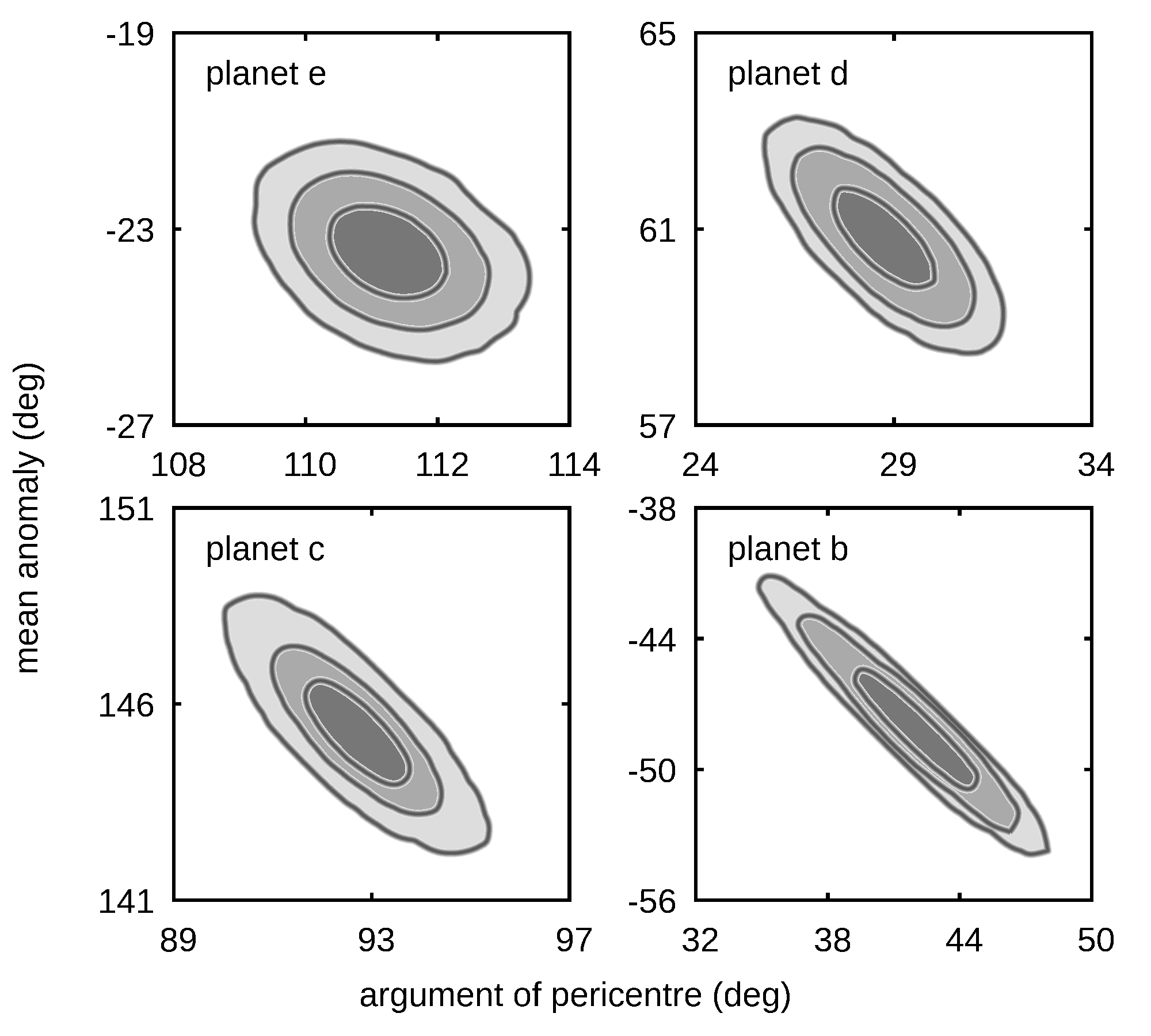}}
}
\caption{
Posterior probability distributions presented in diagrams of semi-major axis \textit{vs.} eccentricity (left four panels), as well as the argument of pericentre \textit{vs.} the mean anomaly (right four panels). The shades of grey, from darkest to lightest, represent  $1\sigma, 2\sigma$ and~$3\sigma$ confidence levels, respectively.
}
\label{fig:figA9}
\end{figure*}

\begin{figure*}
\centerline{
\vbox{
\hbox{
\includegraphics[width=0.36\textwidth]{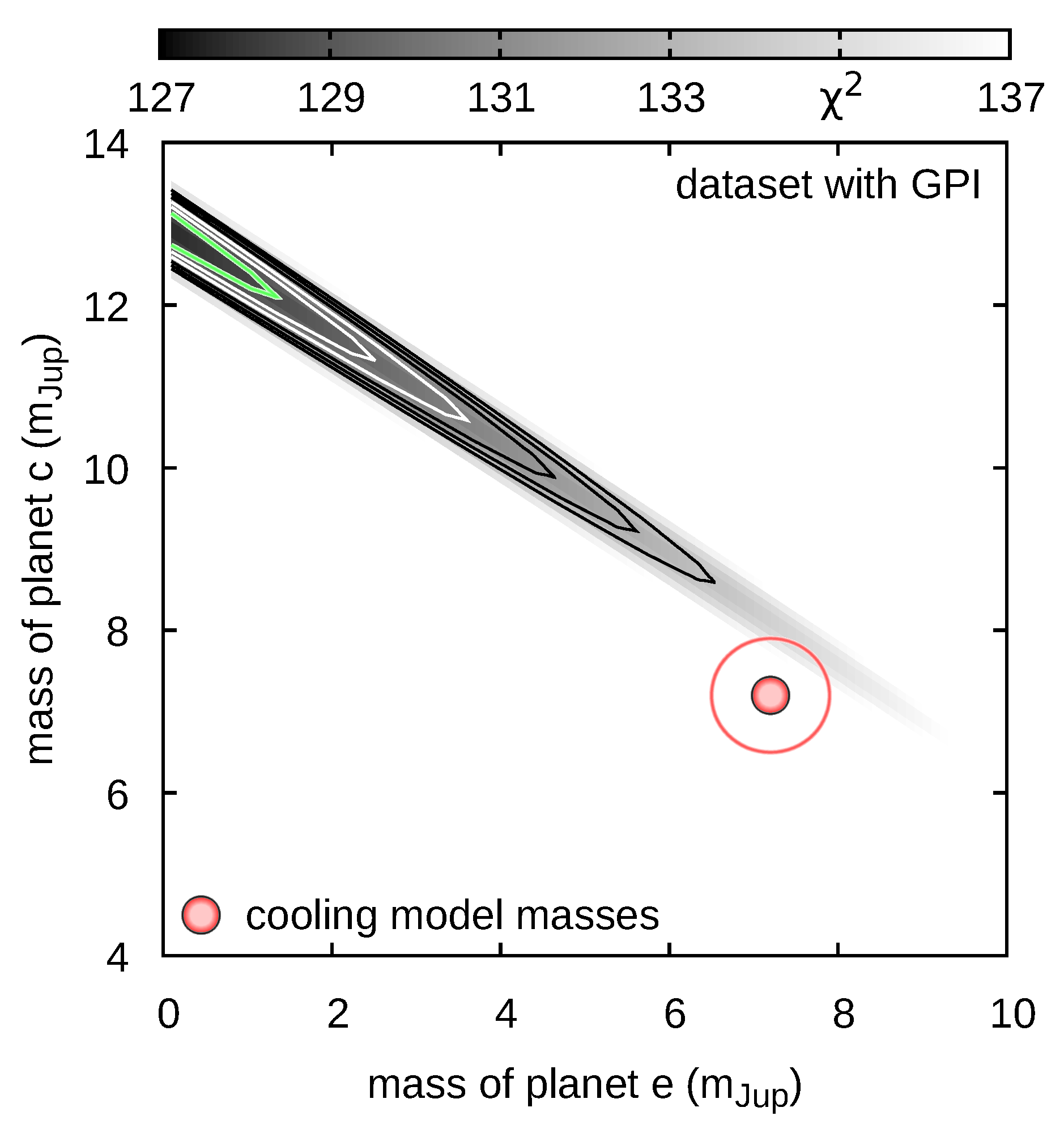}
\includegraphics[width=0.36\textwidth]{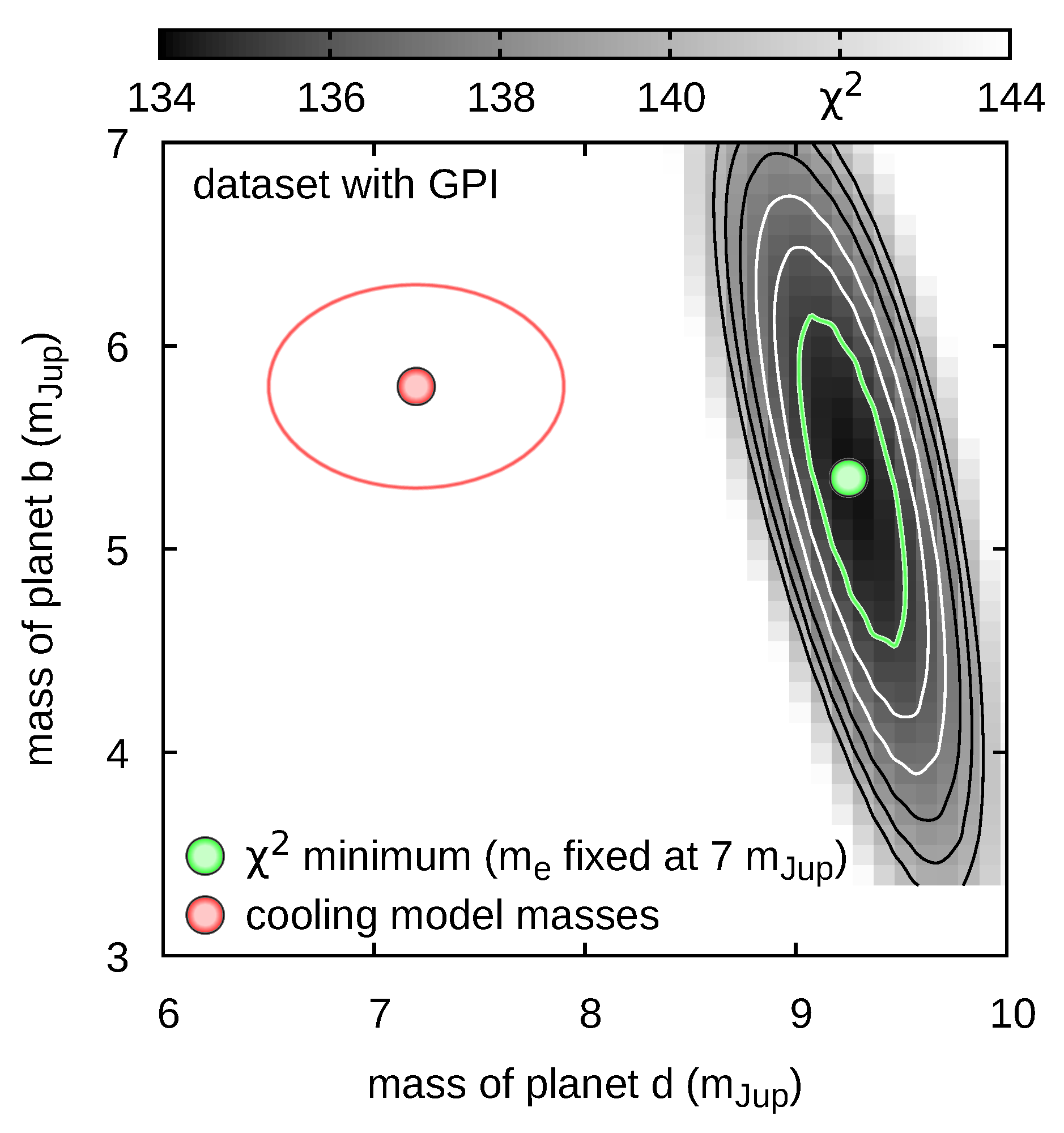}
}
\hbox{
\includegraphics[width=0.36\textwidth]{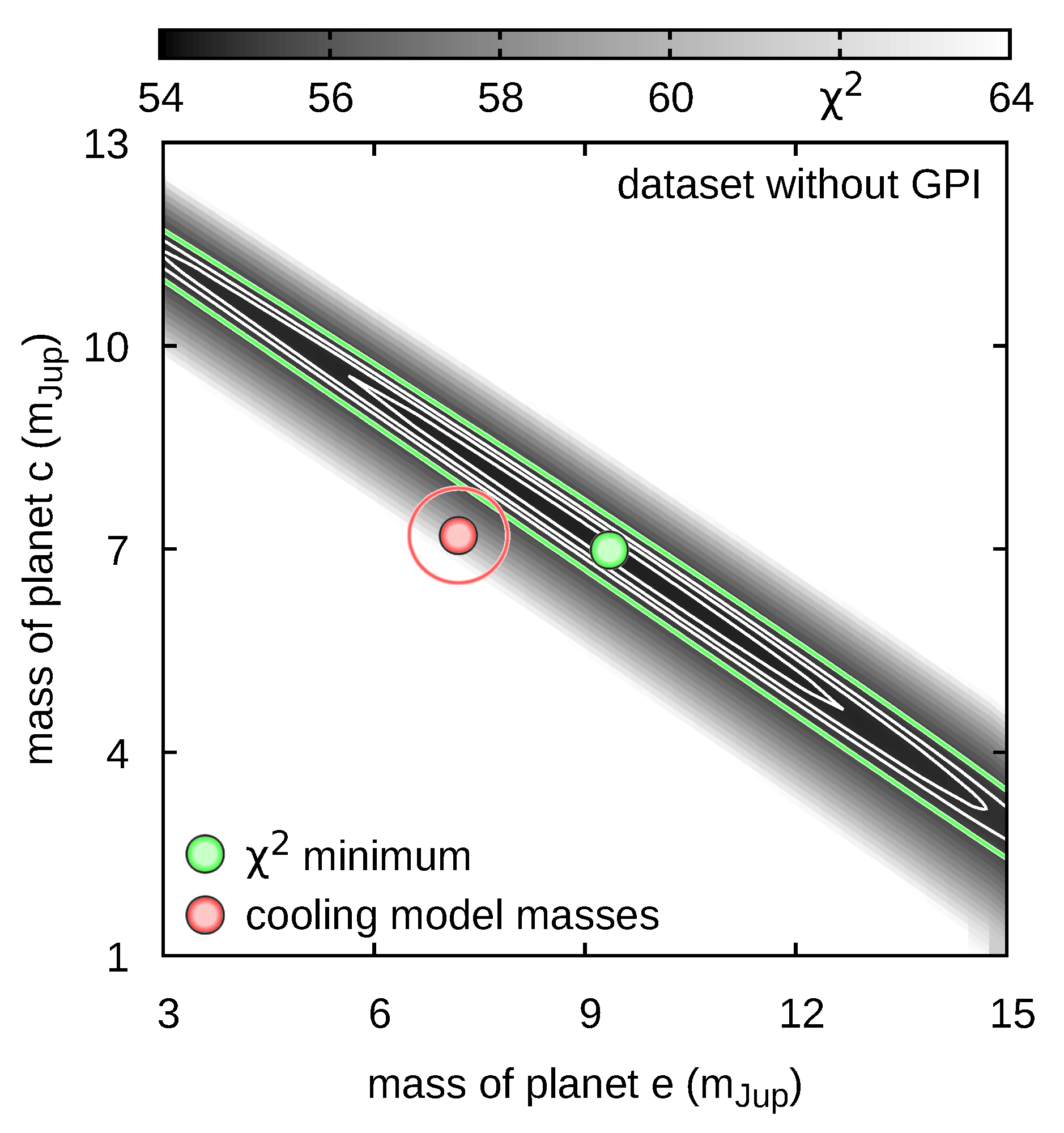}
\includegraphics[width=0.36\textwidth]{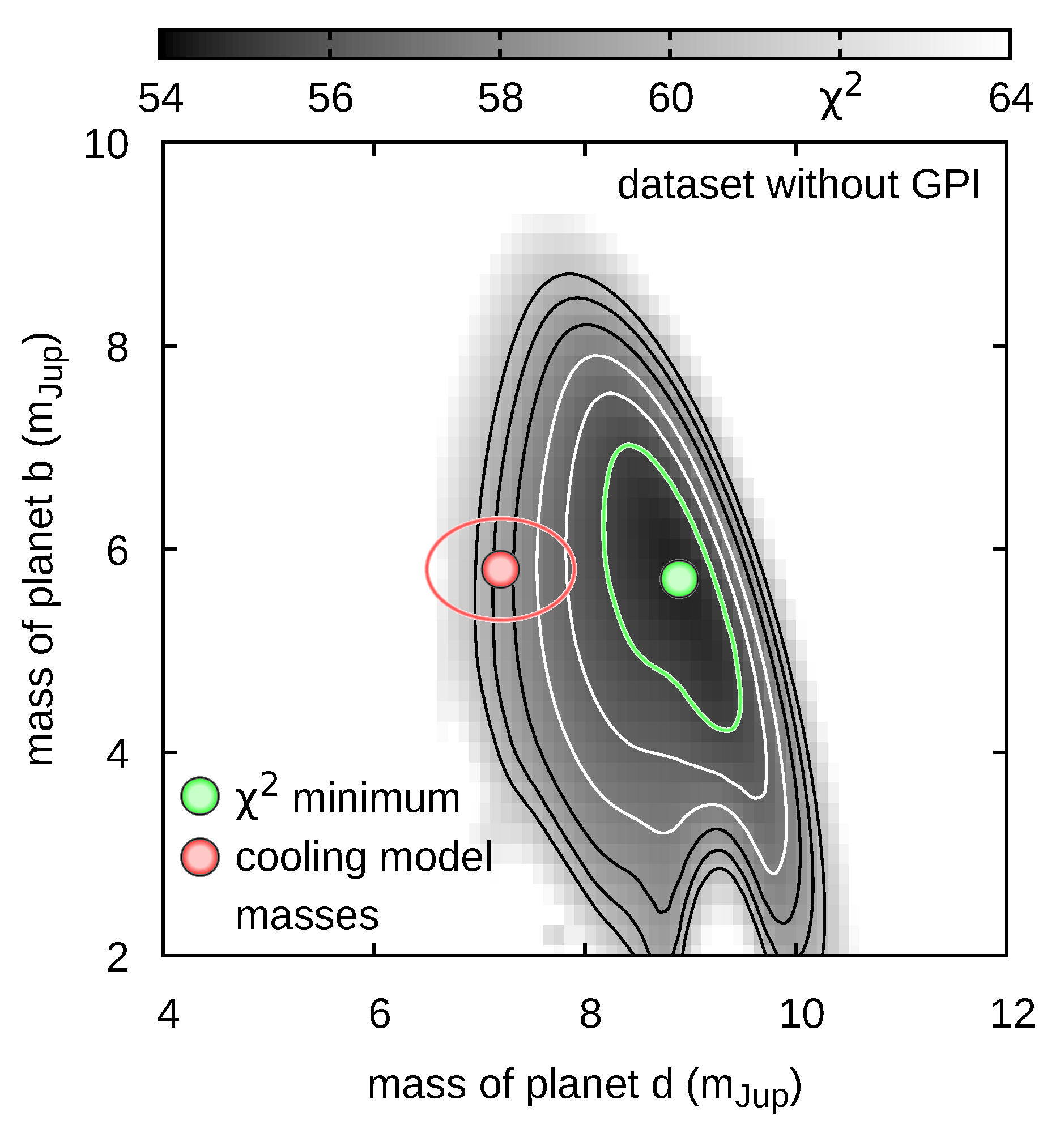}
}
}
}
\caption{
A $\chi^2$-scan of \po{} models in the plane of masses \hr8799{}e and~\hr8799{}c (left panel), and \hr8799{}d and~\hr8799{}b (right panel) for the data set with (top row, ${\cal D}$) and without the \gpi{} measurements (bottom row, ${\cal D}_1$). The red filled symbols and the circle/ellipse mark reference astrophysical masses $\me = (7.2 \pm 0.7)\,\mJ$ and $\mc = (7.2 \pm 0.7)\,\mJ$ (left panel) and $\md = (7.2 \pm 0.7)\,\mJ$ and $\mb = (5.8 \pm 0.5)\,\mJ$ (right panel), following \citep{Wang2018}. The green filled point denotes the position of the minimum of $\chi^2$ function, while the green contour denotes the level of $\min \chi^2 + 1$. The white and black curves denote the levels of $\min \chi^2 +2, \dots, +6$, apart from the bottom-left panel for which the white contours denote the confidence levels of $\min \chi^2 +0.1, +0.3, +0.5$.
}
\label{fig:figA10}
\end{figure*}

\begin{figure*}
\centerline{
\vbox{
\hbox{
\includegraphics[width=0.36\textwidth]{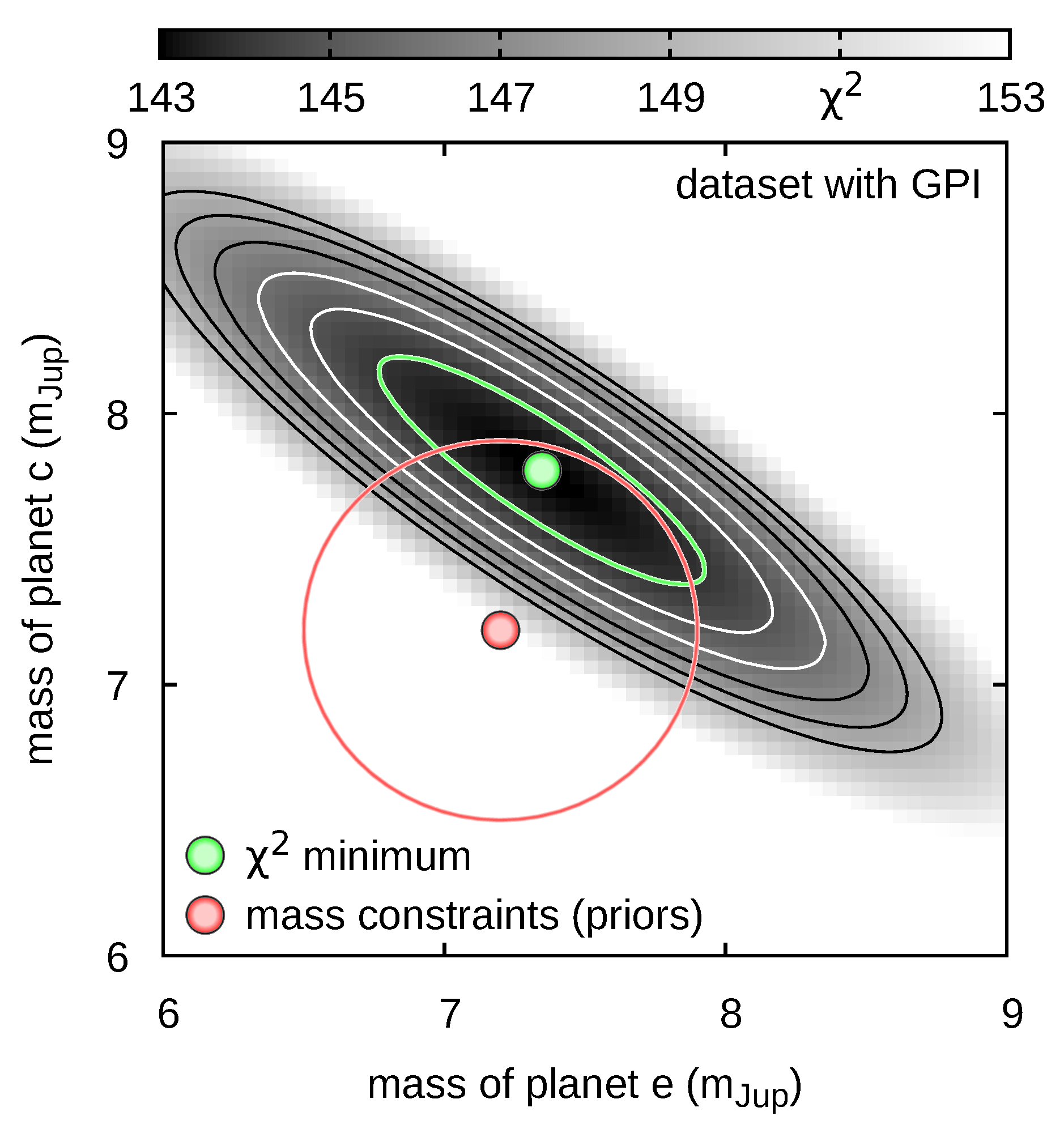}
\includegraphics[width=0.36\textwidth]{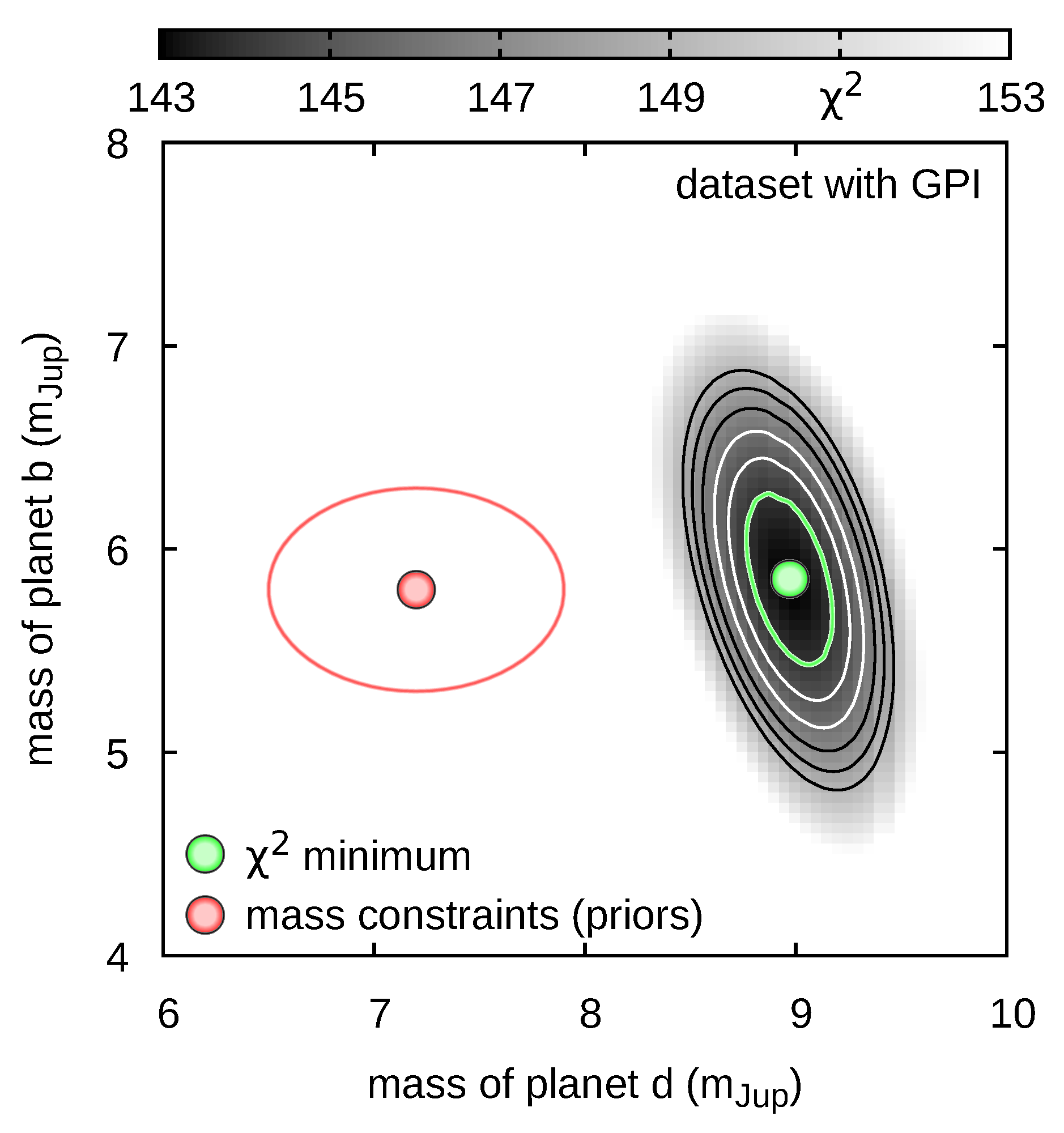}
}
\hbox{
\includegraphics[width=0.36\textwidth]{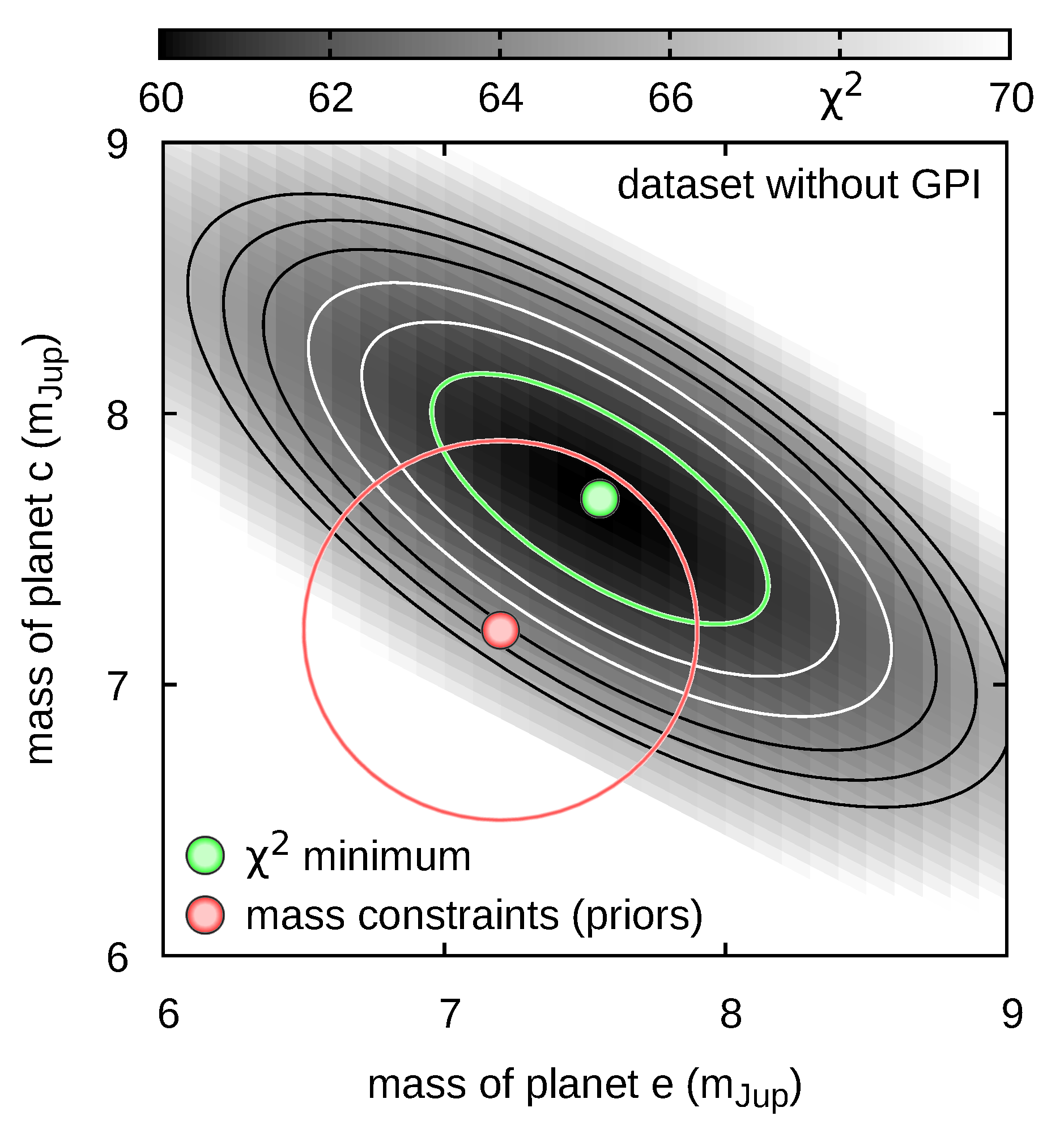}
\includegraphics[width=0.36\textwidth]{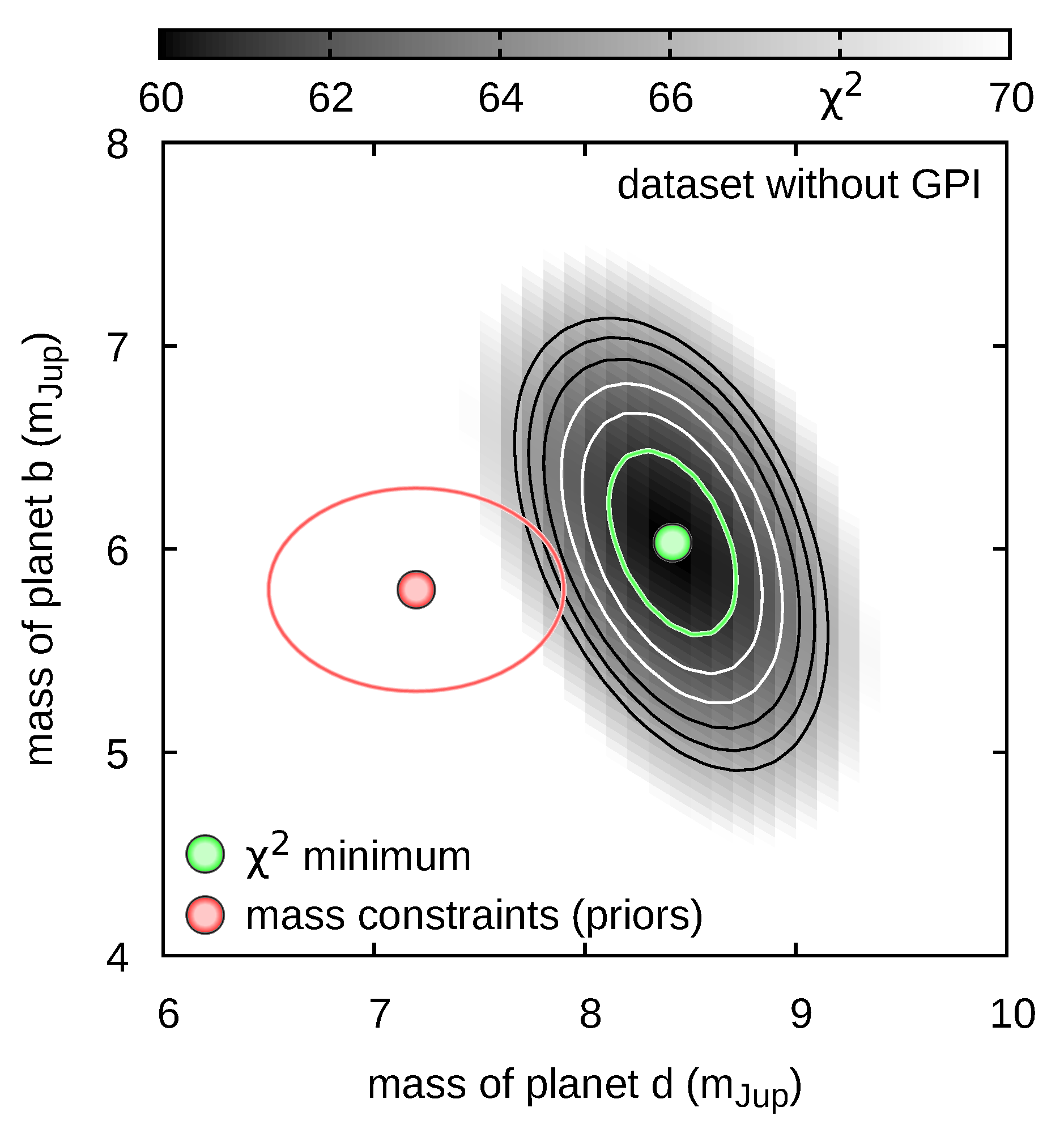}
}
}
}
\caption{
A $\chi^2$-scan of \po{} models in the masses plane of \hr8799{}e and~\hr8799{}c (left panel), and \hr8799{}d and~\hr8799{}b (right panel) for the data set with (top row, ${\cal D}$) and without the \gpi{} measurements (bottom row, ${\cal D}_1$), as well as with the the hot-start cooling theory priors. The red filled symbols and the circle/ellipse correspond to $\me = (7.2 \pm 0.7)\,\mJ$ and $\mc = (7.2 \pm 0.7)\,\mJ$ (left panel) and $\md = (7.2 \pm 0.7)\,\mJ$ and $\mb = (5.8 \pm 0.5)\,\mJ$ (right panel), following \cite{Wang2018}. The green filled point denotes the position of the minimum of $\chi^2$ function, while the green contour denotes the level of $\min \chi^2 + 1$. The white and black curves denote the confidence levels of $\min \chi^2 +2, \dots, +6$.}
\label{fig:figA11}
\end{figure*}

\begin{figure*}
\centerline{
\hbox{
\includegraphics[width=0.4\textwidth]{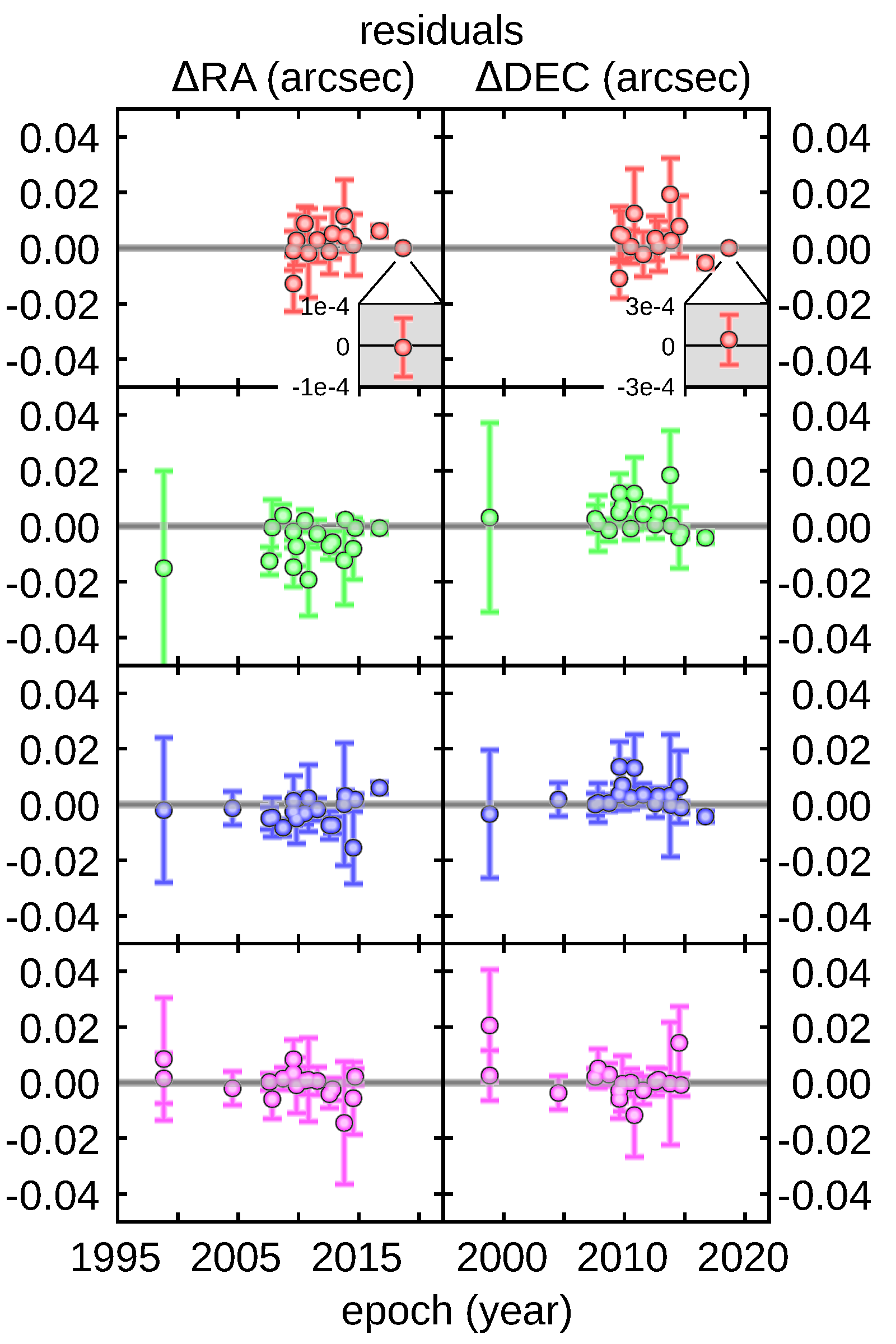}
\qquad
\includegraphics[width=0.4\textwidth]{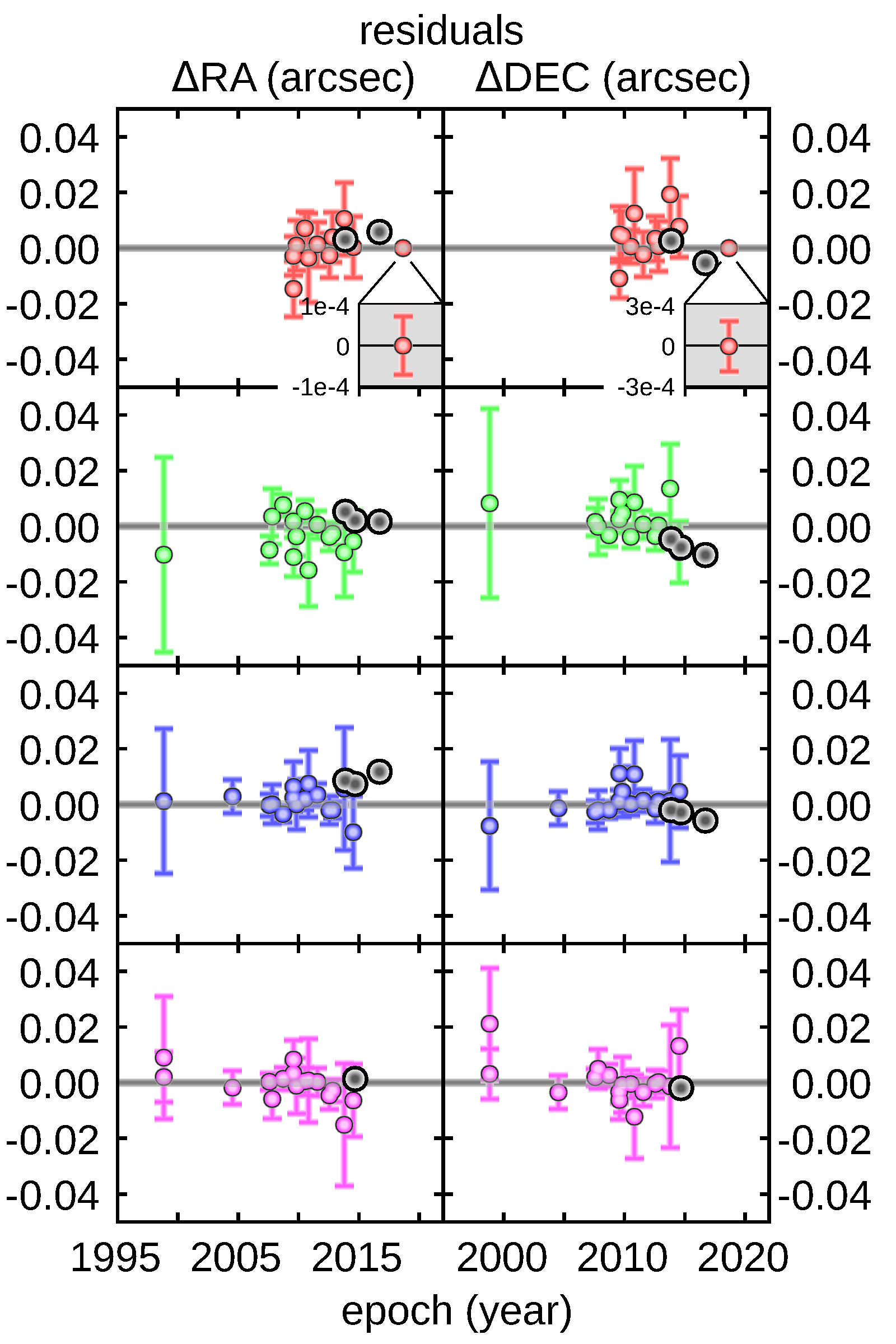}
}
}
\caption{
Residuals of the best-fitting model to the data set with (left panel, ${\cal D}$) and without  the \gpi{} observations (right panel, ${\cal D}_1$). In the right-hand panel the \gpi{} data, not included in the fitted data set, are marked with big grey symbols without error bars. The \gravity{} datum for \hr8799{}~e is additionally enlarged in the top row (grey rectangles).
}
\label{fig:figA12}
\end{figure*}

\begin{figure*}
\centerline{
\vbox{
\hbox{
\includegraphics[width=0.36\textwidth]{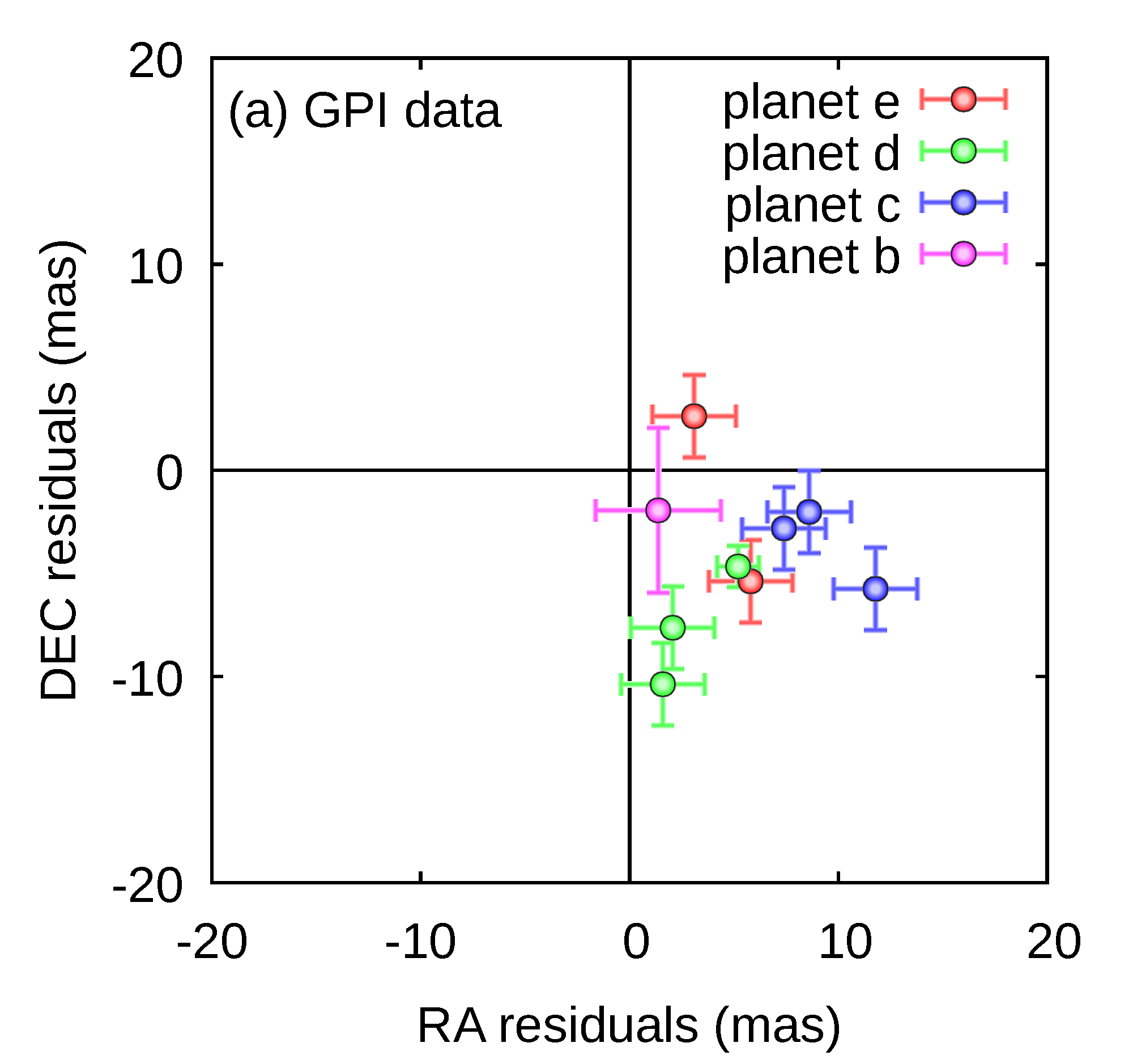}
\includegraphics[width=0.36\textwidth]{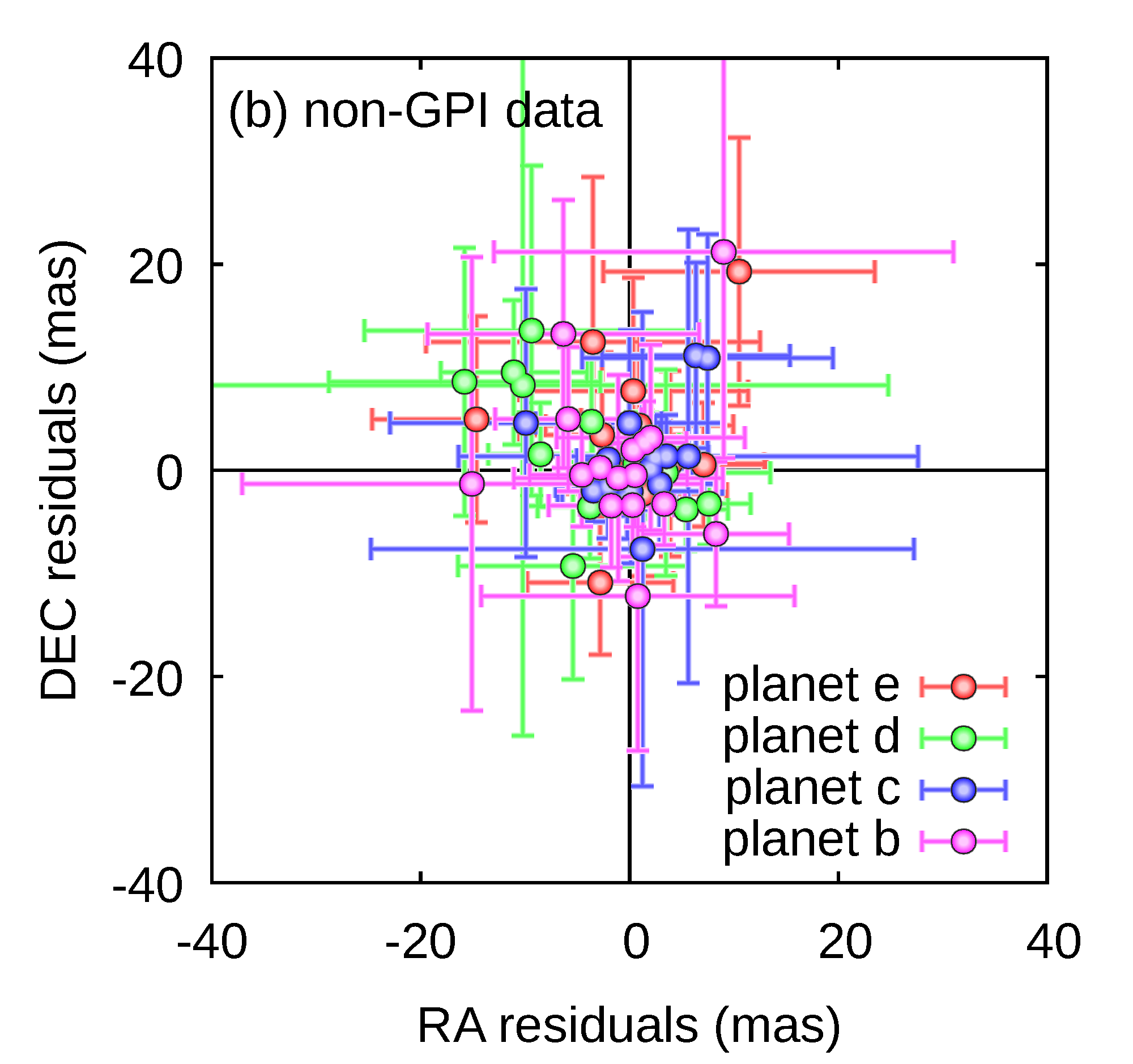}
}
\hbox{
\includegraphics[width=0.36\textwidth]{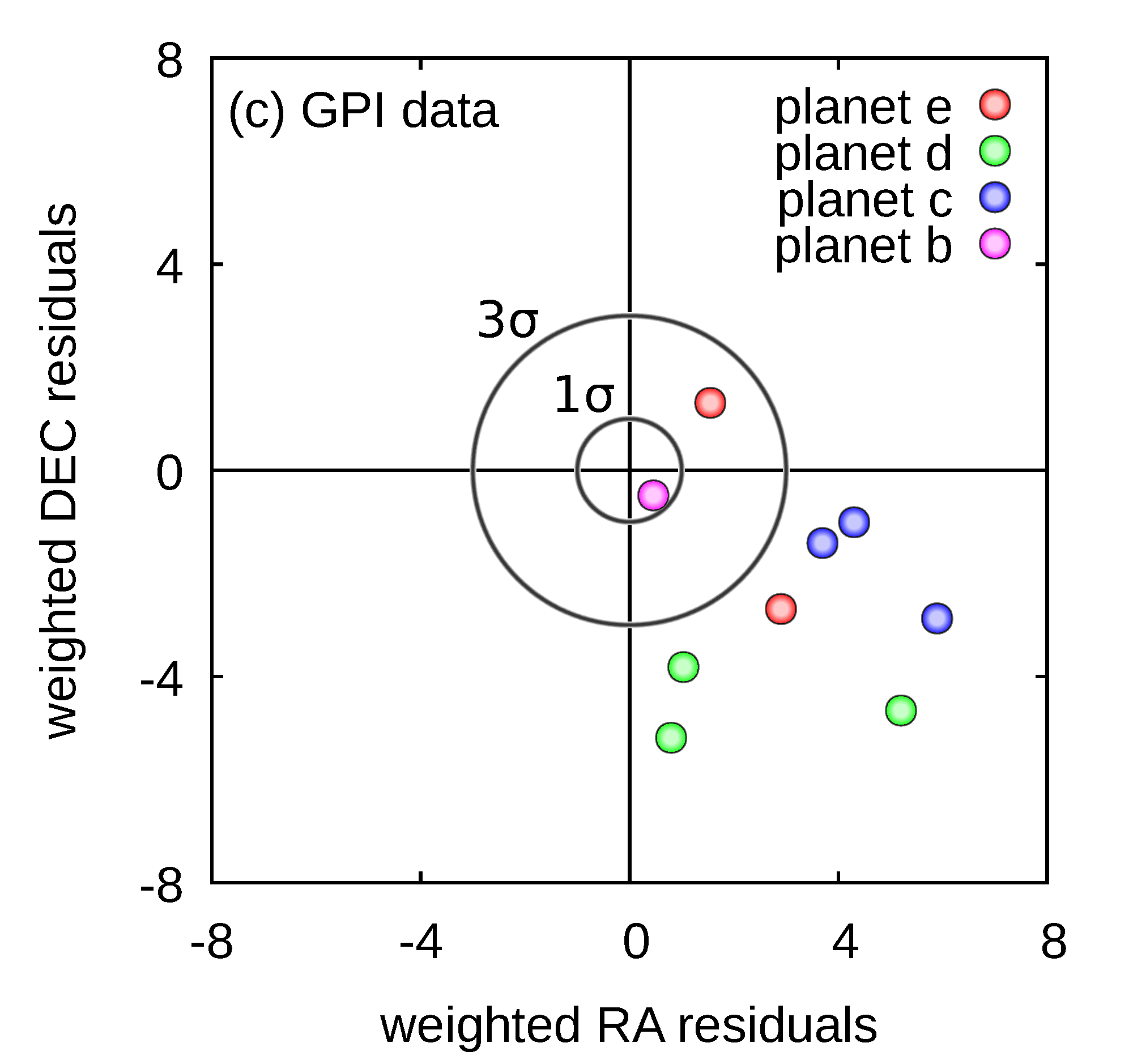}
\includegraphics[width=0.36\textwidth]{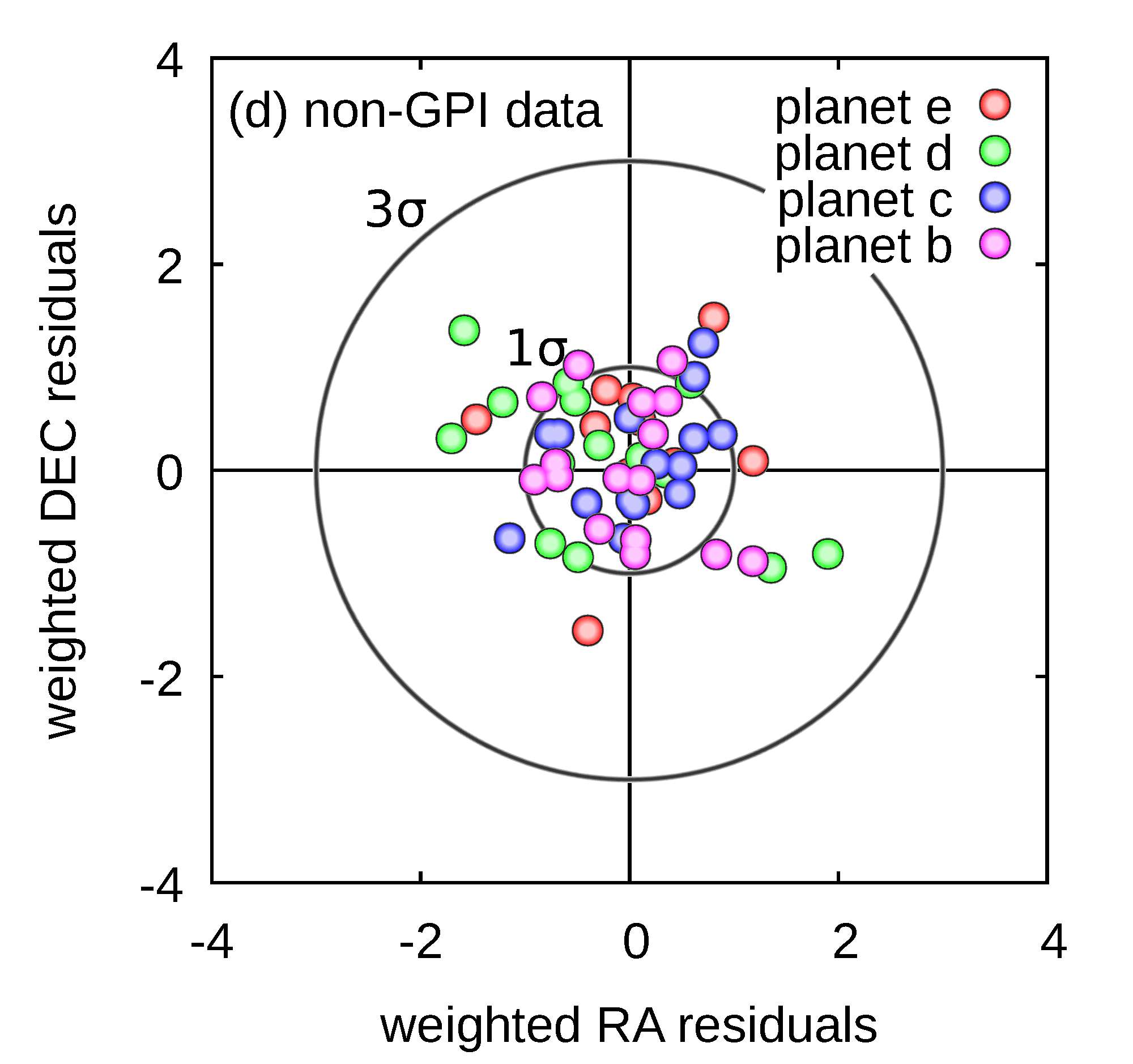}
}
}
}
\caption{
Residuals of the best-fitting model without the mass priors  and without the \gpi{} data. The top panels show the residuals together with their uncertainties, while the bottom panels show the residuals weighted with the uncertainties. Circles of radii equal to $1$ and $3$ marks $1\sigma$ and $3\sigma$ deviations of a given data point from the model. In the right-hand column, the fitted astrometric measurements are illustrated. Deviations of the \gpi{} data from the model are presented in the left-hand panels.
}
\label{fig:figA13}
\end{figure*}

The obtained  $\chi^2$ minima with mass priors for data set ${\cal D}$, presented in top row of Fig.~\ref{fig:figA11} overlap with the results of the DE-MC sampling around the best-fitting model illustrated in Figs.~\ref{fig:figA8} and \ref{fig:figA9}. As noted above, the best-fitting mass of planet \hr8799{}d is the only one significantly inconsistent with the priors from thermodynamical tracks by $\simeq 2\,\mJ$ (top row of Fig.~\ref{fig:figA11}). However, when the \gpi{} measurements are excluded from the data set, then the difference reduces by factor of $\simeq 2$, making the astrometric model results marginally consistent with the \hr8799{}d mass determined from the cooling theory (bottom row of Fig.~\ref{fig:figA11}). All masses become constrained much better for the reduced data set ${\cal D}_1$ than ${\cal D}$ and are marginally consistent with the astrophysical values, although their uncertainties are still significant. This experiment demonstrates the sensitivity of the astrometric model to the most accurate data points.  We also recall, that with added just one \gravity{}-like measurement for each planet close to the present epoch, the astrometric data alone might fully constrain the masses (see the main part, and Fig.~\ref{fig:figA1}).

%
\section{The debris discs simulation}
%
\label{sec:calibration}

\subsection{The $N$-body model of the debris discs}

Given the ongoing discussion in the literature, as summarized in the main part, we aim to resolve the dynamical structure of the debris disks composed of small, Kuiper-belt like objects. Such a structure may reflect unique characteristics implied by the strictly resonant motion of the planets. Here, we essentially follow the approach in GM18. The numerical model relies in determining the orbital stability of small-mass particles in the \hr8799{} system through resolving the chaotic or regular character of their motion with the \megno{} $\Ym$ fast indicator \citep{Cincotta2003}. We dubbed it the $\Ym$-model. As we found in GM18 with the long-term, direct $N$-body integrations, the $\Ym$-model reproduces closely the dynamical structure of the debris discs found with the direct integrations, yet in much shorter computation time. 

Here, we conducted an extensive $\Ym$-model simulation of the debris discs co-planar with the planets involved in the exact Laplace resonance (Tab.~\ref{tab:tab1}).  We considered three mixed fractions of asteroids with masses of $10^{-15}\mj$, similarly to GM18, as well as $10^{-10}\mj$ and $10^{-6}\mj$. As the initial Keplerian osculating elements, we randomly draw the semi-major axis $a_0 \in [10,400]$\,\au{}, the pericenter longitude and the mean  anomaly $\varpi_0,{\cal M}_0 \in [0^{\circ},360^{\circ})$. For the inner part of the disk ($a_0<100$~\au{}), we selected $e_0 \in [0,0.9]$, and for the outer part, beyond planet~\hr8799{}b, $e_0<1-(a_0/a_{\idm{b}})(1+e_{\idm{b}})$, i.e., under the collision curve of asteroids with \hr8799{}b in the $(a_0,e_0)$-plane.  We integrated the equations of motion and the variational equations for the whole $N$-body system of the observed planets (Table \ref{tab:tab1}, primaries), updated by a test particle, with the Bulirsh-Stoer-Gragg (BGS) integrator. The local and absolute accuracy of the integrator set to $\epsilon=10^{-13}$ provided the relative energy error as small as $10^{-9}$ for the total integration time of 10~Myrs. The BGS algorithm has been proved reliable for collisional and chaotic dynamics, which may be anticipated on the basis of previous works  (GM18). 

Concerning the appropriate integration time required to reliably characterize the orbits of asteroids, we note that the four, massive planets are locked deeply in the Laplace resonance (Fig.~\ref{fig:figA2}), and each planet is located in the center of the stability zone (Fig.~\ref{fig:figA3}). The system stability is robust to perturbations of quite massive additional companions (see also simulations in GM18 for ``asteroid'' masses as large as $1$--$2\mj$). Therefore, the $\Ym$ integrations of the best-fitting initial condition extended by the elements of a test asteroid reveal the dynamical character of its motion, and the orbits of the primaries are not affected. 

The geometric structure of the debris discs is illustrated in Figs.~\ref{fig:fig4}, \ref{fig:figA14} and \ref{fig:figA15}. In the numerical experiment, we collected $\simeq 3.3\times 10^6$ $\Ym$-stable orbits. Astrocentric positions of asteroids are marked at  at the end of integration time (top-left panel of Fig.~\ref{fig:figA15}) and at the initial epoch (top-right panel in Fig.~\ref{fig:figA15}), and colour-coded, according to their osculating eccentricity. Such snapshots represents a population of quasi-periodic and resonant orbits of the asteroids with various orbital phases and eccentricity, while their semi-major axes may overlap. We note, following GM18, that the orbits might be potentially present in the real system, but the actual population of asteroids may depend on the formation history of the whole system, its migration history, as well as locally variable density of asteroids.

In regions interior to, and beyond orbit of planet \hr8799{}b, the test orbits are extremely chaotic besides particular resonant solutions. Such $\Ym$-unstable orbits are also strongly unstable in the Lagrangian (geometric) sense --- particles are ejected or collide with the primaries in the time scale of a~few ~Myrs only.  We found this after testing the semi-major axis--eccentricity evolution in time for orbits selected in a strip of 1000 initial conditions marked with red filled points in  the right panel of Fig.~\ref{fig:figA14}. It shows the  proper (canonical) elements  \citep{Morbidelli2001} of dynamically stable asteroids in the semi-major axis--eccentricity plane $(a_0,e_0)$, see the right panel. \corr{In order to study unstable motions, test particles were randomly placed} under the collision curve with planet \hr8799{}b.  The initial eccentricity of their orbits is  slightly larger than the respective limit of $\Ym$-stable motions, and the initial semi-major axes $a_0 \in [100,400]$\,\au{} as well as initial phases are also random. We investigated closely orbits of all these test asteroids by integrating them for 10~Myrs. In this set, 466 asteroids collided with planet \hr8799{}b, 128 objects collided with the star, and 351 asteroids were ejected from the system beyond 5000~\au{}, leaving the radius of $800$\,\au{} typically in a few~Myrs, and less than the maximum interval of 10~Myrs. Only $\simeq 50$ objects located in stable, resonant regions survived for the maximum integration interval.

Moreover, with the Modified Fourier Transform or the Fundamental Frequency Analysis \citep{Nesvorny1996} of the canonical Jacobi elements $z_{i} = e_{i} (\cos \varpi_{i} +\sqrt{-1} \sin \varpi_{i})$, ($i=\mbox{b},\mbox{c},\mbox{d},\mbox{e}$), illustrated in the bottom-left panel in Fig.~\ref{fig:figA2}, we computed the frequency spectrum of planet pericenters rotation $\dot \varpi(t)$. Since the motion of the planets is strictly periodic, the $z_i(t)$ signals involve common leading frequency  $f_{\varpi} = -440.418''$~yr$^{-1}$ equivalent to the retrograde rotation of the system with the period $P_{\varpi} \simeq 2942.66$~years, i.e., only $\simeq 6$ orbits of planet \hr8799{}b. Besides the leading frequency there are a~few even larger, with periods smaller than $1000$~years.

Since the dynamics is governed by short-term MMRs and, possibly, secular resonances, we could fix the same integration time of $10$~Myr across the whole
disk. That integration time corresponds to $\simeq$20,000 orbital periods of the outermost planet and roughly 12,000 orbits at $\simeq 100$~\au{} which is sufficient to resolve the dynamical character of asteroid orbits. Particles marked as $\Ym$-stable for that interval of time should persist for more than 10~times longer interval in Lagrange-stable orbits, roughly $100$--$160$~Myr (see GM14, GM18), which is much longer than typical estimates of the parent star lifetime, $\simeq 42$~Myrs \citep{Wang2018}, in the 30--60~Myr range earlier adopted in \citep{Marois2010}. At the outer edge of the disk $\simeq 430$~\au, as determined by \cite{Booth2016}, the integration interval translates to a~few thousands of orbital periods, which is still meaningful to determine the stability border in the $(a_0,e_0)$-plane, as we justified above.  Moreover, given strong instability generated by the short term interactions, the $\Ym$ integrations may be stopped as soon as $\Ym>5$, sufficiently different from $\Ym\simeq 2$ for stable systems. That makes it possible to examine large sets of a few $10^7$ test orbits, orders of magnitude larger than they could be sampled with the direct $N$-body integrations. The most complex and interesting parts of the debris discs may be then mapped in detail with the $\Ym$-model. 

 
\subsection{Dynamical structure and features of the debris discs}

\begin{figure*}
\centerline{
\hbox{
\hbox{\includegraphics[width=0.5\textwidth]{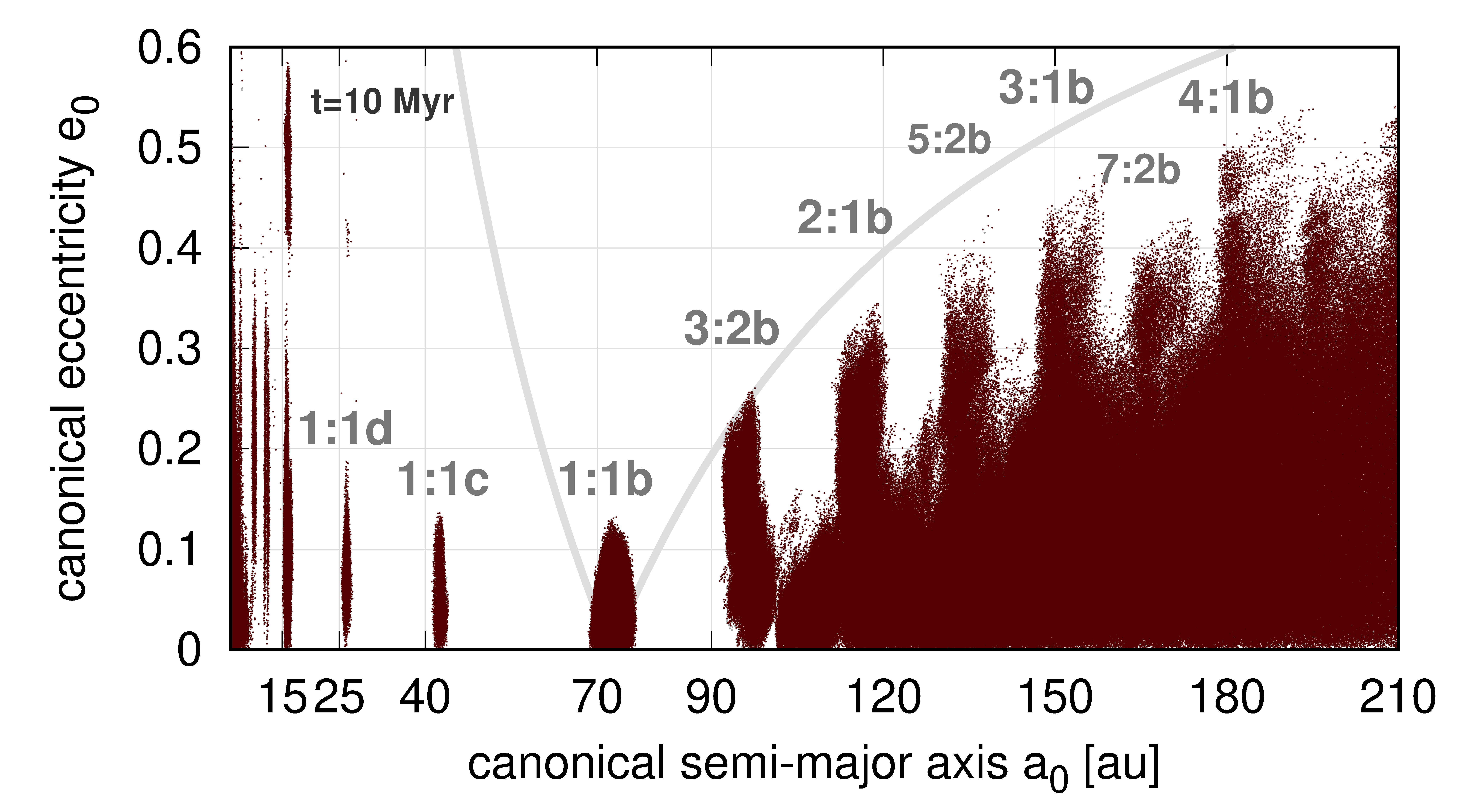}}
\hbox{\includegraphics[width=0.5\textwidth]{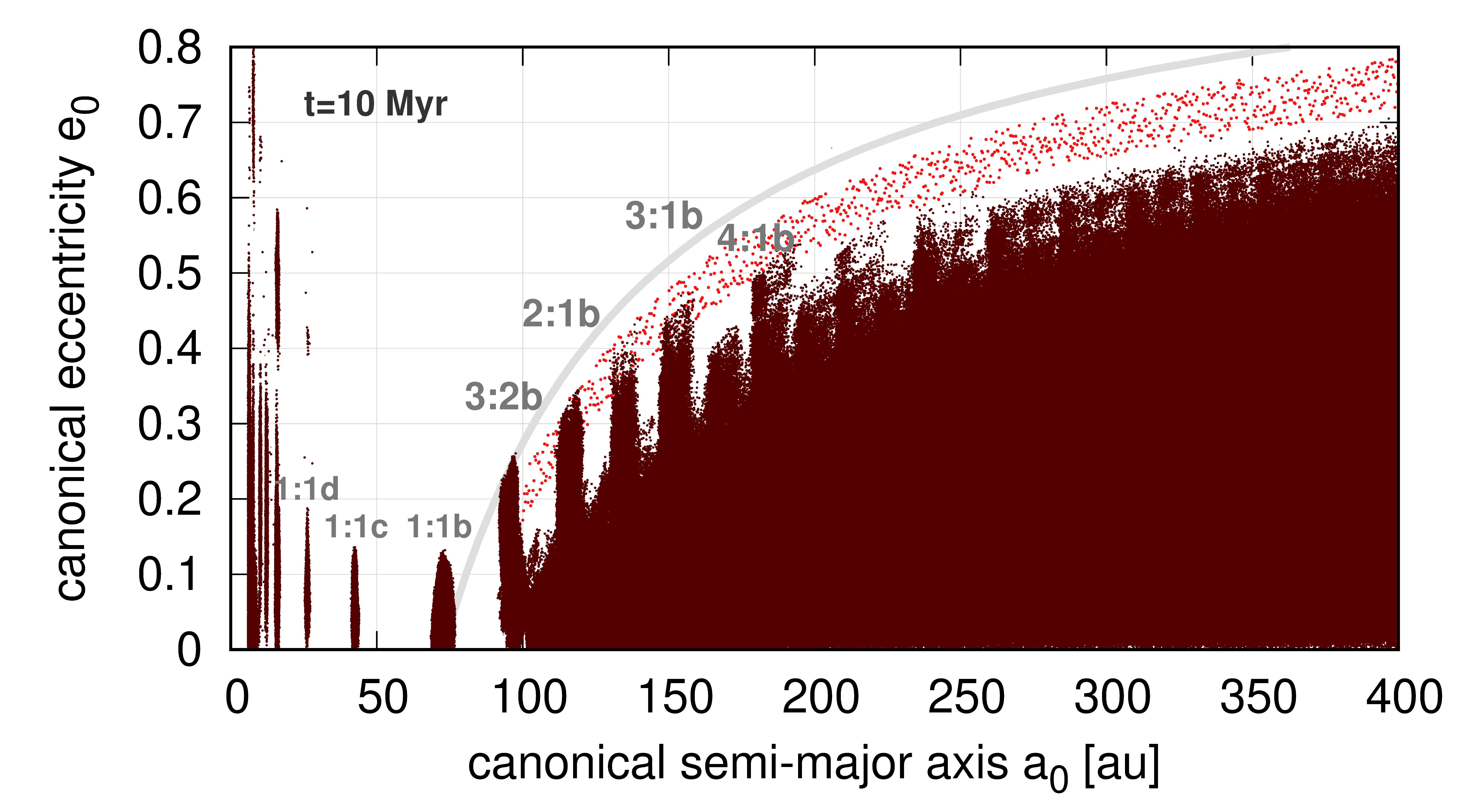}}
}
}
\caption{
Canonical, Poincar\'e elements $(a_{\rm 0},e_{\rm 0})$ of $\Ym$-stable solutions at the end of the integration interval of 10~Myr.  Grey lines are for the collision curve of orbits with planet \hr8799{}b. Approximate positions of a few low-order MMRs with planets \hr8799{}b and \hr8799{}c are labeled. The left-hand panel is for the inner part of the
disk, and the right panel is for the whole simulation. Red filled circles in the right-hand panel illustrate 1,000 initial condition of test orbits, analysed in order to explain wide instability zone below the collision curve.
}
\label{fig:figA14}
\end{figure*}

Regarding the inner part of the system, we found the same irregular inner boundary of the outer disk, similarly to simulations in GM18. In order to understand this feature, we analyse the $(a_0,e_0)$--diagram, shown in Fig.~\ref{fig:figA14}. Comparing the left-hand panel of Fig.~\ref{fig:figA14} with the disk structure illustrated in
Fig.~\ref{fig:fig4}, we find that the inner edge of the outer disk is significantly asymmetric due to low and moderate eccentricity orbits in the 1:1 and 3:2~MMR with planet \hr8799b{}. Low density of asteroids around $\simeq 110$~\au{} appears due to unstable 2:1~MMR and higher order resonances forming a~kind of thickening ``comb'' with increasing semi-major axis. It forms a border of stable orbits shifted below the collision curve with planet \hr8799{}b by a substantial value of $\sim 0.1$. We can now understand and interpret the strongly unstable orbital evolution of the tested asteroids in this zone (Fig.~\ref{fig:figA14}, the right panel). The strong instability is caused by overlapping two-body MMRs, multi-body MMRs and (possibly) mixed secular--mean motion resonances. The pericenter frequency of the system is commensurate with the mean motion of asteroids $n_0$ in this zone, for instance, regarding absolute values of the frequencies, $1f_{\varpi}:1 n_0$ at $\simeq 235$\,\au{}, $3f_{\varpi}:2 n_0$ at $\simeq 310$\,\au{}, $2f_{\varpi}:1 n_0$ at $\simeq 375$\,\au{}, $3f_{\varpi}:1 n_0$ at $\simeq 490$\,\au{}. However, the resonances are retrograde for the disk rotating with the same spin direction, as the planets, therefore we did not observe their direct or clear dynamical influence on the asteroids. A streaking feature of stable zone beyond \hr8799b{} is the presence of low density rings, which could be identified with higher-order resonances with this outermost planet, such as 2:1, 3:2, 3:1 and 5:2, extending up to $\simeq 200$\,\au{} (Fig.~\ref{fig:figA14} and Fig.~\ref{fig:fig4} in the main part).

A stable 1:1~MMR with planet \hr8799{}b forms huge, symmetric Lagrangian areas of low eccentricity objects extending for 70--80\,\au{} and $\simeq 10$\,\au{} across. 
The Lagrangian 1:1 MMRs governed by inner planets are non-symmetric in respective pairs. There are also islands of the 2:1~MMR and 3:2~MMR with  \hr8799{}d and \hr8799{}e. In these islands, eccentricity of the asteroids reaches $e_0 \simeq 0.8$ (yellow colour in Fig.~\ref{fig:fig4}). The outer, continuous edge of the inner debris disk appears at $\simeq 8$\,\au{}. (The dynamical structure of the inner disk was investigated in more detail in GM18).

\begin{figure*}
\centerline{ 
\vbox{
\hbox{
\includegraphics[width=0.48\textwidth]{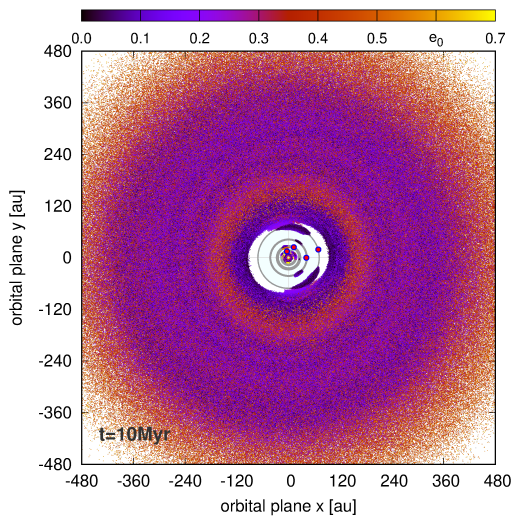}
\includegraphics[width=0.48\textwidth]{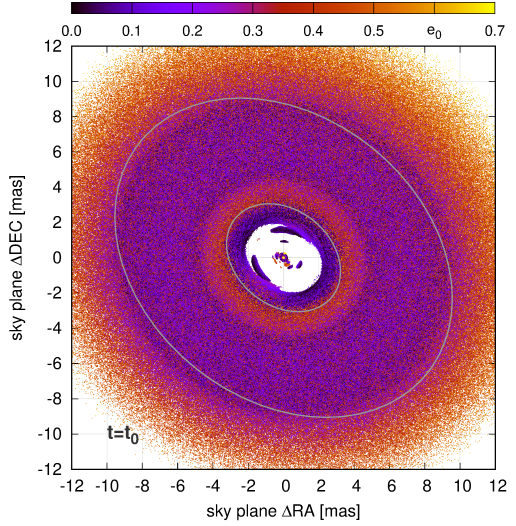}
}
\hbox{
\includegraphics[width=0.48\textwidth]{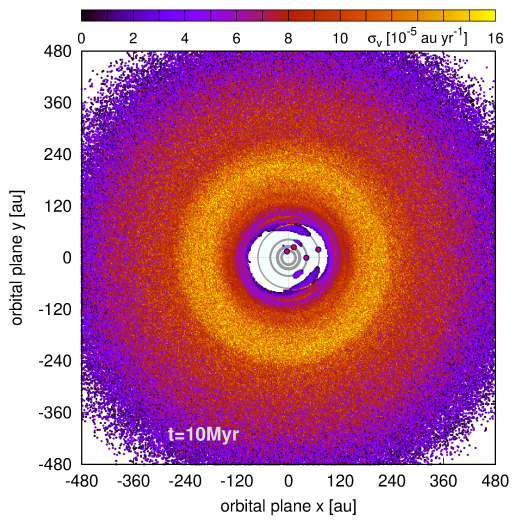}
\includegraphics[width=0.48\textwidth]{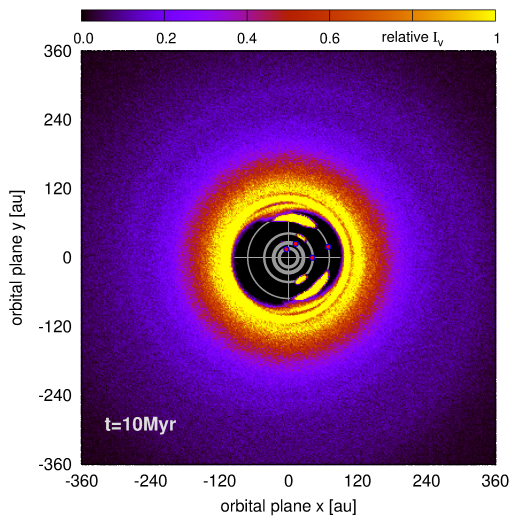}
}
}
}
\caption{
Top-left panel: the global view of debris disk revealed by $\simeq 3.3\times 10^6 \Ym$-stable orbits of the whole simulation illustrated as a snapshot of astrocentric coordinates $(x,y)$ of asteroids at the end of integration interval of 10~Myrs. Osculating orbital eccentricities $e_{\rm 0}$ of these orbits are color-coded and labeled in the top bar. Initial positions of planets are marked with filled circles. Gray rings illustrate their orbits integrated for 10~Myr.
Top-right panel:  similar to Fig.~\ref{fig:fig4} and the top-left panel, 
but the initial astrocentric $(x,y)$-coordinates of asteroids in $\Ym$-stable orbits are rotated by the inclination and nodal angles of the initial condition in Tab.~\ref{tab:tab1}.  Gray ellipses illustrate the disk
boundaries $r_{\idm{in}}=145\,\au$ and $r_{\idm{out}}=429\,\au$, respectively, fitted in \citep{Booth2016}, and rotated by inclination $I=41^{\circ}$ and nodal angle $\Omega=50^{\circ}$ derived in this paper.
Bottom-left panel: velocity dispersion in the debris discs evaluated at $2\,\au\times2\,\au$ bins and color-coded over initial astrocentric $(x,y)$-coordinates of asteroids in $\Ym$-stable orbits.
Bottom-right: The relative Planck intensity of the outer disk, $I_\nu \sim K \Sigma(r) r^{-1/2}$, as in \cite{Read2018}, where $\Sigma(r)$ is proportional to the number of asteroids in $2\,\au\times 2\,\au$ bins, the same as calculated in the velocity dispersion plot in
the bottom-left panel. 
}
\label{fig:figA15}
\end{figure*}

In the top panels of Fig.~\ref{fig:figA15}, we present the global view of the 
debris discs revealed by $\simeq 3.3\times 10^6 \Ym$-stable orbits in the 
whole simulation. Similarly to Fig.~\ref{fig:fig4}, the panels represent snapshots of astrocentric coordinates $(x,y)$ of  asteroids and their osculating orbital eccentricities $e_{\rm 0}$  (color-coded and labeled in the top bar), at the initial epoch (the right panel) and at the end of the integration interval of 10~Myrs (the left panel).  We selected the end epoch in order to illustrate a saturation of asteroids after a substantial interval of thousands of orbital periods. Initial positions of the planets are marked with filled circles. For a reference, gray rings illustrate their orbits integrated for the same interval of 10~Myr, with the initial conditions in Table~\ref{tab:tab1}, independently of the disk integrations. 

\corr{The top-left panel of Fig.~\ref{fig:figA15} shows the debris disks in the orbital plane at the final epoch $t=10$~Myr, and the top-right panel is for the sky-view of the disks at the initial epoch $t_0$, rotated by the inclination and nodal angle in the initial condition (Tab.~\ref{tab:tab1}), respectively.}
These global representations for the outer disk reveal a ring of  high-eccentric orbits between $\simeq 140$\,\au{} and $\simeq 200$\,\au{} and broad outer ring forming a diffuse outer edge of the disk.  We note that the inner ring is substantially shifted with respect to the inner edge of the disk found at $\simeq 90$\,\au{}. The dynamical structure of the whole disk is also illustrated in the $(a_0,e_0)$-plane of the canonical Poincar\'e elements in the right plot of Fig.~\ref{fig:figA14}. 
In the top-right panel of Fig.~\ref{fig:figA15}
we also marked the inner and outer boundary of the disk model in \cite{Booth2016}, according with their estimate of the inclination $I=41^{\circ}$ and nodal angle $\Omega=50^{\circ}$. These values appear substantially different from the inclination and nodal angle of the best-fitting elements of the planetary system orbital plane (Tab.~\ref{tab:tab1}). 

In order to interpret the ring structure around $\simeq 150$\,\au{}
we plotted (not shown here) the canonical, osculating eccentricity $e_0$ vs. the astrocentric radius of particles at the epoch $t_0$, and also at $t_0+10$~Myr. They reveal that the excess of particles with high eccentricity seems to be a real feature, unlikely due to a particular sampling or plotting order of the particles. It is also clear that $e_0$ is a very steep function of the radius $r_0$ at the innermost part of the outer disk. 

Given the variation of eccentricity across the disk,  we computed  the Keplerian velocity dispersion of the particles. We binned asteroids in the region covering the whole
disk, $x,y \in [-480,480]$\,\au{} in square bins of $2\,\au{}\times2$\,\au{}. In each box, with non-zero number of particles, we computed 
$
 \sigma^2_v = \sum_i^{n} (v_i-\overline{v})^2/n,
$
where $v_i$ is the velocity module of a particle $i$ in the given bin, $n$ is the counted number of particles in this bin and $\overline{v}$ is the mean velocity module.
The results are illustrated in the bottom-left panel of Fig.~\ref{fig:figA15}. The ring structure associated with high eccentricity asteroids and the gradient of $e_0(r_0)$ implies the  velocity dispersion $\sigma_v$ a few times larger than in the inner parts of the disk. It could imply more intense dust production due to both locally larger density of objects and higher velocity during their collisions. We may note that the inner disk boundary fitted by \cite{Booth2016} seems to overlap with the eccentricity ring edge, which could suggest a systematic shift of the detected emission w.r.t. the actual dynamical border of the disk at $\simeq 100$\,\au{}. It might actually confirm the results of \cite{Wilner2018}, in their more recent model of the disk predicting the inner edge also at $\simeq 100$\,\au{}. Such border is better consistent with our updated orbital model of the \hr8799{} system, regarding the present parallax estimate in the GAIA DR2 catalogue, and the resulting, true linear dimensions of the system.

Finally, we simulated the relative intensity image of the disk. The relative intensity is defined the same as in \citep{Read2018},
$
I_{\nu(r)} \sim K \Sigma(r) r^{-1/2},
$
where $\Sigma(r)$ is the surface density and $K$ is the scaling factor. In order to estimate $\Sigma(r)$, we used the counts of asteroids in the same $2\times2$\,\au{} bins used for computing the velocity dispersion. The results are illustrated in the bottom-right panel in Fig.~\ref{fig:figA15}. The bright rings are associated with fractions of stable asteroids in the 3:2~MMR and 2:1~MMR with planet \hr8799b{}. 

While interpretation of the results needs more work, we might briefly conclude that the disk simulation reveals features related to the resonant character of the system. They
consits of asymmetry of the inner edge of the outer debris disk, and highly variable density of asteroids in its inner part, due to low-order MMRs with planet \hr8799{}b, including large Lagrangian clouds. There are also possible two rings of high-eccentricity asteroids around $140$--$160$\,\au{} and at the outer edge $\simeq 430\,\au$. These features may influence the intensity images used for modeling the emission in different wavelengths, and likely they should be accounted for in order to avoid biases in the emission models.

\end{document}